%% file: delaunay04.tex
\documentclass[]{elsart}

\usepackage{epsfig}
\usepackage{graphics}
\usepackage{graphicx}
\usepackage{amsmath}
\usepackage{url}
\usepackage{subfigure}

\newtheorem{proposition}{Proposition}
\newtheorem{finding}{Finding}

\begin{document}

\begin{frontmatter}

\title{Excitable Delaunay triangulations}

\author{Andrew Adamatzky}

\address{University of the West of England, Bristol, United Kingdom}

\date{\today}

\begin{abstract}
In an excitable Delaunay triangulation every node takes three 
states (resting, excited and refractory) and updates its state in discrete 
time depending on a ratio of excited neighbours. All nodes update their states 
in parallel. By varying excitability of nodes we produce a range of 
phenomena, including reflection of 
excitation wave from edge of triangulation, backfire of excitation, 
branching clusters of excitation and localized excitation domains. 
Our findings contribute to studies of propagating perturbations and 
waves in non-crystalline substrates.

\vspace{0.5cm}

\end{abstract}


\maketitle

\end{frontmatter}

\section{Introduction}

Given a finite set of planar points Delaunay triangulation is a planar proximity graph 
which subdivides the space onto triangles with  nodes in the given set such that the 
circumcircle of any triangle contains no points of the given set other than the 
triangle's vertices~\cite{delauanay}. Delaunay triangulation is a graph-theoretic 
dual of Voronoi diagrams~\cite{voronoi_1907}. It represents connectivity of Voronoi cells.  
Voronoi diagram and its dual Delaunay triangulation are widely used in studies related 
to filling a space with connected structural units. Voronoi diagram and Delaunay triangulation 
are used to approximate arrangements of discs~\cite{gervois_1995}, 
sphere packing~\cite{lochman_2006,filatovs_1998,luchnikov_2002}, 
to make structural analysis of liquids and gases~\cite{anikeenko_2004}, and 
protein structure~\cite{poupon_2004}, and to model dense gels~\cite{zarzycki_1992} and 
inter-atomic bonds~\cite{hobbs_1995}.  

We are interested in studying excitable Delaunay triangulation because they may provide
a good alternative to existing approaches of modelling unstructured unconventional computers~\cite{teuscher_2005}. 
Experimental research in novel and emerging computing paradigms and materials shows a great progress in designing laboratory 
prototypes of spatially extended computing devices. In these devices computation is implemented
by excitation waves and localisations in reaction-diffusion chemical 
media~\cite{adamatzky_rdc}, geometrically constrained and compartmentalized excitable substrates~\cite{gorecki_2006a,gorecka_2007,ben_gun,kaminaga_2006}, organic molecular 
assemblies~\cite{bandyo_2010}, and gas-discharge systems~\cite{astrov_2010}.
These unconventional computing  substrate can be formally represented by Delaunay triangulations 
with excitable nodes. Thus it is important to uncover most common  types of excitation dynamics 
on the Delaunay diagrams.

Despite being a ubiquitous graph representation of wide range of natural phenomena the Delaunay triangulation 
was not studied from automaton point of view. Will excitable Delaunay triangulation behave as a conventional
excitable cellular automata or there will be some unusual phenomena? We answer the question by slightly 
modifying classical Greenberg-Hasting model~\cite{greenberg_1978} and considering not only a threshold of 
excitation but also a ratio of excited neighbours as an essential factor of nodes' activation.    

The paper is structured as follows. We introduce automata on triangulations and excitation rules 
in Sect.~\ref{definitions}. In Sect.~\ref{properties} we discuss structural properties of automata 
triangulations. Sections~\ref{absoluteexcitability} and ~\ref{relativeexcitability} present classification 
of space-time dynamics of excitation for absolute (based on a number of excited neighbours) and 
relative (based on a ratio of excited neighbours) rules of excitation. Results are discussed in 
Sect.~\ref{discussion}.

\section{Delaunay automata and excitations}
\label{definitions}

Given a planar finite set ${\bf V}$ the Delaunay triangulation~\cite{delauanay} 
${\mathcal D}({\bf V})=\langle {\bf V}, {\bf E} \rangle$ 
is a graph  subdividing the space onto triangles with vertices  in ${\bf V}$ and edges in $\bf E$ 
where the circumcircle of any  triangle contains no points of $\bf V$ other  than its vertices. 
Neighbours of a node $v \in \bf V$ are nodes from $\bf V$ connected with $v$ by edges from $\bf E$. 
 
 The set $\mathbf V$ is constructed as follows. 
 We take a disc-container of radius 480 and fill it with up to 15,000 disc-nodes. We assume 
 that each disc-node has radius 2.5, thus a minimal distance between any two nodes is 5. 
 The Voronoi diagram, and its dual triangulation, are appropriate representations of such
 identical sphere packing on 2D surface, where  planar points of $\mathbf V$ represent 
 centres of the spheres. 

\begin{figure}[!tbp]
\centering
\subfigure[]{\includegraphics[width=0.45\textwidth]{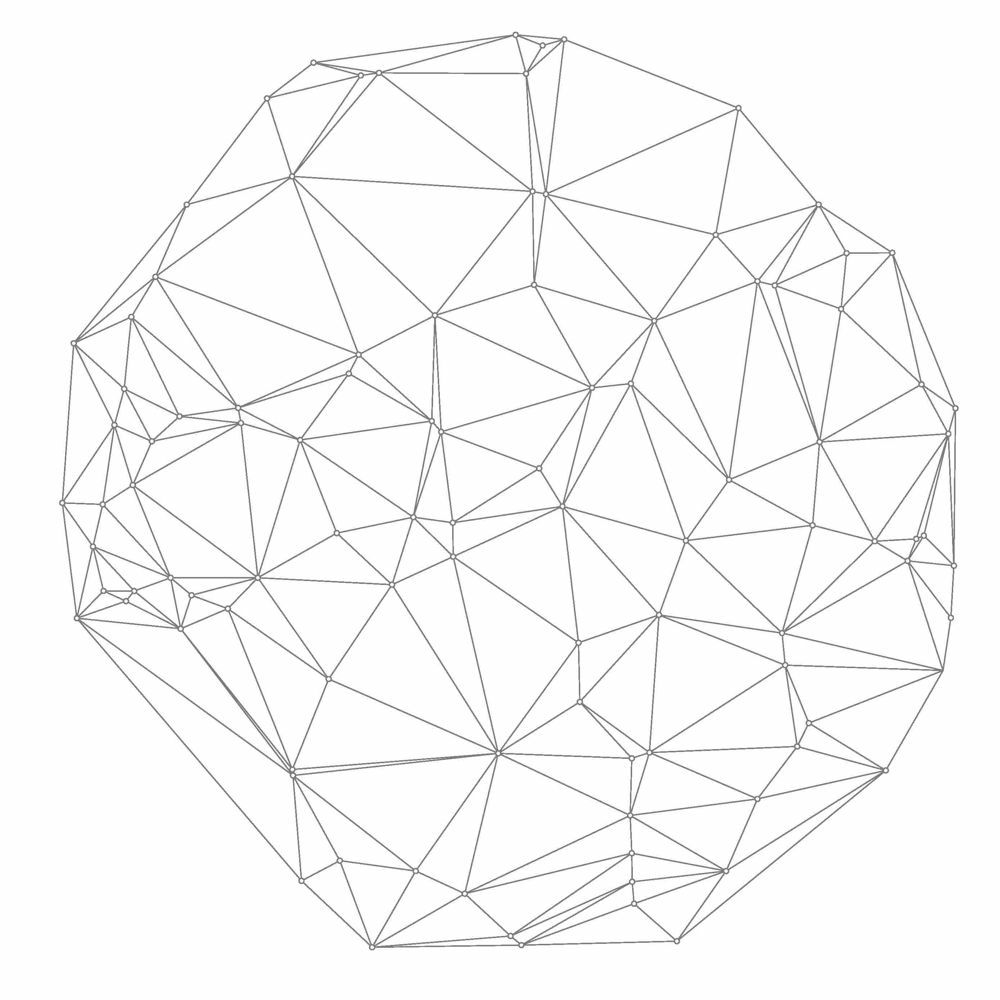}}
\subfigure[]{\includegraphics[width=0.45\textwidth]{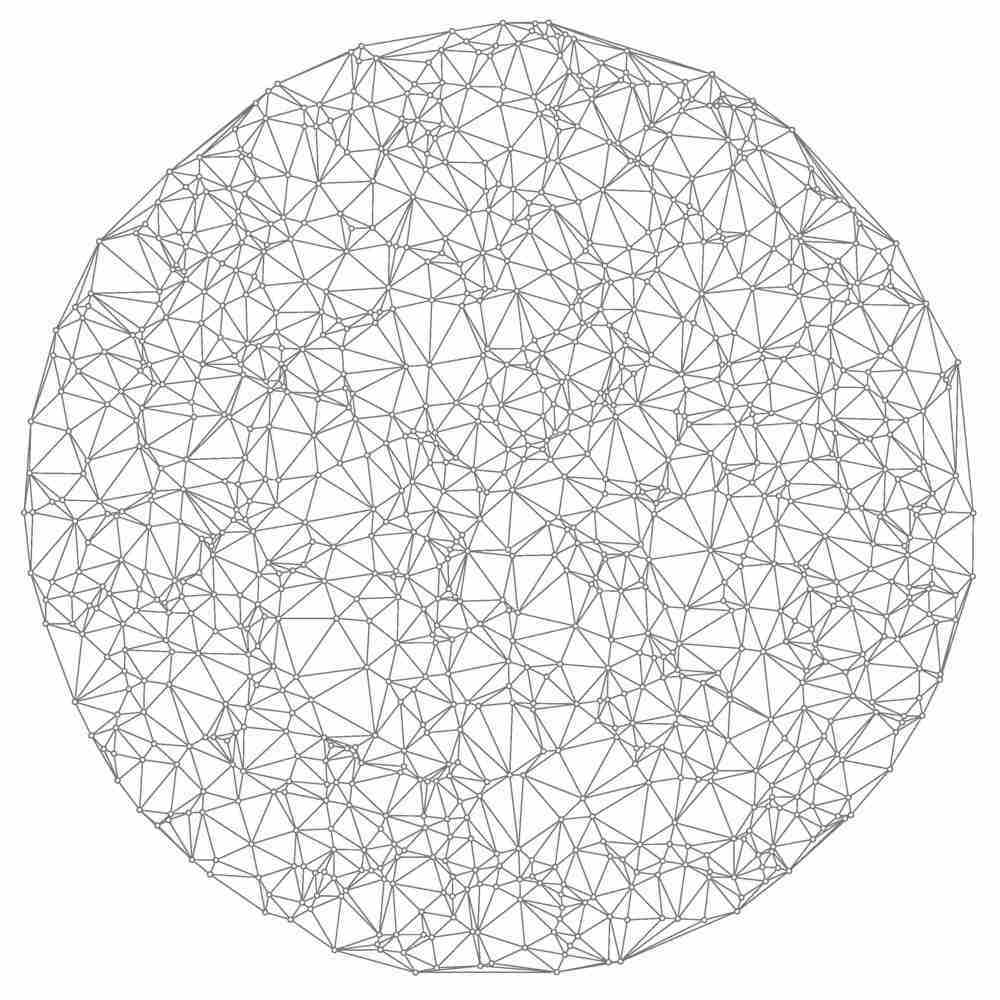}}
\subfigure[]{\includegraphics[width=0.45\textwidth]{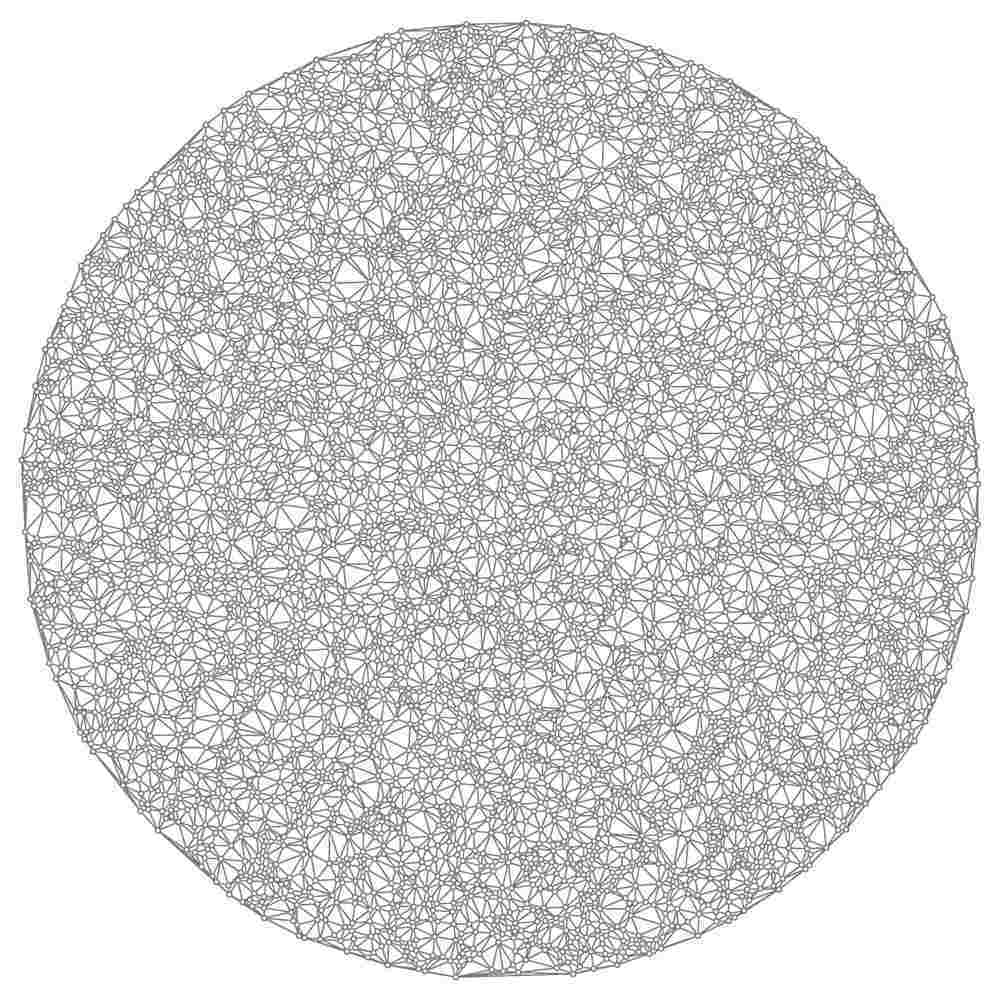}}
\subfigure[]{\includegraphics[width=0.45\textwidth]{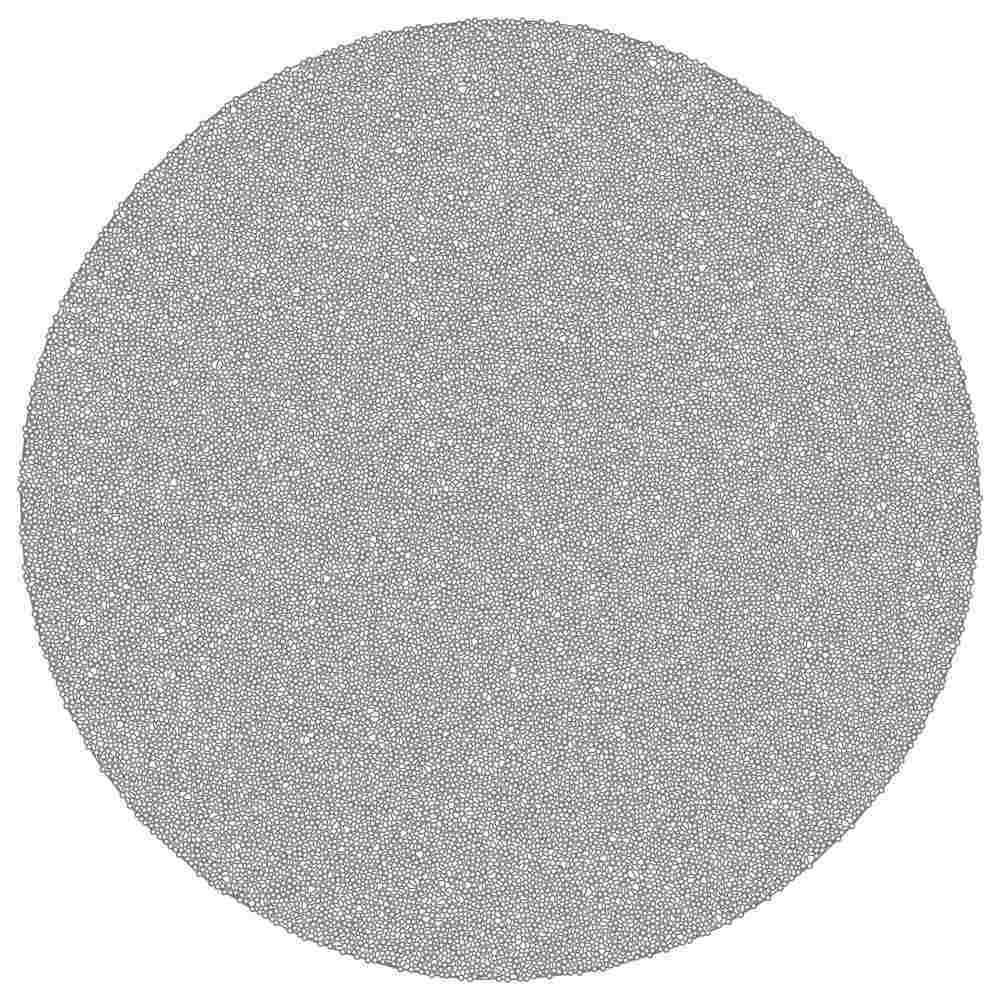}}
\subfigure[]{\includegraphics[width=0.45\textwidth]{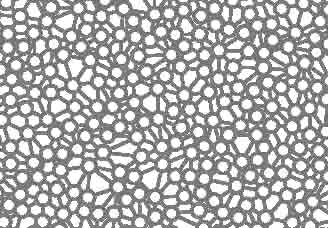}}
\caption{Examples of triangulations for densities (a)~$\phi=0.0027$,
(b)~$\phi=0.027$, (c)~$\phi=0.136$, (d)~$\phi=0.407$, (e)~fragment of triangulation
(a) is zoomed on.} 
\label{examples}
\end{figure}

We define density $\phi$ as a ratio of areas occupied by discs to area of the disc container. 
Examples of triangulations for  densities $\phi=0.0027, 0.027, 0.136$ and $0.407$, corresponding to 
number of nodes packed 100, 1000, 5000, and 15000, are shown in Fig.~\ref{examples}. 

The Delaunay automaton is defined in the following way.  A node $v \in \bf V$ is a finite state machine. 
Every node updates its state in discrete time depending on states of its neighbours. All nodes update 
their states simultaneously.  Nodes can have different number of neighbours therefore we better use 
totalistic node-state update function, where a node updates its state depending on just the 
numbers of different node-states in its neghbourhood.  

Here we are concerned only with modelling excitation on Delaunay triangulations. Thus we assign 
three states --- resting ($\circ$), excited ($+$) and refractory ($-$) --- to nodes of ${\bf V}$. 
We assume that a resting node excites depending on a number of excited neighbours. If a node is 
excited at time $t$ the node takes refractory state at time step $t+1$, independently on states of 
its neighbours. Transition from refractory to resting state is also unconditional.

Let $\nu(v)=\{ u \in {\bf V}: (vu) \in {\bf E} \}$ be node $v$'s neighbourhood, 
$v^t$ a state of node $v$ at time step $t$, $\sigma^t(v)$  a number of excited neighbours of $v$ at step $t$, 
and $d(v)$ be a degree, or a number of neighbours $|\nu(v)|$, of node $v$. Then the node-state 
transition functions can be defined as follows.
\begin{itemize}
\item \emph{Absolute excitability}: 
\begin{equation}
v^{t+1}=
\begin{cases}
+, \text{ if } \sigma^t(v) \in {\bf S}\\
-, \text{ if } v^t=+\\
\circ,  \text{ if } v^t=- 
\end{cases}
\label{eq1}
\end{equation}
Node $v$ excites if $\sigma^t(v) \in {\mathbf S}$, where $\mathbf S$ is a 
set of natural numbers. For example, ${\mathbf S} = \{ 2, 4 \}$ means a resting node excites if it has two or 
four excited neighbours. The rule includes threshold excitation $\sigma^t(v)  \geq \theta$, $\theta$ is a natural number. 
For example, $\theta=2$ means a resting node excites if it has at least 2 excited neighbours, 
i.e. ${\mathbf S}=\{2, 3, 4, \cdots \}$.
\item \emph{Relative excitability}: 
\begin{equation}
v^{t+1}=
\begin{cases}
+, \text{ if } \frac{\sigma^t(v)}{d(v)} > \epsilon\\
-, \text{ if } v^t=+\\
\circ,  \text{ if } v^t=- 
\end{cases}
\label{eq2}
\end{equation}
Node $v$ excites if $\rho^t(v)=\frac{\sigma^t(v)}{d(v)} > \epsilon$, where
$0 \leq \epsilon \leq 1$.   This condition bring more `fairness' in excitation process. Some nodes can have less neighbours
than other nodes, thus measuring excitation of neighbourhood just by number of excited neighbour would not be `fair'. 
It is feasible to calculate a ratio $\rho^t(v)$ of excited neighbours $\sigma^t(v)$ to a total number of neighbours $d(v)$.   
\end{itemize}

\section{Structural properties of Delaunay automata}
\label{properties}
 
 \begin{figure}[!tbp]
\centering
\subfigure[]{\input{figs/dist7.tex}}
\subfigure[]{\input{figs/deviation3.tex}}
\caption{Distribution of degree $d$ probabilities $p(d)$~(a) and 
standard deviation $\sigma$ of distribution $p(d)$~(b) for several 
densities $\phi$ of disc-nodes.}
\label{degreedistribution}
\end{figure}
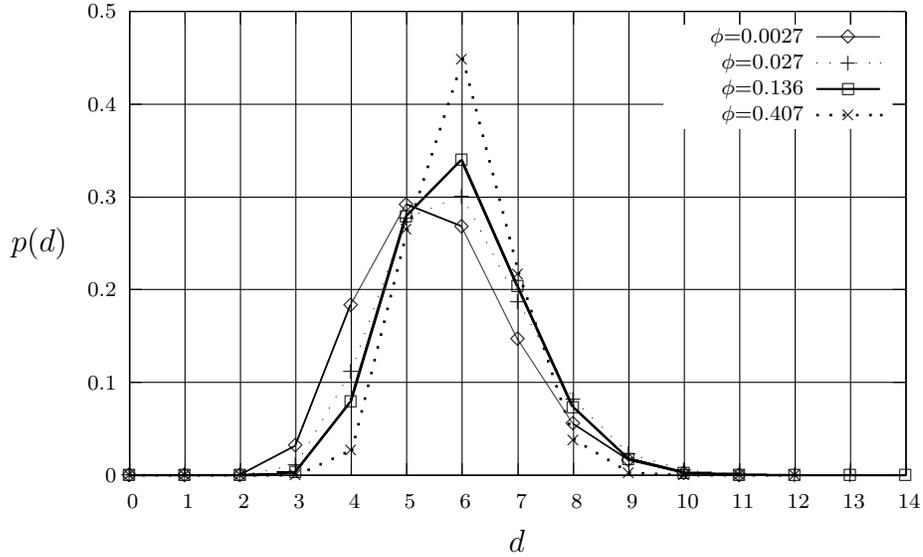
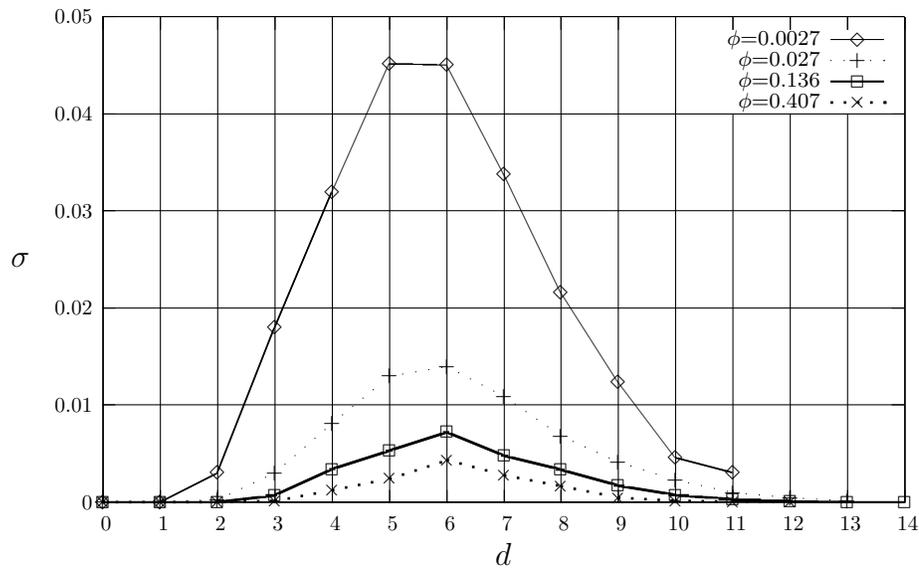

 Propagation of excitation in Delaunay automata depends on 
 structural properties of their graphs.  We found that in case of sparse distribution 
 of disc-nodes (Fig.~\ref{examples}a), $\phi=0.0027$, majority of nodes have five 
 then six and four neighbours each (Fig.~\ref{degreedistribution}a). With increase of 
 the density (Fig.~\ref{examples}b--d) maximum of degree distribution is 
shifted to six neighbours per node  (with probability $p(6)=0.45$), five nodes ($p(5)=0.27$) and 
seven nodes ($p(7)=0.22$) (Fig.~\ref{degreedistribution}a). The deviation of the distribution 
significantly decreases with increase of the density (Fig.~\ref{degreedistribution}b).

The degree distributions found experimentally conform to classical results on 
packing of equal spheres which is between 5.5 for sparse packing and 6.4 for a 
close yet random packing~\cite{bernal_1960}, and mean degree of 6 for randomly packed
sphere with density 0.64~\cite{gotoh_1974}.

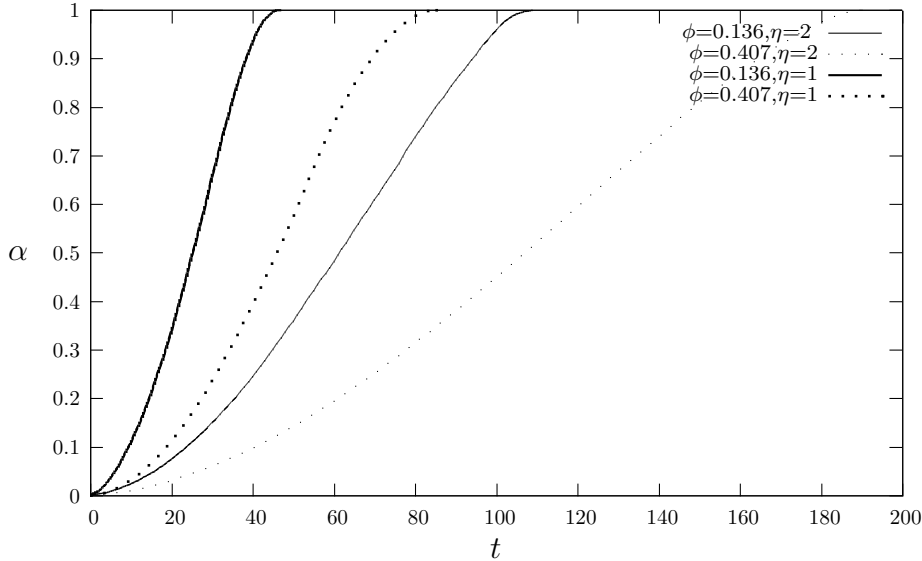
\begin{figure}[!tbp]
\centering
\input{figs/activation1.tex}
\caption{Dynamics of occupation $\alpha$ of triangulations from single-node perturbation;
$\alpha$ is a ratio of occupied nodes at time step $t$ to a total number of nodes in the triangulation. 
Graph presents dynamics of occupation for  densities $\phi=0.136$ and $\phi=0.407$, and occupation 
thresholds $\eta=1$ and $\eta=2$.}
\label{activation}
\end{figure}

Let every node of a triangulation takes just two states: '0' (unoccupied) and '1' (occupied). A 
resting node becomes occupied if it has at least $\eta$ neighbours in state '1'. Initially 
just one (for $\eta=1$) or two (for $\eta=2$) nodes are assigned state '1'. Dynamics of occupation, 
measured in a ratio $\alpha$ of nodes occupied by time step $t$, is shown in Fig.~\ref{activation}.

A speed of propagation of state '1' measured in a ratio of nodes occupied in step $t$ 
increases with decrease of occupation threshold $\eta$. If we measure speed in discrete 
time steps, we will see that it also decreases with increase of the density $\phi$. However, the 
distance covered in any period of time remains comparable between triangulations with different 
number of nodes.

\begin{figure}[!tbp]
\centering
\subfigure[]{\includegraphics[width=0.49\textwidth]{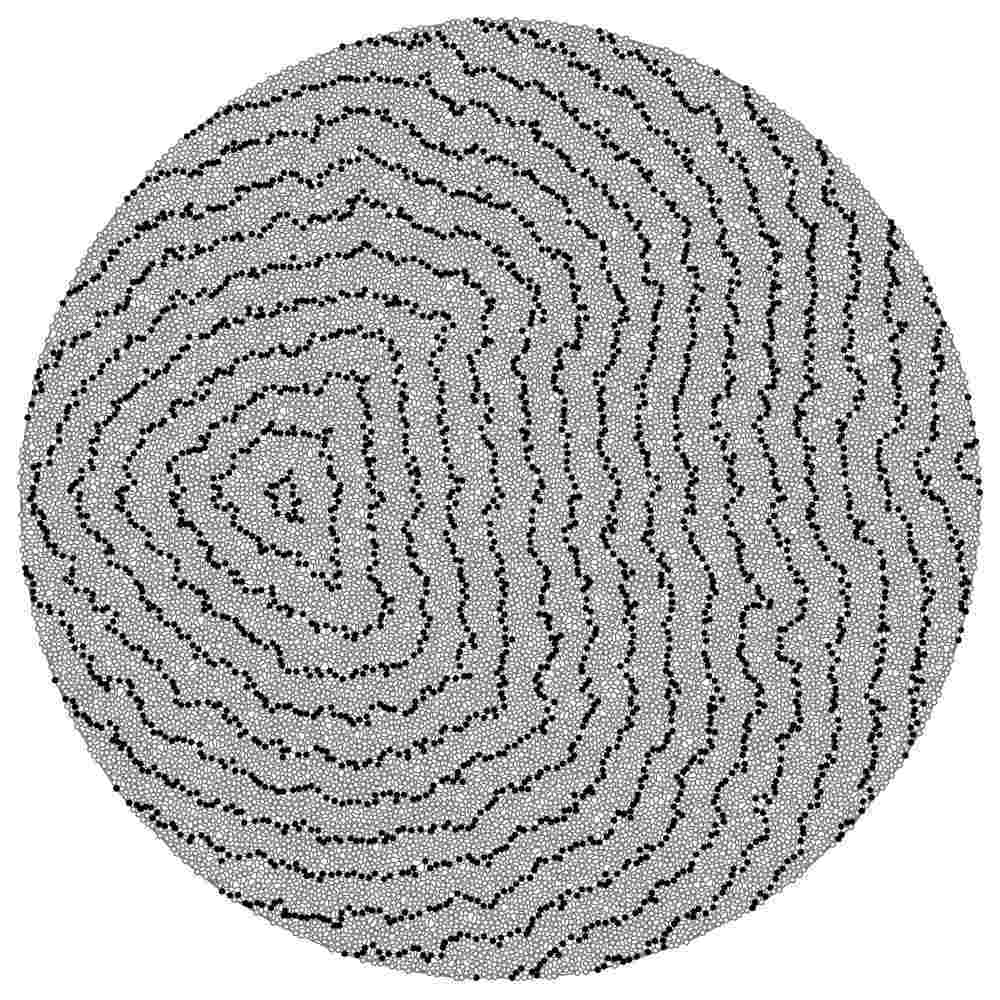}}
\subfigure[]{\includegraphics[width=0.49\textwidth]{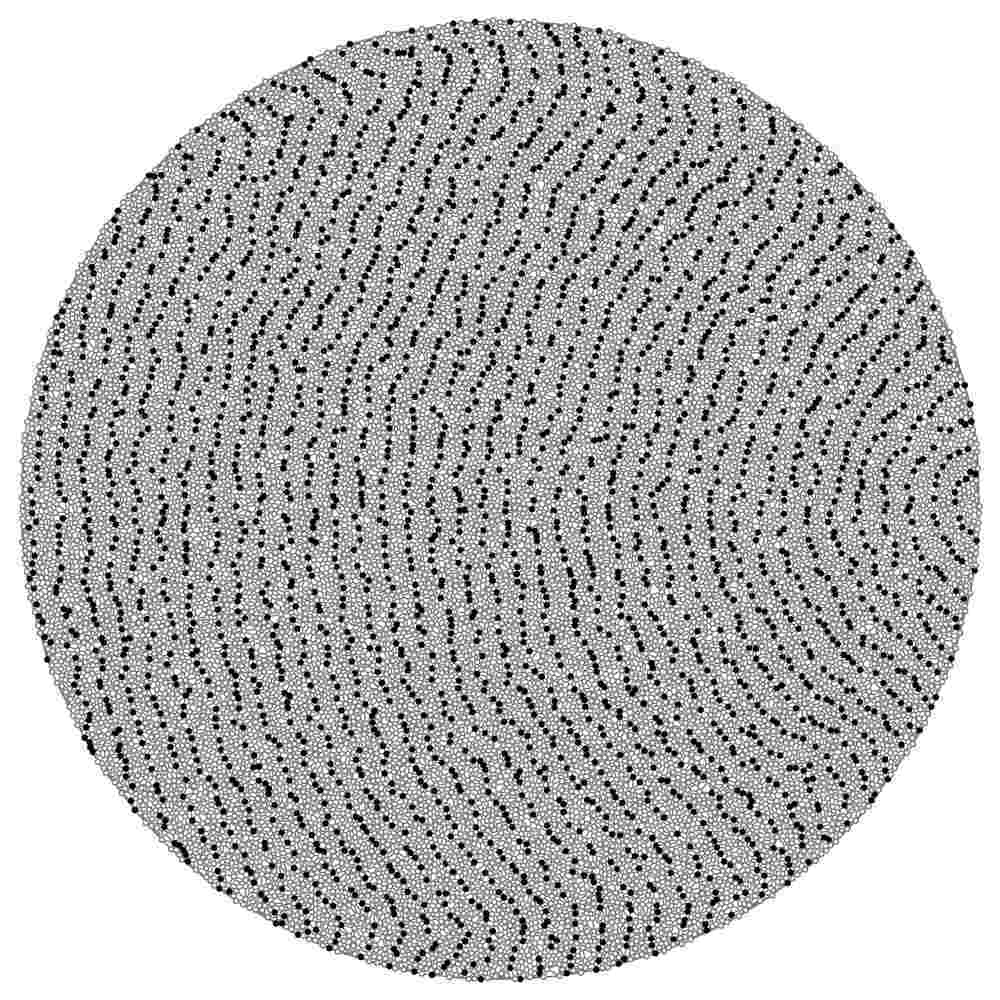}}
\caption{Time lapsed contours of activation in triangular lattices with density $\phi=0.407$ and 
activation threshold (a)~$\eta=1$ and (b)~$\eta=2$. Propagating wave front is converted to a contour every
10th step of simulation. Initially eastmost nodes are activated.} 
\label{contoursofpropagation}
\end{figure}

\begin{finding}
Threshold $\eta$ of node occupation determines a shape of propagating occupation fronts.
\end{finding}

For $\eta=2$ occupation wave-front is convex, shaping into planar by the end of propagation  (Fig.~\ref{contoursofpropagation}b). 
In case of lower threshold of occupation, $\eta=1$, propagation wave near the edges of triangulation moves quicker 
then inside the triangulation core  (Fig.~\ref{contoursofpropagation}a). Thus wave front becomes concave. The part of wave front travelling along edges of triangulation reaches the side of triangulation opposite to the initial perturbation side quicker then the front propagating inside the triangulation. Thus wave front becomes closed (Fig.~\ref{contoursofpropagation}a). The domain surrounded by a  nested group of target waves, in the western part of the triangulation in Fig.~\ref{contoursofpropagation}a, is the place where the occupation waves traveling from east collapse.

\begin{finding}
Perturbation propagates faster along edges of triangulation when threshold of node perturbation is low.
\end{finding}

\begin{figure}[!tbp]
\centering
\subfigure[]{\includegraphics[width=0.24\textwidth]{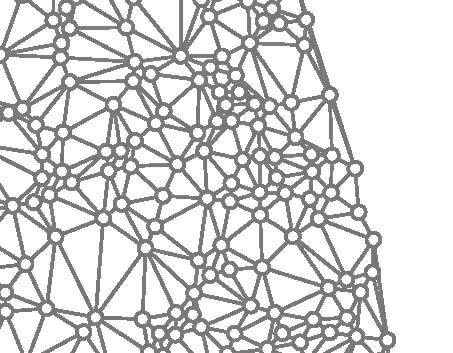}}
\subfigure[]{\includegraphics[width=0.64\textwidth]{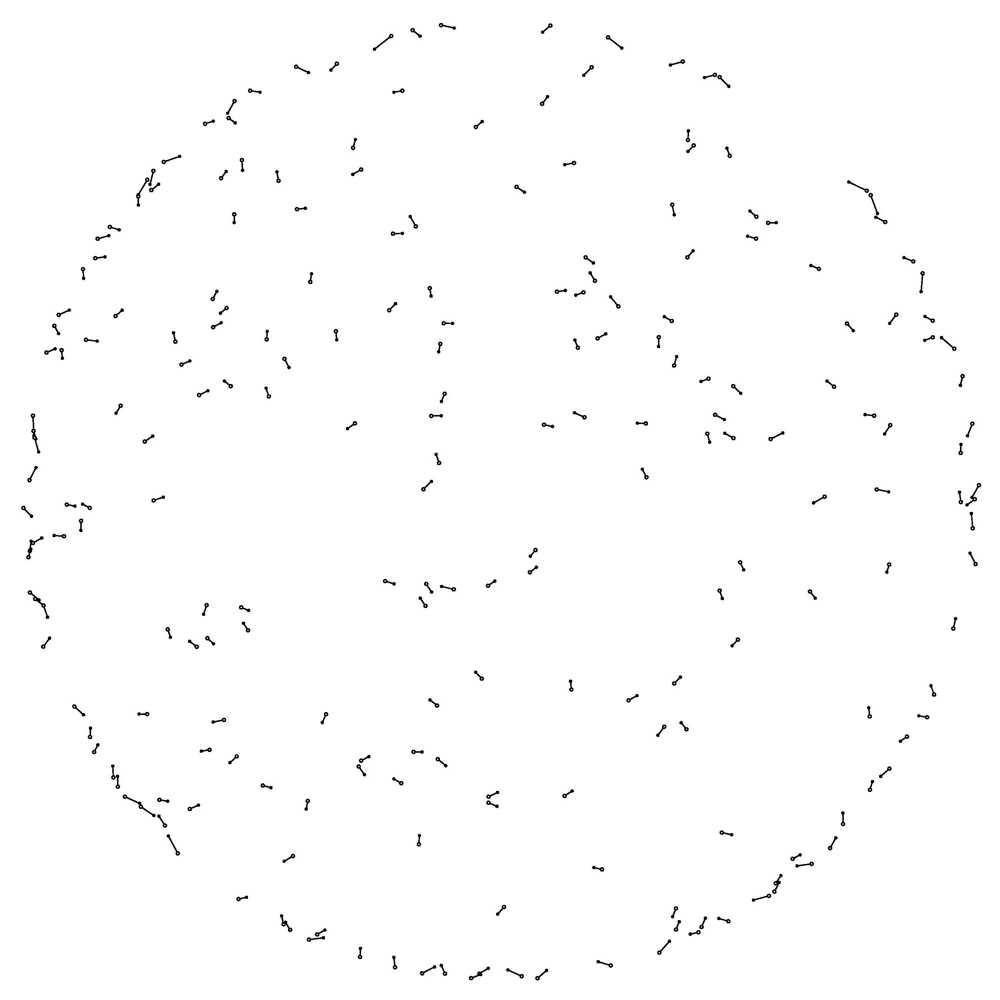}}
\caption{Geometry of neighbourhoods:
(a)~a fragment of triangulation, density $\phi=0.136$, with visible neighbourhoods of edge nodes, 
edge of triangulation is on the right,
(b)~vectors from nodes to their geometric centres, only vectors which length exceeding eight units are shown.
Density is $\phi=0.136$.} 
\label{vectors}
\end{figure}

An edge of triangulation is a set of nodes lying on segments of the convex hull of ${\bf V}$.
Spatial structures of node neigbourhoods at the edge of triangulation may be responsible for the 
particulars of propagations described above.

Typically neighbours of each node are distributed more or less equally around the node while 
nodes belonging to the edge of triangulation have their neighbours located towards inside 
of the triangulation or on the edge  (Fig.~\ref{vectors}a).   
We can integrate spatial distribution of neighbours of node $v$ as a vector $p=\overline{vg}$ from the node $v$ 
to geometric centre $g$ of the node $v$'s neighbourhood. Edge nodes of triangulation have usually
longer vectors $p$, see Fig.~\ref{vectors}b. This is why a perturbation propagates faster on or near 
edges of triangulation  for a low threshold of occupation  (Fig.~\ref{contoursofpropagation}a). Increase 
of occupation threshold is disadvantageous for edge nodes 
which is reflected in changed shape of the growing pattern  (Fig.~\ref{contoursofpropagation}b).

\section{Absolute excitability}
\label{absoluteexcitability}

\begin{figure}[!tbp]
\centering
\subfigure[]{\includegraphics[width=0.45\textwidth]{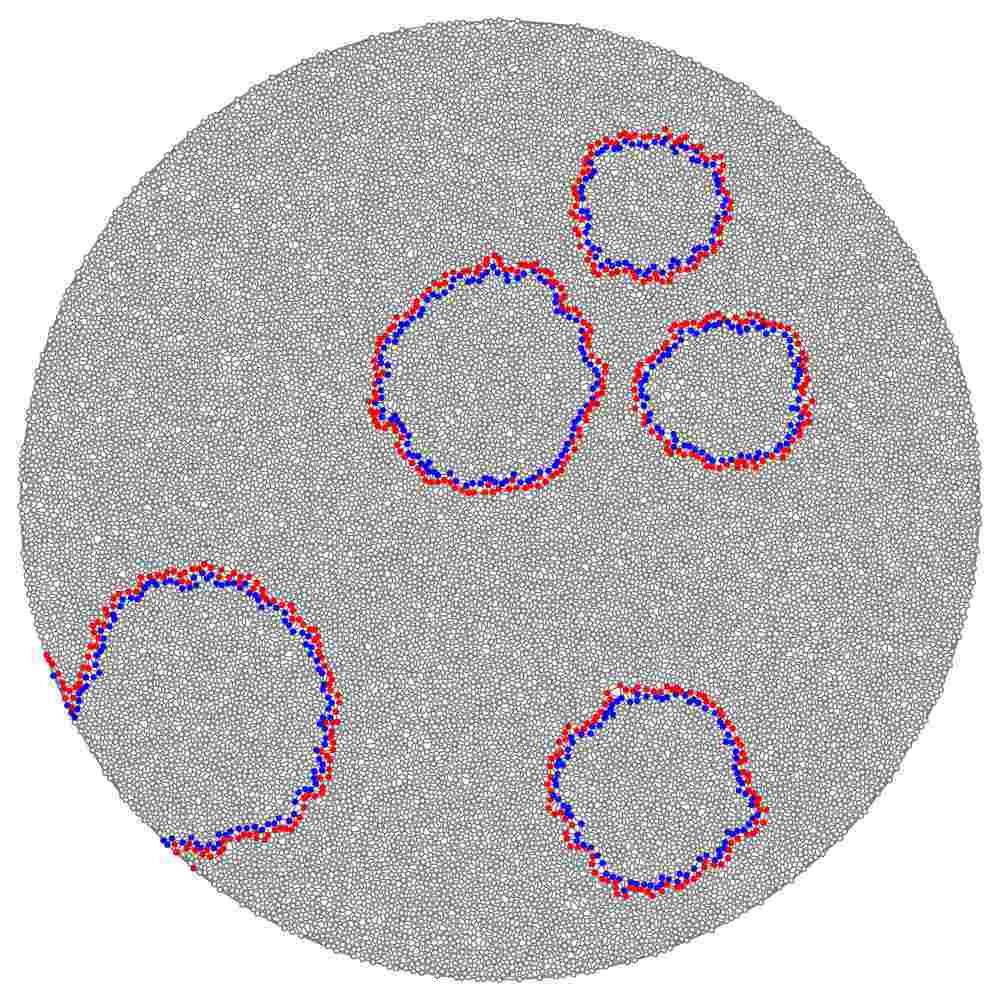}}
\subfigure[]{\includegraphics[width=0.45\textwidth]{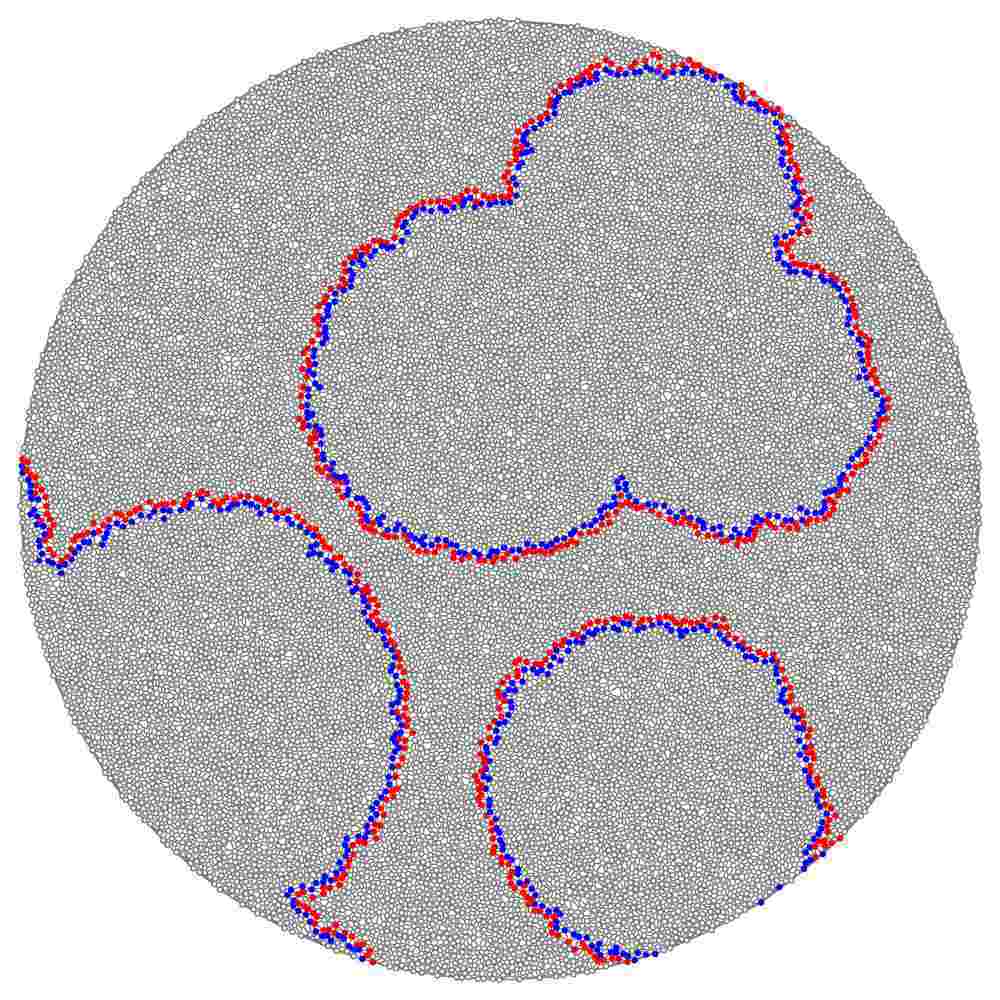}}
\subfigure[]{\includegraphics[width=0.45\textwidth]{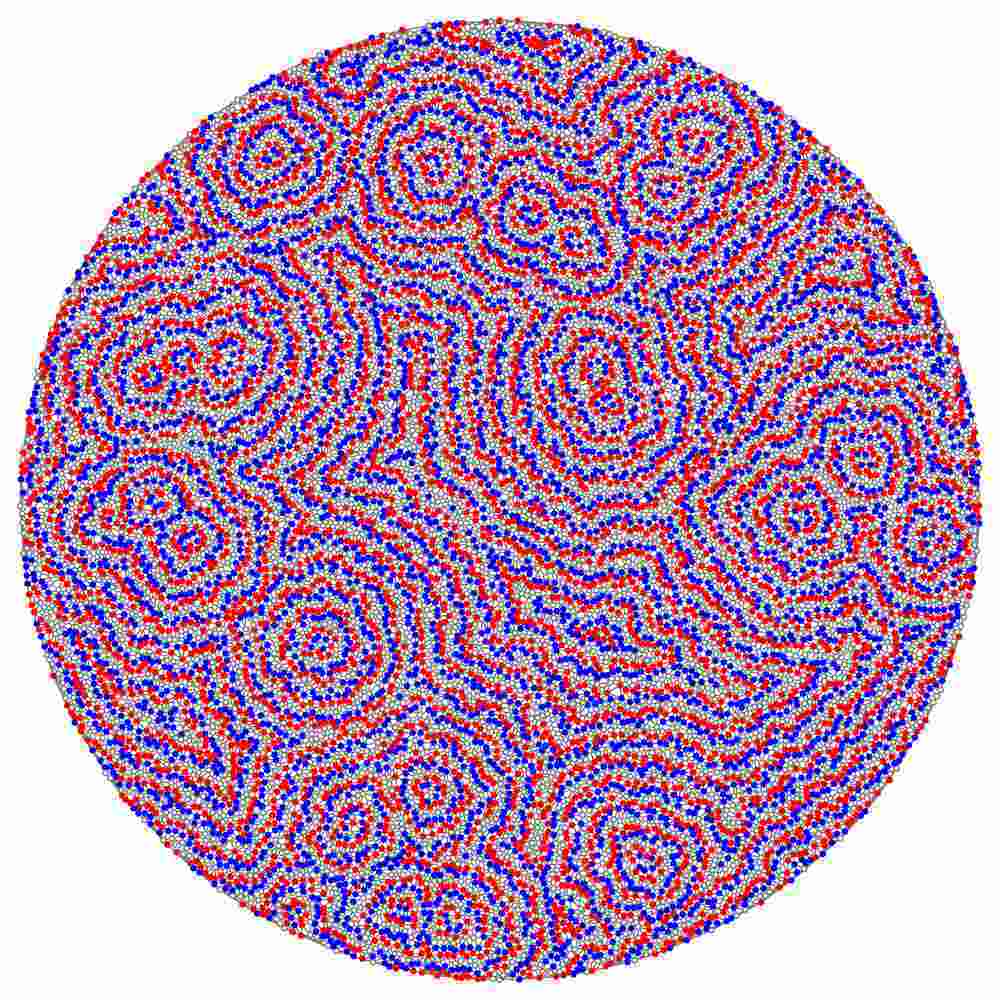}}
\subfigure[]{\includegraphics[width=0.45\textwidth]{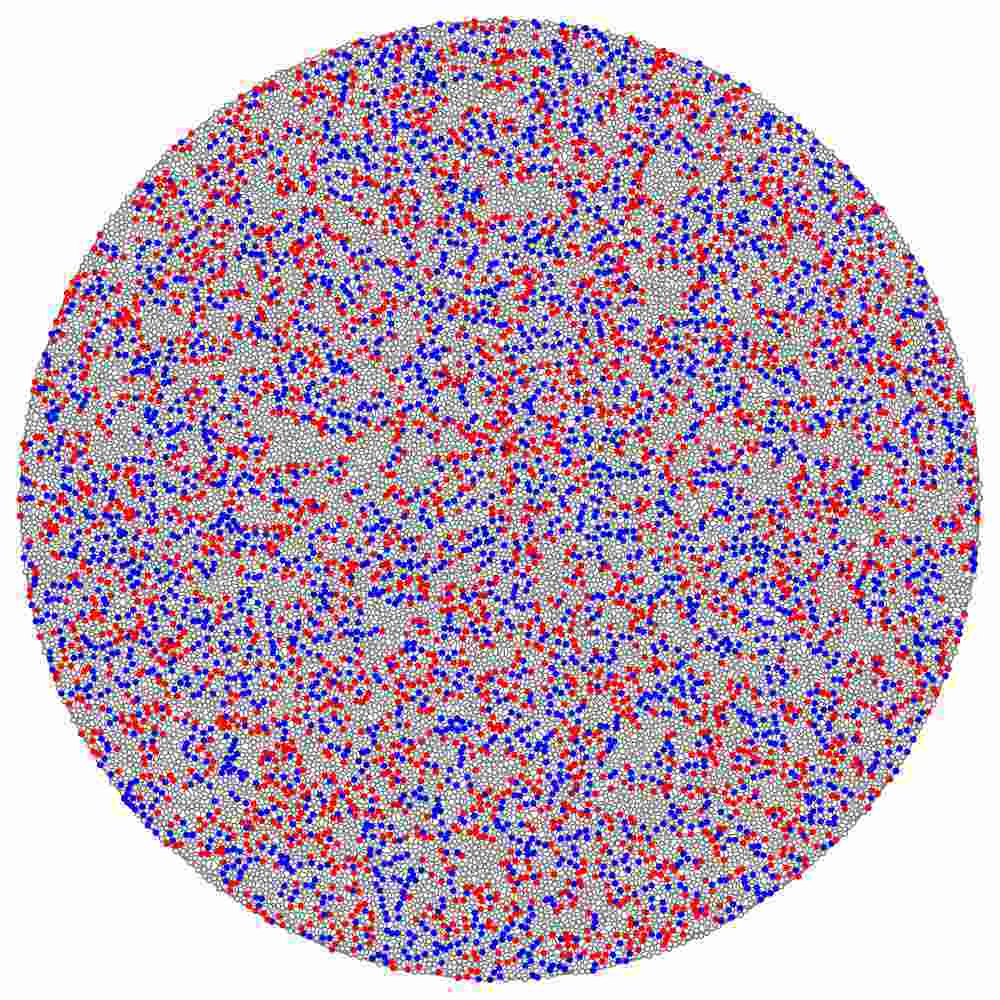}}
\caption{Examples of excitation dynamics for rule (\ref{eq1}) $\sigma^t(v) \geq 1$~(a)--(c) and 
rule (\ref{eq1}) $\sigma^t(v)=1$~(d): 
(ab)~waves generated by singular excitations merge when collide, 
(b)~random initial configuration (at the beginning of simulation a node got excited state with probability 0.1, refractory 
state with probability 0.1, and resting state with probability 0.8) develops into a configuration 
of several target-wave generators; (c)~configuration of excitation developed in 300 time steps from 
initial configuration where all but one nodes are resting. Density of disc-nodes in triangulation is $\phi=0.407$.} 
\label{theta2}
\end{figure}

If a  resting node $v$ excites when it has at least one excited neighbour, rule (\ref{eq1}) $\sigma^t(v) \geq 1$, `classical' 
excitation waves are observed (Fig.~\ref{theta2}ab). By `classical' we mean that excitation waves annihilate
 when they reach edges of triangulation, two waves merge when they collide one with another, a single-site perturbation initiates a circularly propagating excitation wave, and refractory state is a necessary component in seeds of target-wave generators.

The excitable triangulation behaves less conventionally when nodes are selective in their
excitability. Consider the rule (\ref{eq1}) ${\bf S}=\{ 1 \}$: a resting node excites if it has exactly one excited neighbour. 
Single site excitation leads to formation of circular waves. The waves do merge when collide with each other 
and also they may form generators of target waves at the sites of their collision (Fig.~\ref{theta2}c). 

\begin{finding}
Let a resting node of Delaunay triangulation excite if exactly one neighbour is excited, then 
a generator of target-waves can be produced by a single-site excitation.
\end{finding}

For $\sigma^t(v) \geq 2$ no excitation persists. However when resting node excites if exactly two or three 
of its neighbours are excited we obtain a quasi-chaotic excitation dynamics (Fig.~\ref{theta2}d). 
This happens when nodes follow state update rules (\ref{eq1}) for ${\bf S}=2$ and ${\bf S}=\{ 2, 3 \}$. 

\begin{figure}[!tbp]
\centering
\subfigure[$t=35$]{\includegraphics[width=0.45\textwidth]{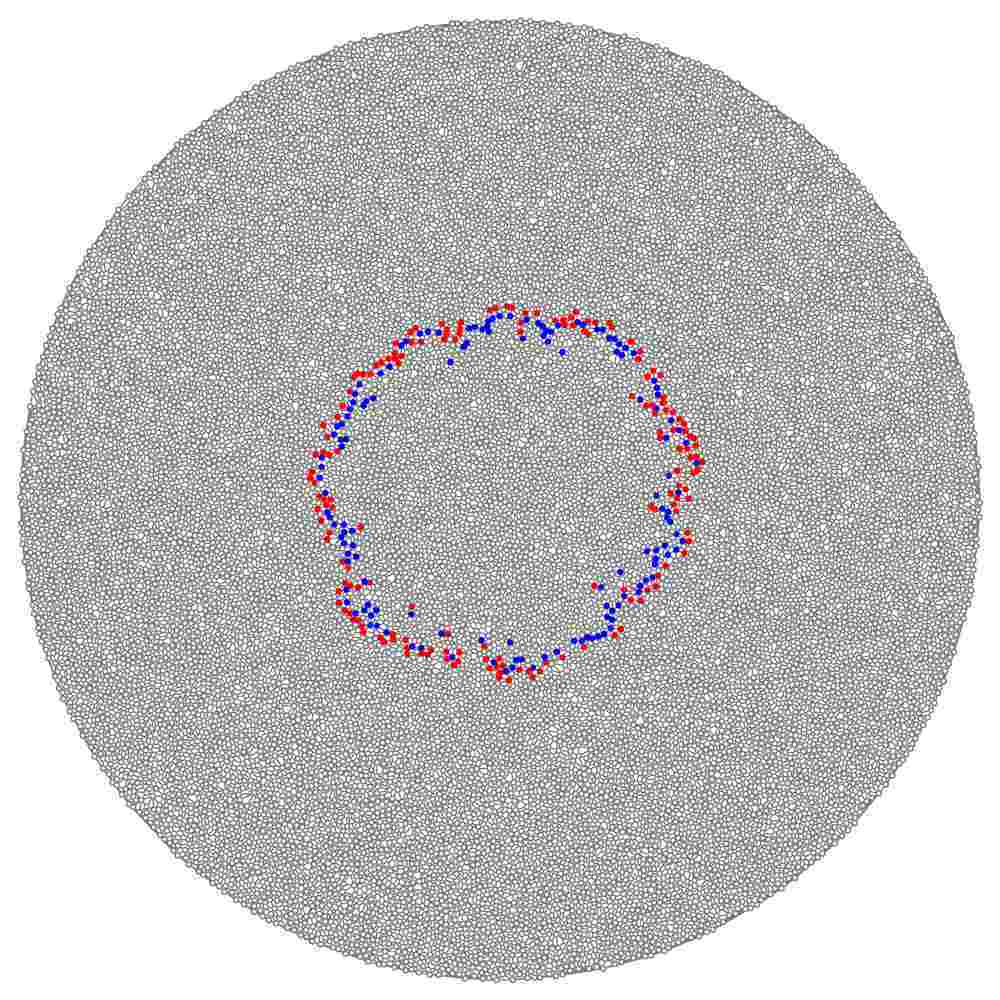}}
\subfigure[$t=52$]{\includegraphics[width=0.45\textwidth]{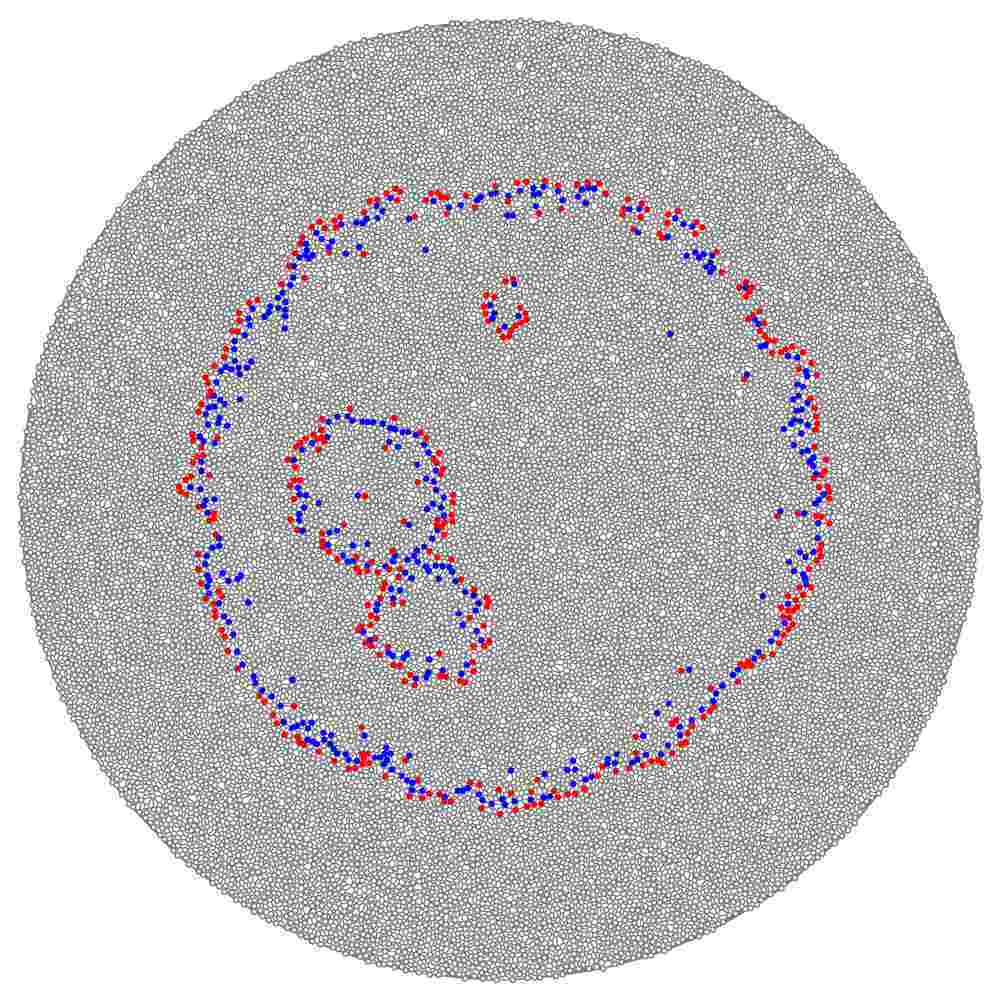}}
\subfigure[$t=65$]{\includegraphics[width=0.45\textwidth]{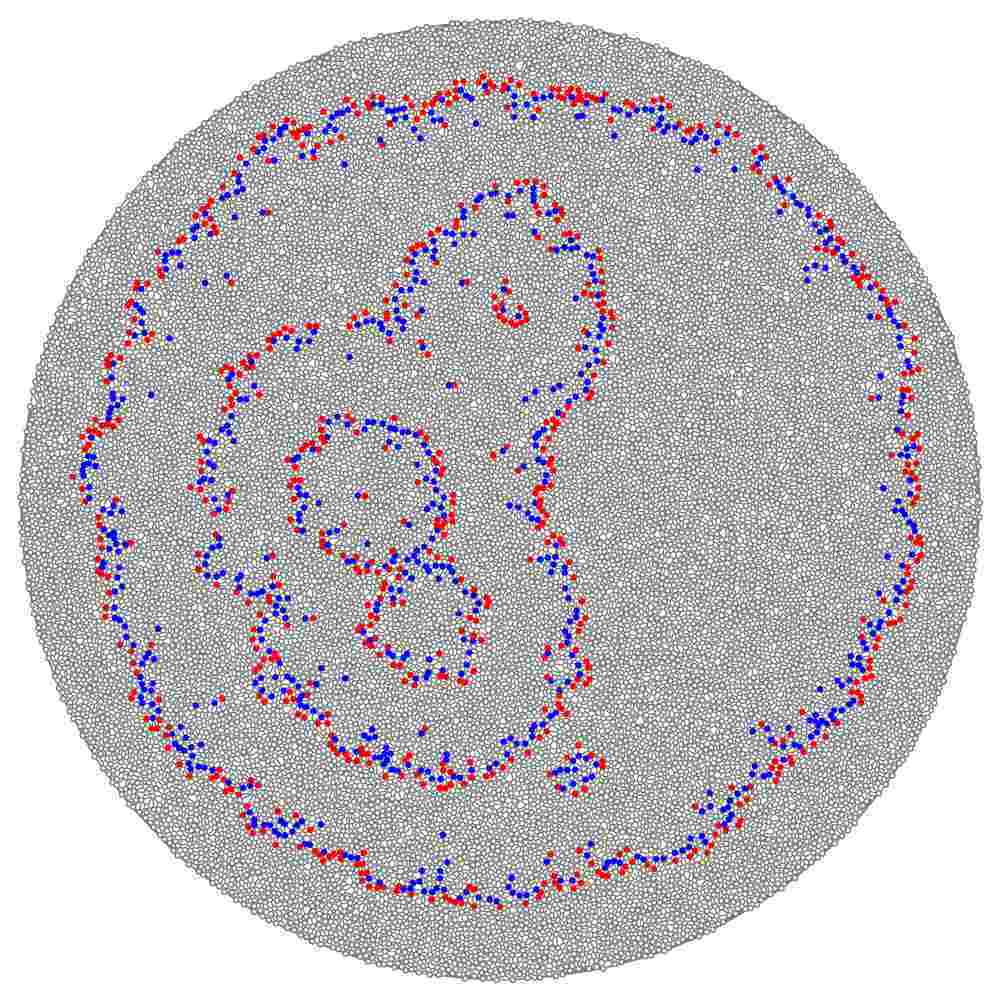}}
\subfigure[$t=148$]{\includegraphics[width=0.45\textwidth]{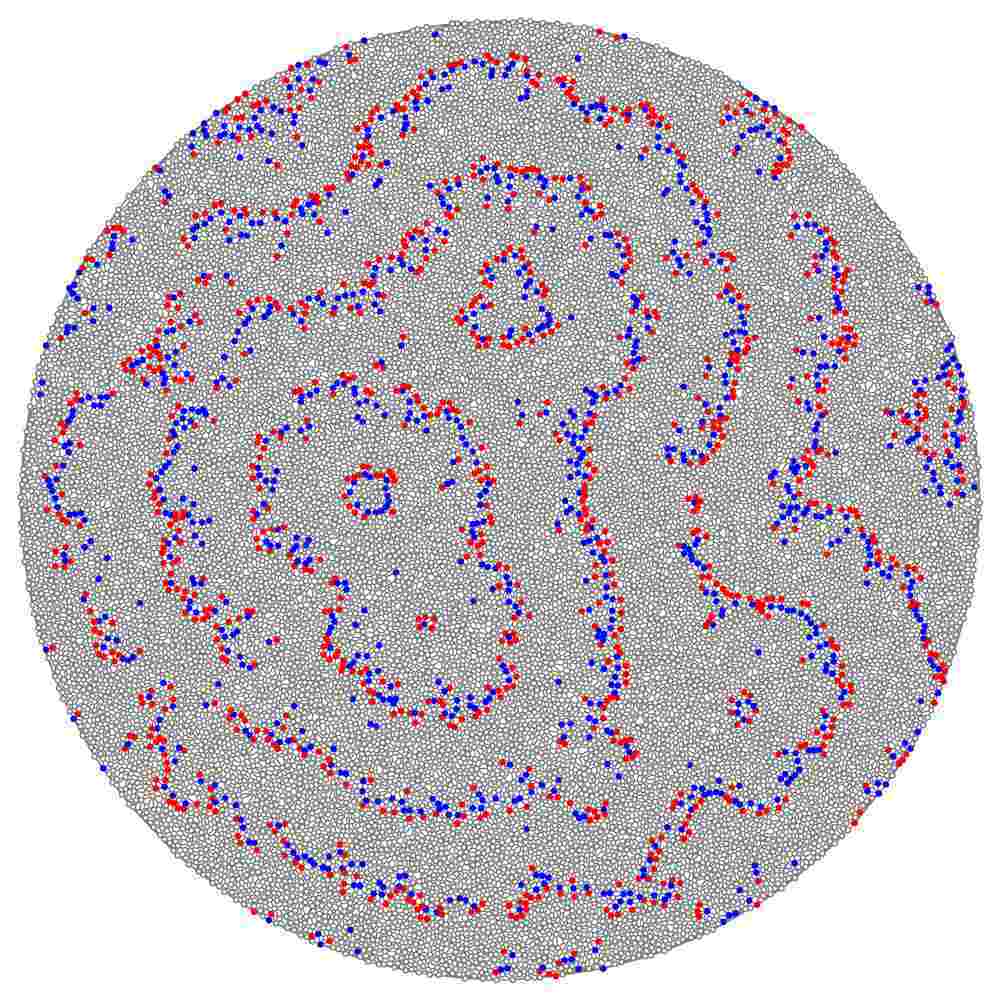}}
\caption{Example of wave generators formed due to backfiring of excitation from a single propagating wave front.
Rule $\sigma^t(v) \geq 1$ with delayed (by 10 time steps) recovery from refractory state. Density of disc-nodes in triangulation 
is $\phi=0.407$. } 
\label{backfiring}
\end{figure}

A delay in recovery from refractory states modifies space-time dynamics of excitation. 
Assume that if a node took refractory state then the  node stays in the refractory state 
for $\delta$ time steps. In automata governed by rule (\ref{eq1}) with excitability 
$\sigma^t(v) \geq 1$ initial random disturbances --- nuclei of excited and refractory states --- will 
not lead to formation of wave generators (as in rule $\sigma^t(v) \geq 1$ without node-state recovery delay). 
However a phenomenon of excitation backfiring is still observed. Travelling excitation wave can backfire 
with localized excitations. These localized excitations pass through the wave's refractory tail 
and initiate generation of new wave fronts (Fig.~\ref{backfiring}).

\begin{finding}
Delaunay excitable automata governed by rules of absolute excitability exhibit the following phenomena: 
\begin{itemize}
\item threshold activation rules causes formation of classical excitation wave fronts,
\item rules relying on exact number of excited neighbours show formation of target wave generators during collision between
ordinary circular waves,
\item when recovery of a node from refractory state to resting state is delayed propagating wave fronts backfire with 
localized excitations, which cause formation of target-wave generators.
\end{itemize}
\end{finding}

\section{Relative excitation}
\label{relativeexcitability}

We found that we can initiate a persistent excitation patterns for $0 \leq \epsilon < 0.25$. For $\epsilon \geq 0.25$
any initially invoked excitation quickly ceases. Automata excited by rule (\ref{eq2}) with $\epsilon =0$ exhibit 
persistent global oscillations because every node autonomously follows the 
cycle $\circ \rightarrow + \rightarrow - \rightarrow \cdots$. For excitability range 
$0 < \epsilon < 0.09$  Delaunay automata exhibit classical excitation waves. Single excitation generates single circular 
wave, clusters of excited and refractory states may become generators of target wave. When a 
propagating wave front hits edge of triangulation the wave disappears. Two colliding wave-fronts
 merge or annihilate.

\begin{figure}[!tbp]
\centering
\subfigure[$t=20$]{\includegraphics[width=0.32\textwidth]{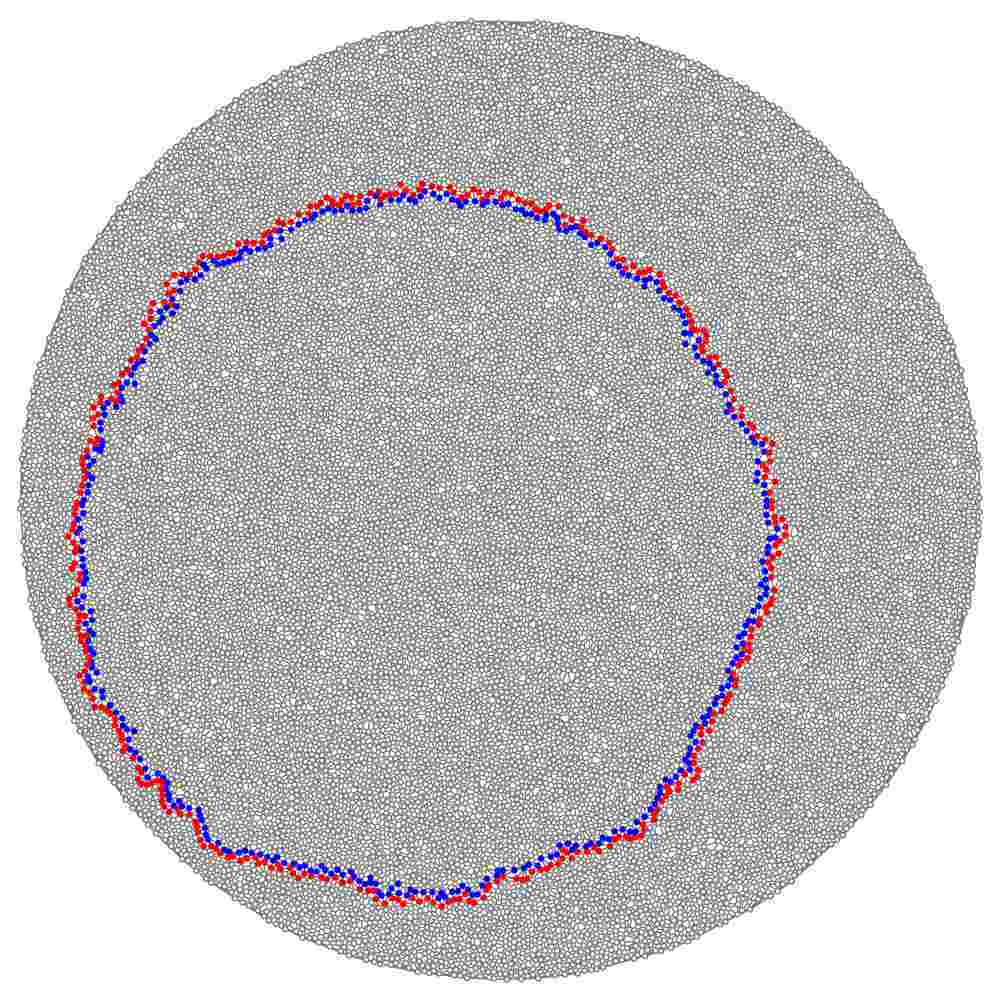}}
\subfigure[$t=30$]{\includegraphics[width=0.32\textwidth]{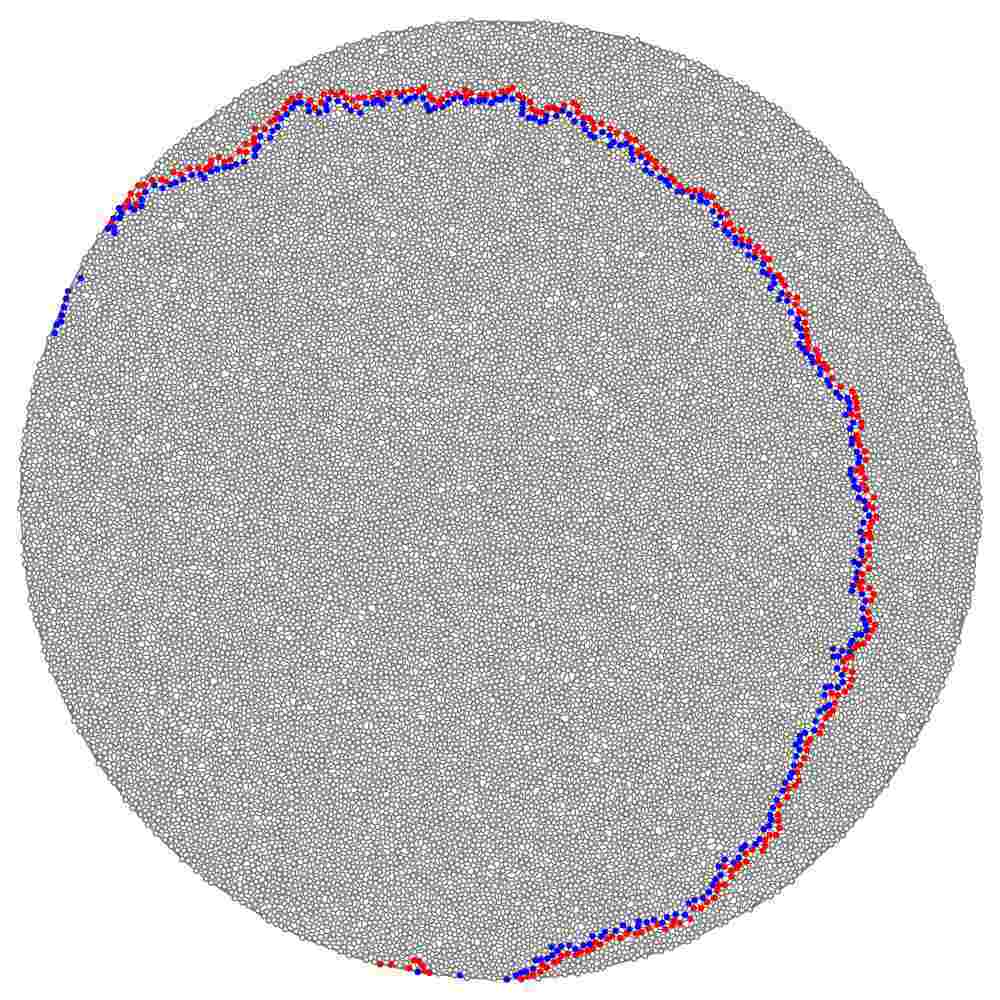}}
\subfigure[$t=41$]{\includegraphics[width=0.32\textwidth]{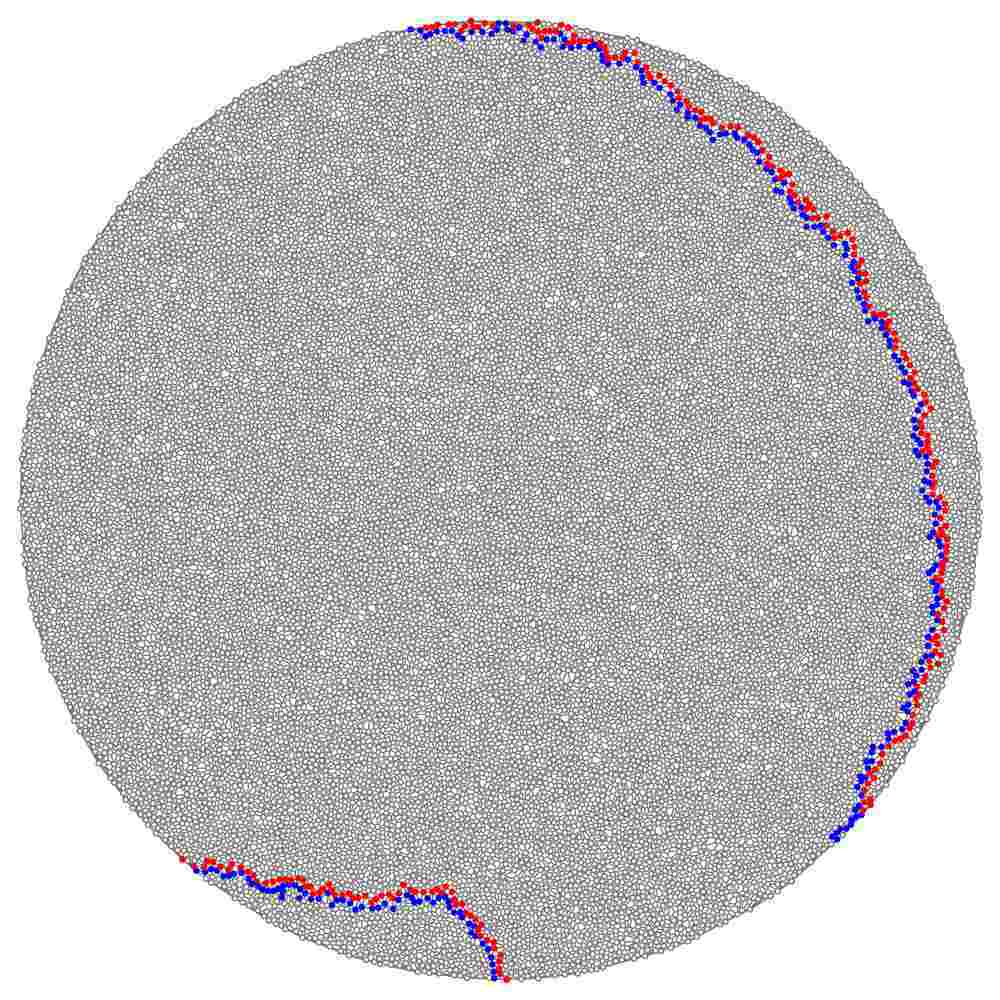}}
\subfigure[$t=67$]{\includegraphics[width=0.32\textwidth]{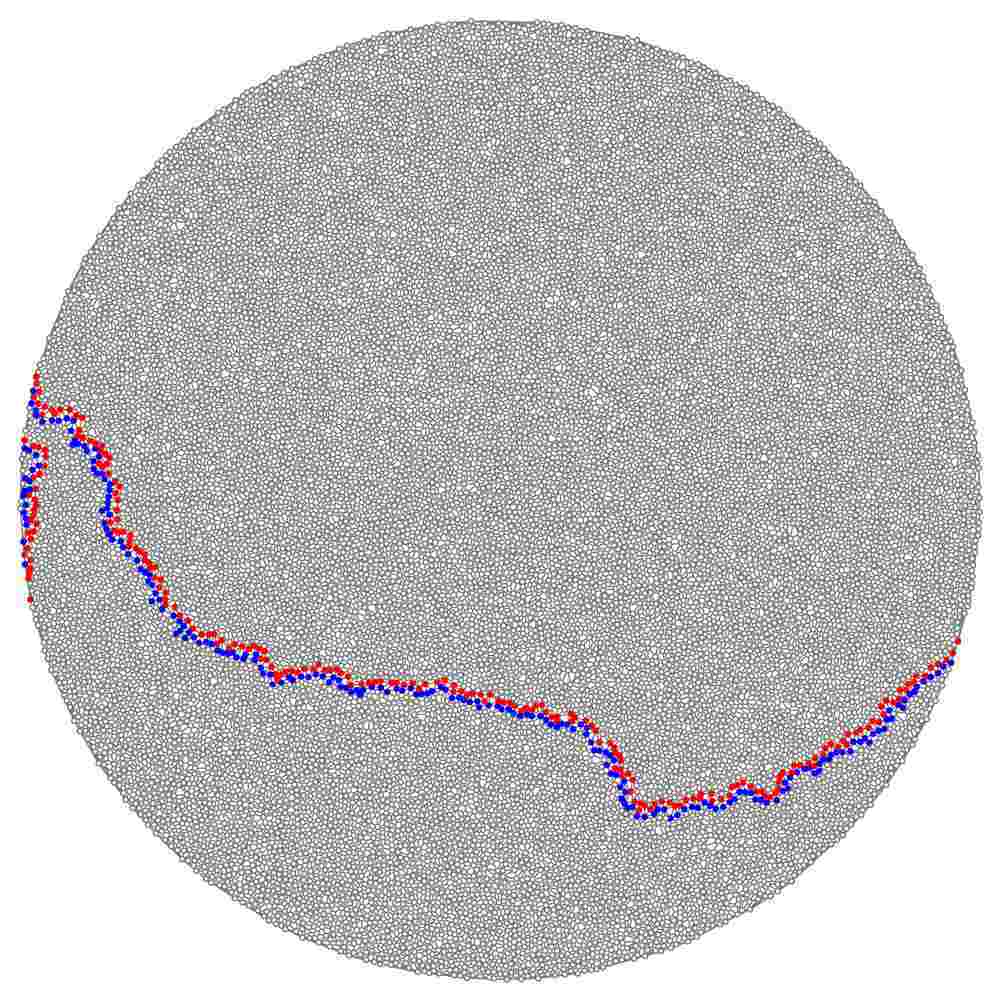}}
\subfigure[$t=100$]{\includegraphics[width=0.32\textwidth]{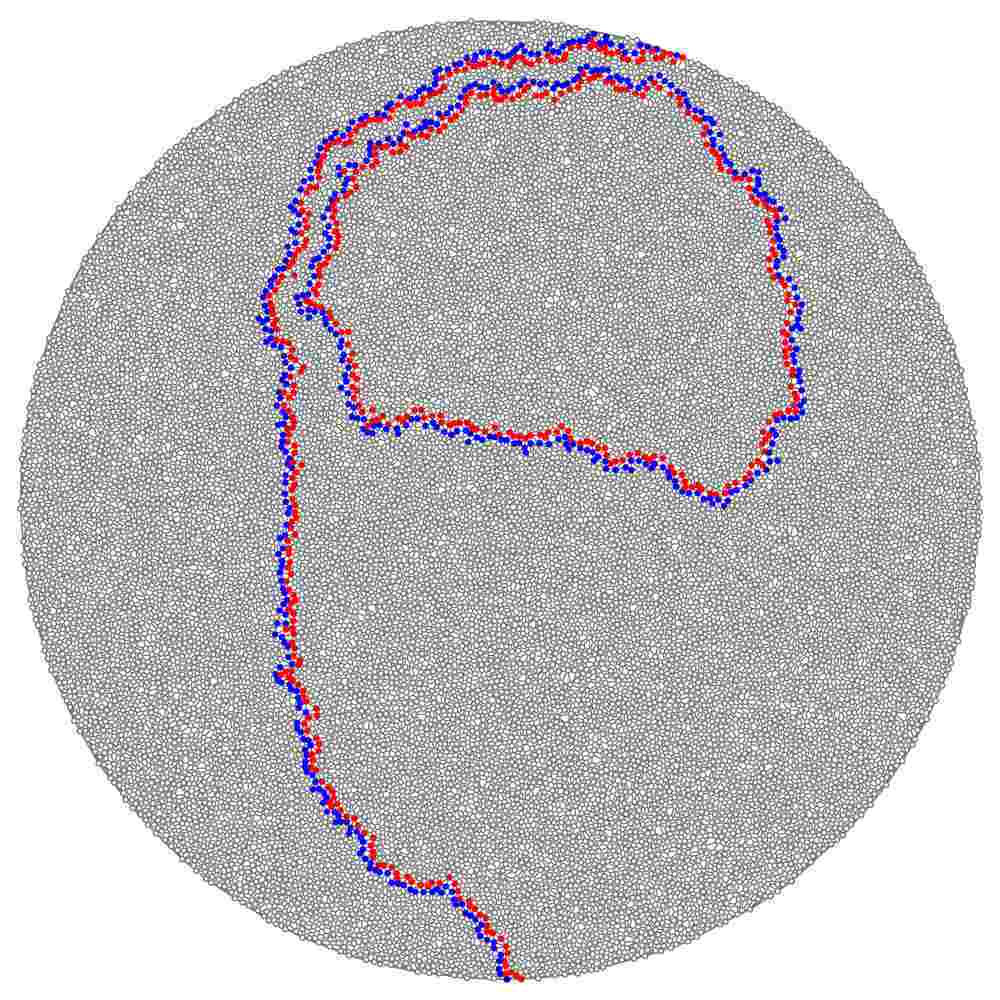}}
\subfigure[$t=132$]{\includegraphics[width=0.32\textwidth]{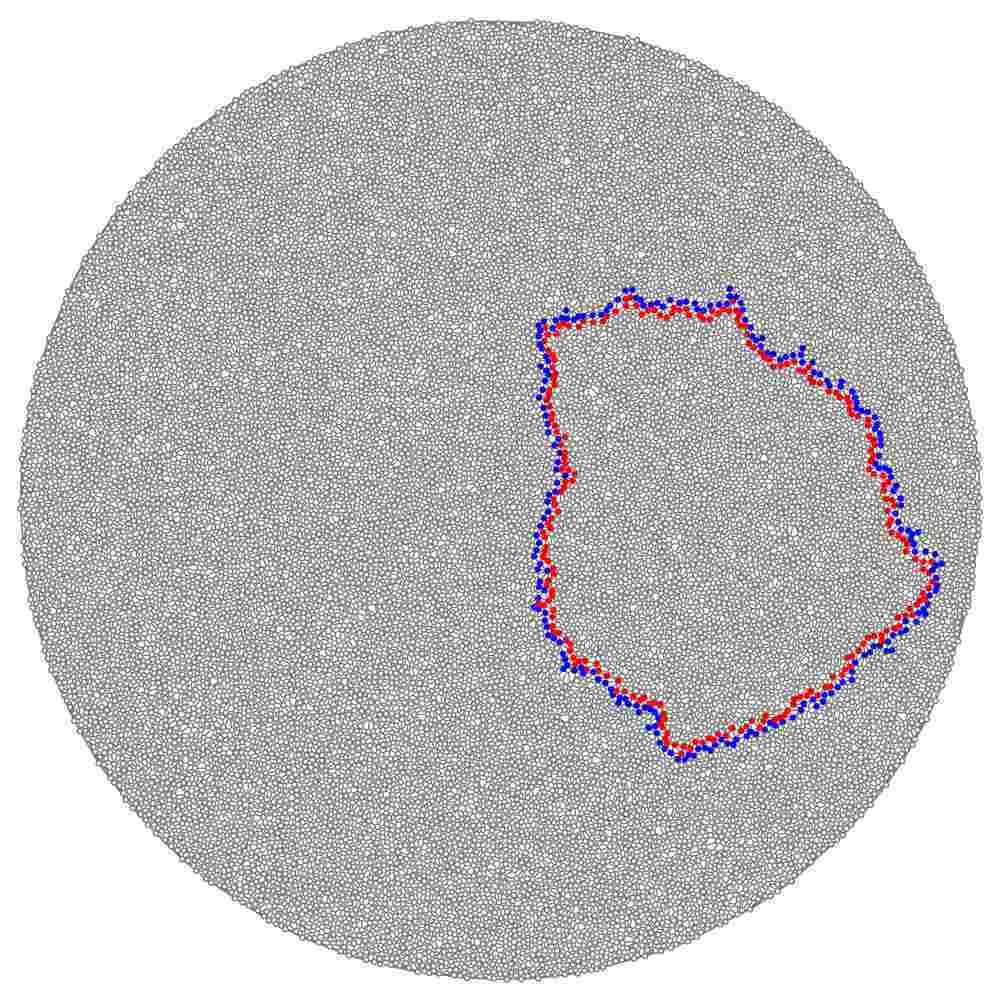}}
\caption{Dynamics of excitable triangulation for rule (\ref{eq2}) $\epsilon= 0.091$. Initially just a 
single node is excited. Density of disc-nodes in triangulation 
is $\phi=0.407$.} 
\label{reflection}
\end{figure}

When threshold of excitation increases to $\epsilon=0.09$ edges of the triangulation 
may become reflective for excitation waves. If an excitation wave front hits an edge 
of the triangulation the front 
reverses it velocity vector and again propagates inside the triangulation. An example is shown
in Fig.~\ref{reflection}. Initially graph is in resting state. We excited one node. A circular wave of excitation 
propagates outwards the initial perturbation (Fig.~\ref{reflection}a). When the wave front reaches edge of the 
triangulation it starts to disappear and almost annihilate. However in some part of triangulation edge 
domains of centrifugal wave-fragments evoke centripetal wave-fragments (Fig.~\ref{reflection}bc). 
Fragments of the centripetal wave closer to edge of triangulation propagate quicker then 
those inside the triangulation. Therefore the wave encircles the triangulation (Fig.~\ref{reflection}de) and 
collapses inside the triangulation (Fig.~\ref{reflection}f). 

The reflection of waves can be observed for values $0.09 < \epsilon < 0.11$. Further increase of excitation
threshold brings up the phenomenon of excitation backfiring again.

\begin{figure}[!tbp]
\centering
\subfigure[$t=18$]{\includegraphics[width=0.32\textwidth]{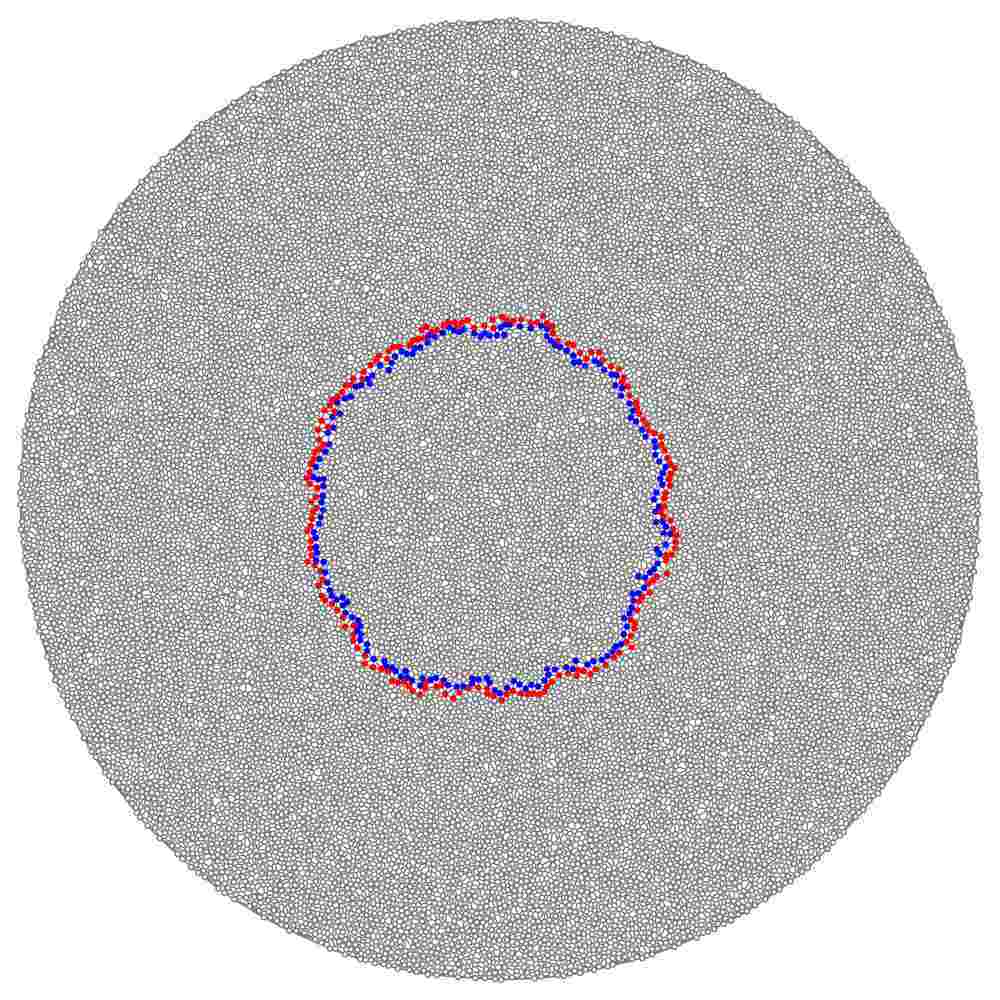}}
\subfigure[$t=28$]{\includegraphics[width=0.32\textwidth]{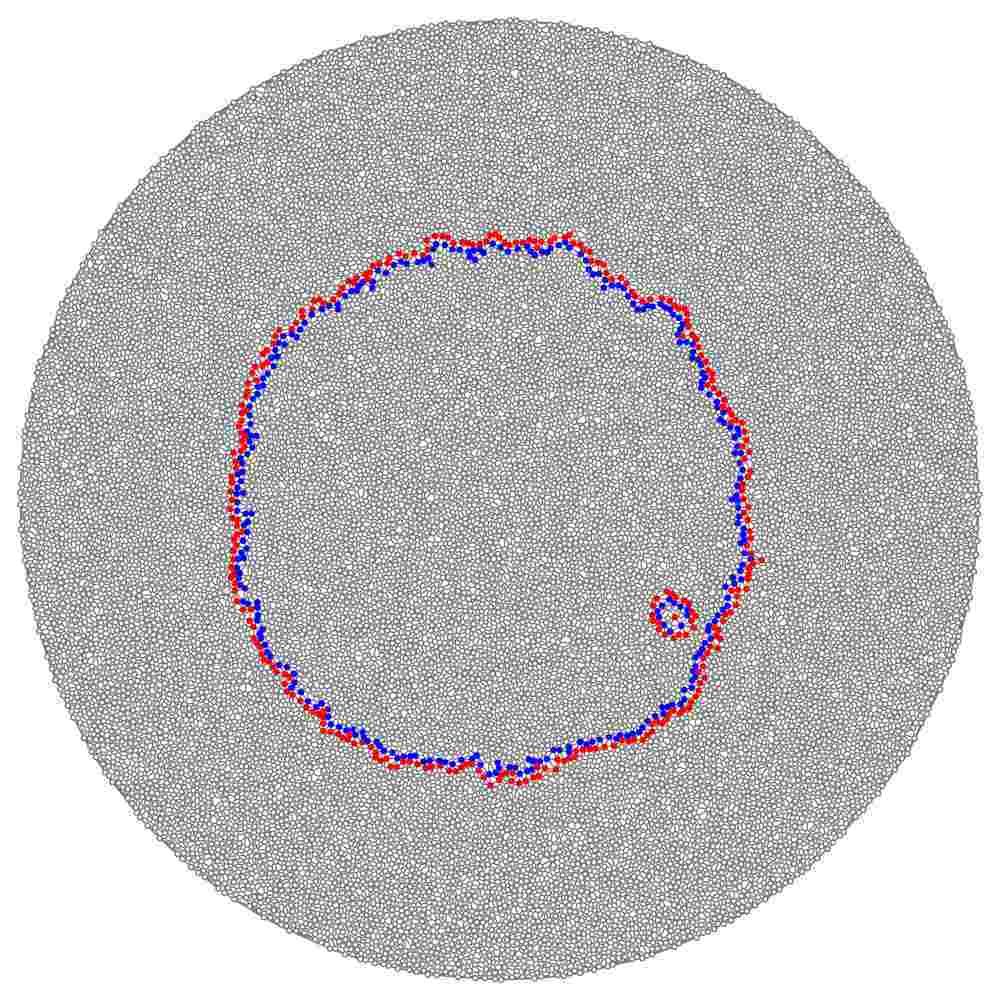}}
\subfigure[$t=37$]{\includegraphics[width=0.32\textwidth]{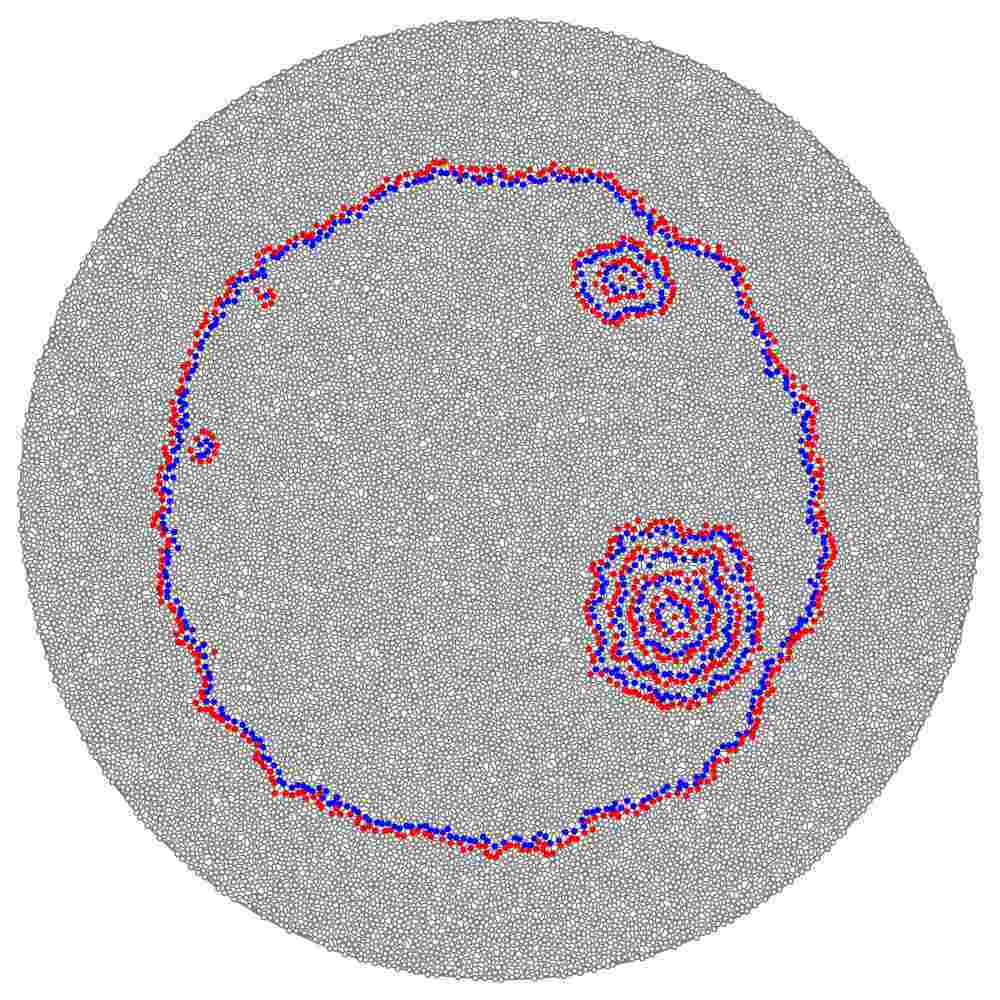}}
\subfigure[$t=45$]{\includegraphics[width=0.32\textwidth]{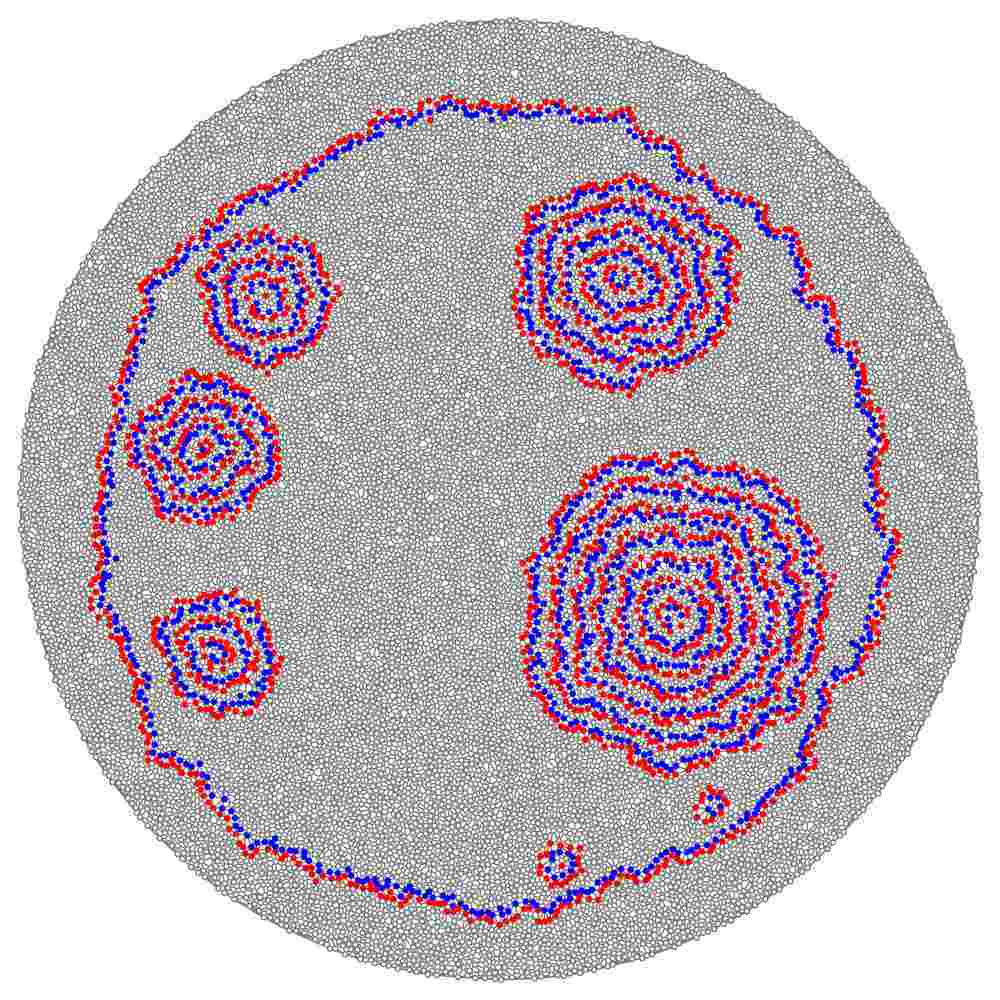}}
\subfigure[$t=53$]{\includegraphics[width=0.32\textwidth]{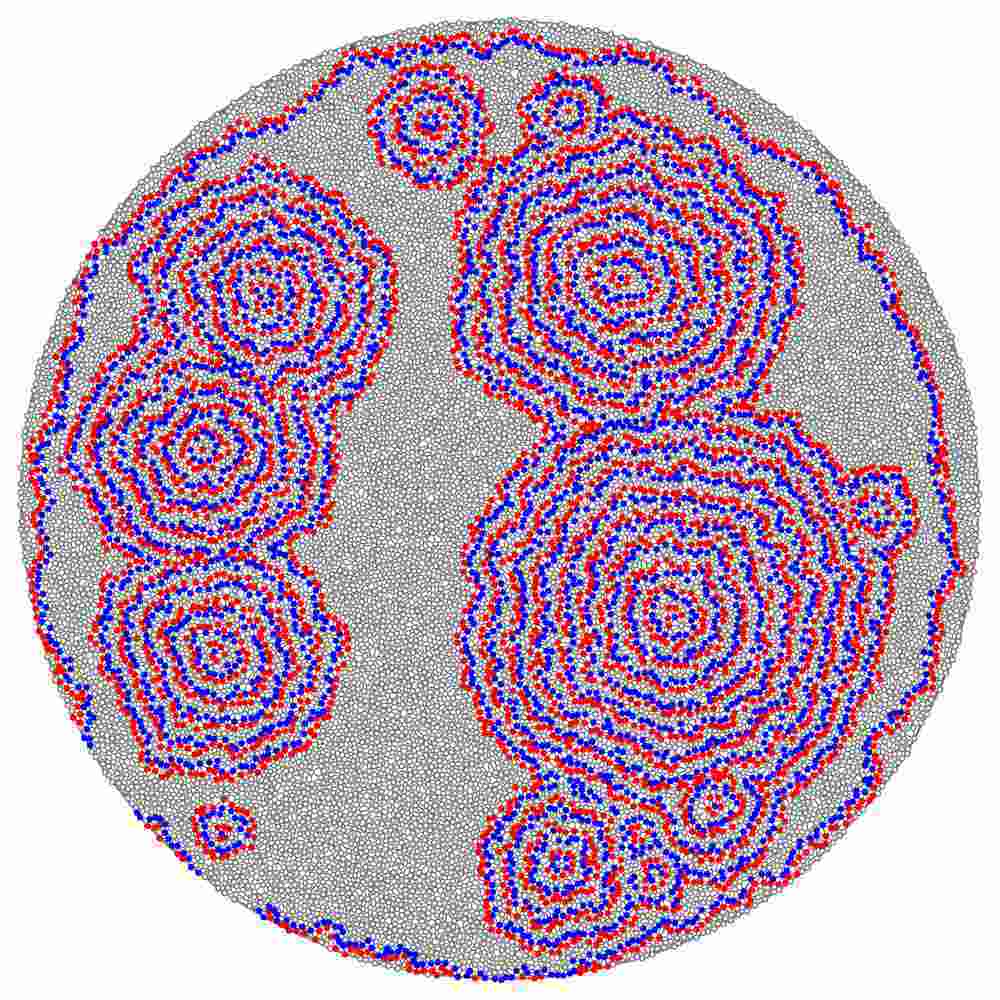}}
\subfigure[$t=83$]{\includegraphics[width=0.32\textwidth]{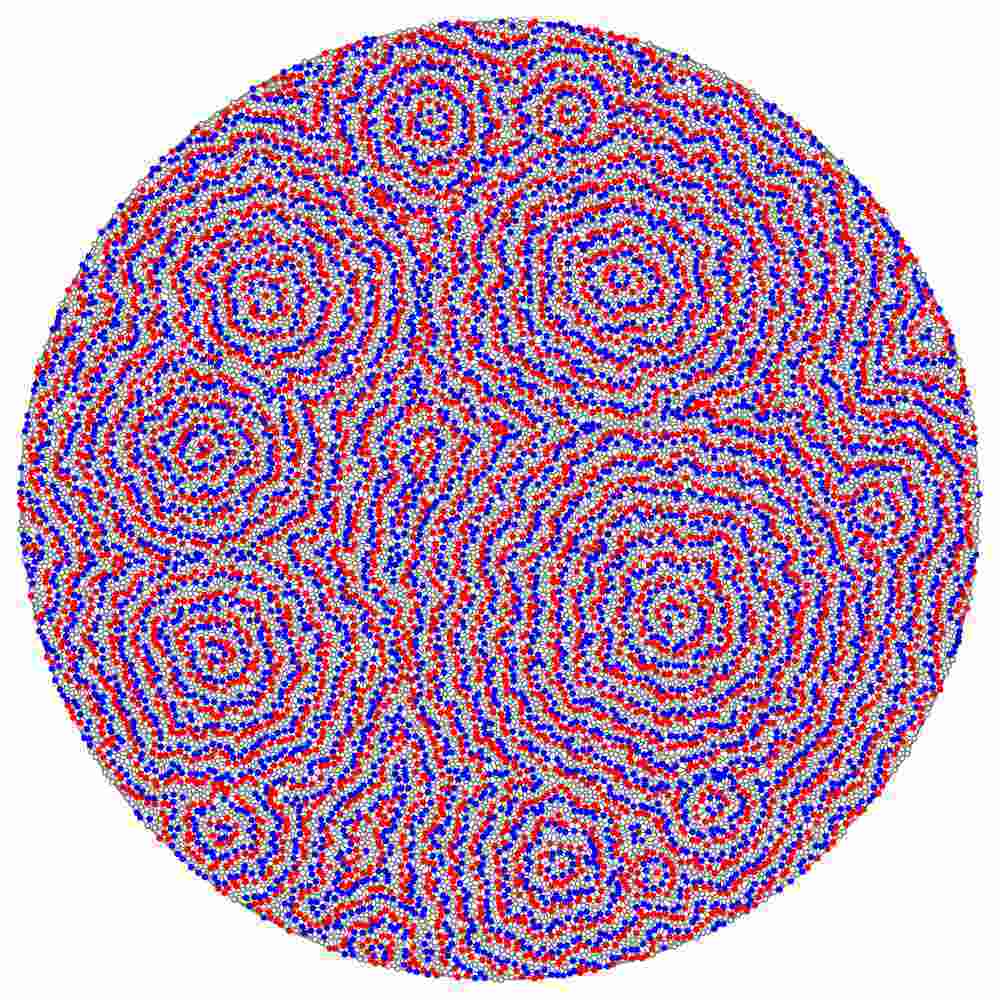}}
\caption{Example of backfiring of excitation wave fronts which leads to formation of generators 
of target waves in excitable triangulation, density $\phi=0.407$, for rule (\ref{eq2}) with  $\epsilon = 0.11$.} 
\label{backfiring0112}
\end{figure}

Triangulations, which nodes are excited if ratio $\epsilon$ is at least 0.11, exhibit backfiring --- 
generators of target waves are formed behind the wave front initiated by a single excitation 
(Fig.~\ref{backfiring0112}). Due to inhomogeneous structures of node neighbourhoods the wave 
front (Fig.~\ref{backfiring0112}a) breaks up (Fig.~\ref{backfiring0112}b). A gap is formed 
and some part of the front folds backward. A spiral wave or waves are formed. They give rise 
to a succession of target waves (Fig.~\ref{backfiring0112}cde). Eventually original wave front 
disappears at the edge of triangulation but the triangulation remains filled with sources 
of target waves (Fig.~\ref{backfiring0112}f). 

\begin{figure}[!tbp]
\centering
\includegraphics[width=0.45\textwidth]{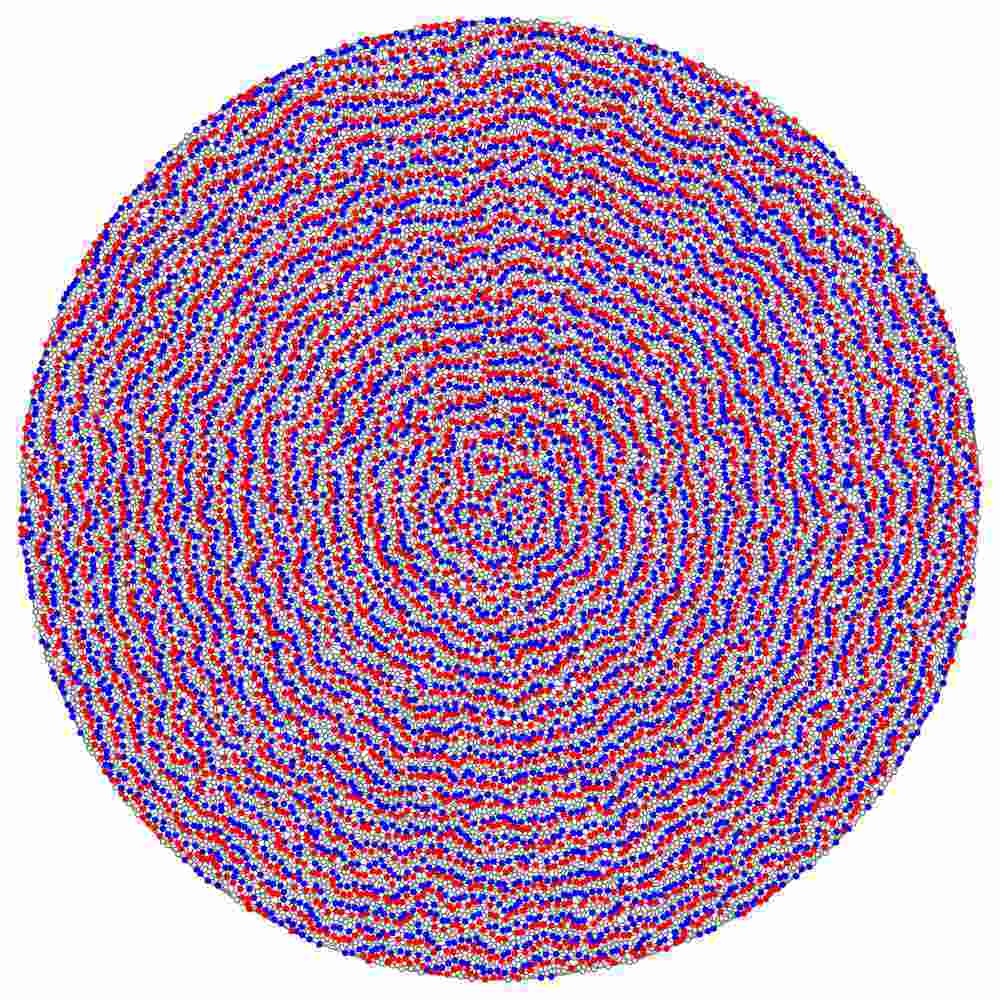}
\caption{Example of generators formed close to a site of original perturbation. Automaton is excited by 
rule (\ref{eq2}) with $\epsilon = 0.150$. Density of disc-nodes in triangulation 
is $\phi=0.407$.} 
\label{singlegenerator}
\end{figure}

With $\epsilon$ increasing from 0.11 to 0.17 a number of wave generators forming behind propagating wave front increases 
considerably. Thus, in networks excited by rule (\ref{eq2}) with $\epsilon > 0.14$ a generator is formed almost immediately after  
single-excitation wave starts its propagation. The earlier in time a generator is born the more likely the generator 
will dominate the triangulation. Therefore in many cases we can observe just few generators close to the site of original excitation (Fig.~\ref{singlegenerator}).

\begin{figure}[!tbp]
\centering
\subfigure[$t=10$]{\includegraphics[width=0.45\textwidth]{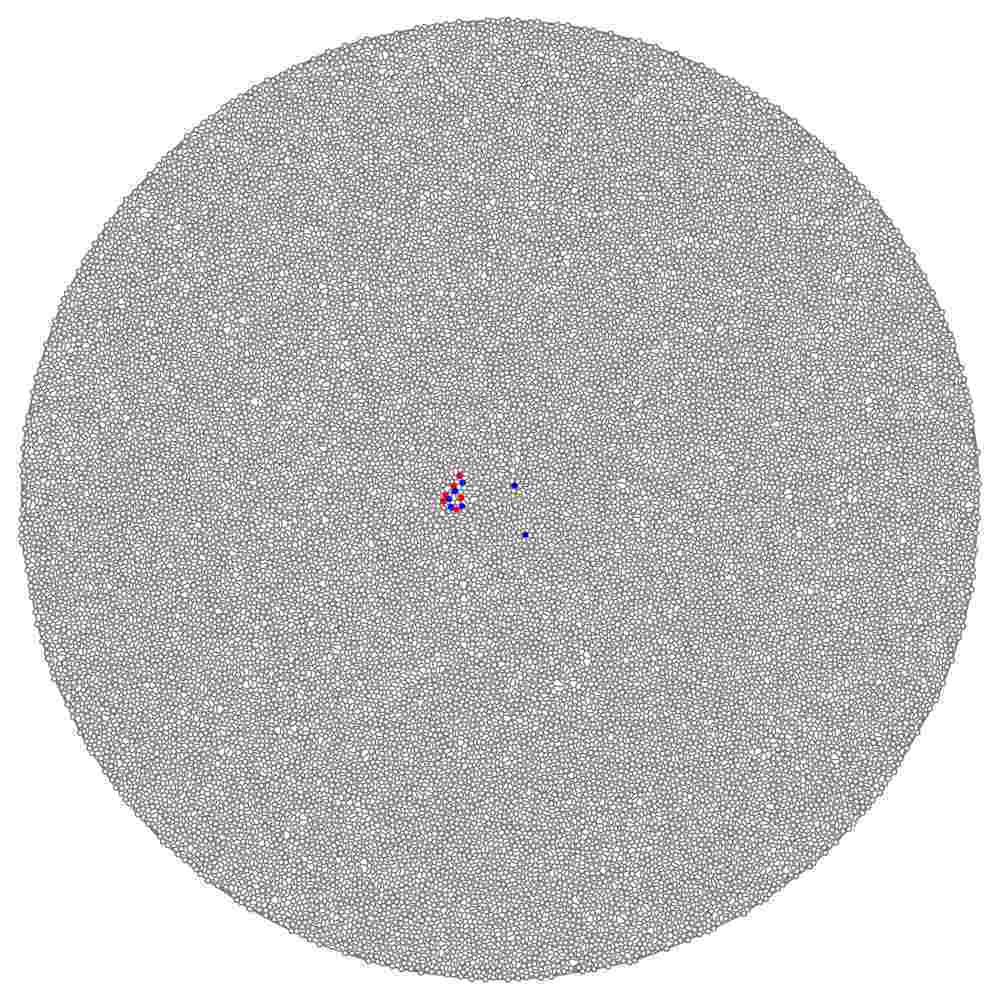}}
\subfigure[$t=16$]{\includegraphics[width=0.45\textwidth]{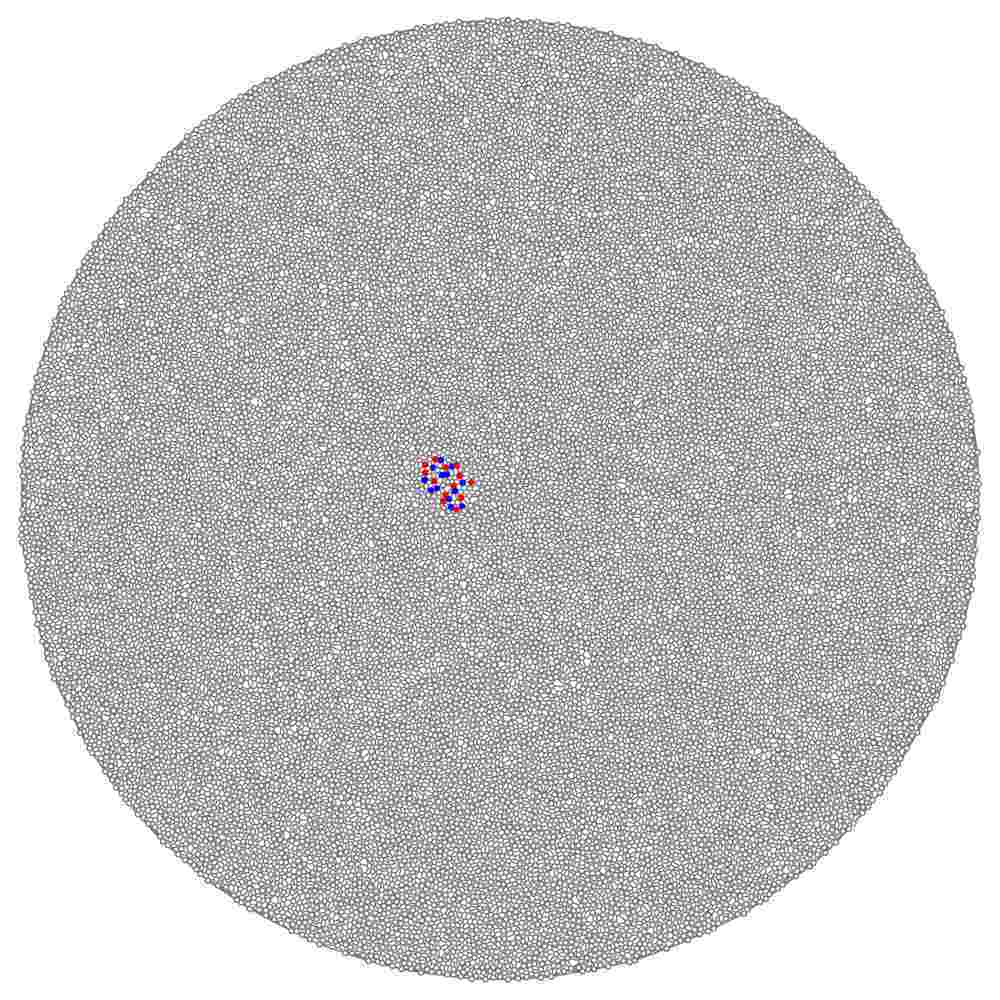}}
\subfigure[$t=40$]{\includegraphics[width=0.45\textwidth]{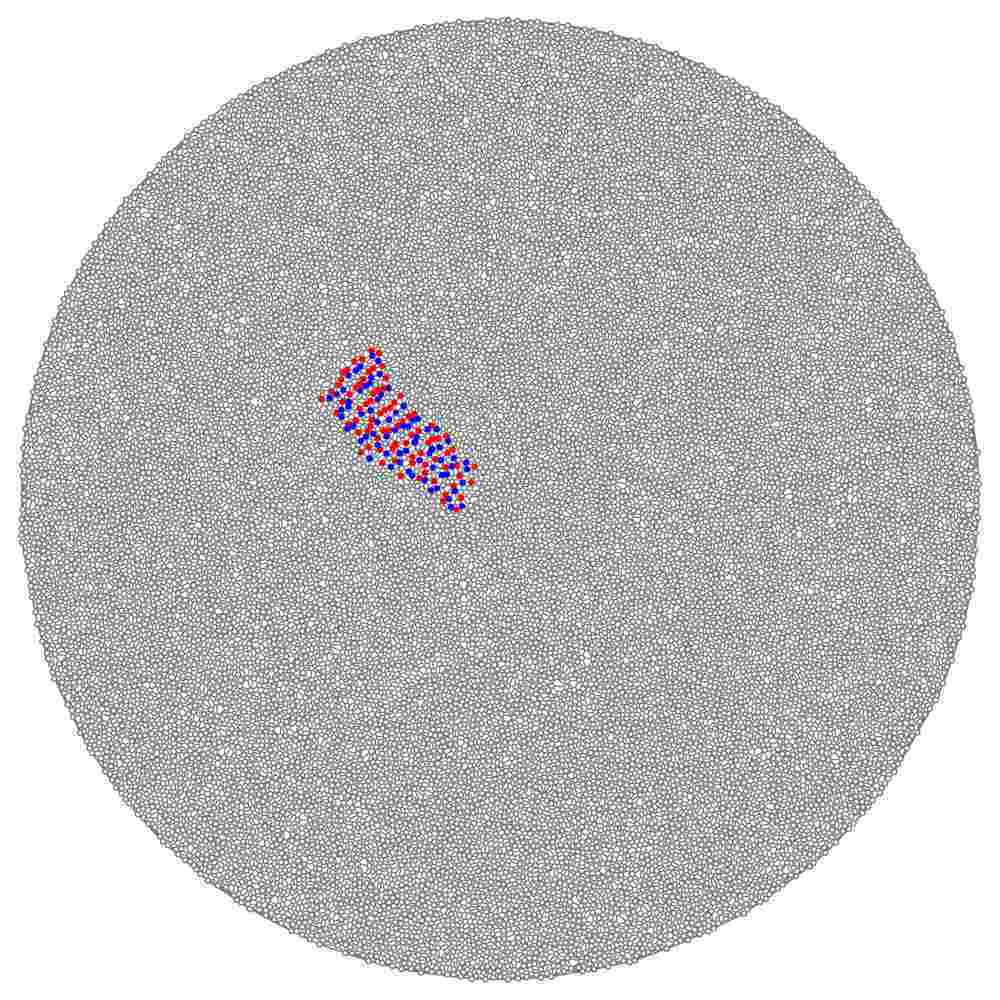}}
\subfigure[$t=142$]{\includegraphics[width=0.45\textwidth]{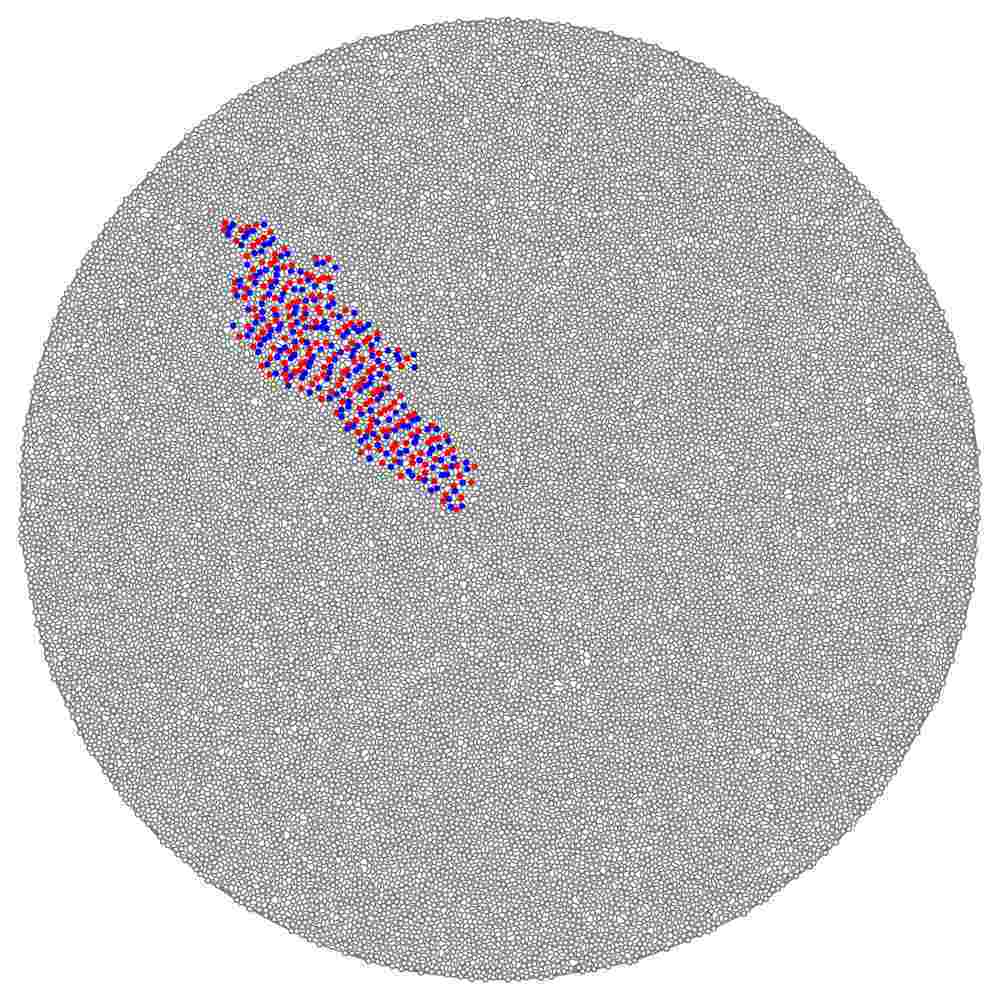}}
\caption{Example of excitation dynamics for $\epsilon=0.17$. A local perturbation leads to localised 
excitations and slowly growing domains. The domain shown in (d) has stationary boundaries
however configuration of excited and refractory states inside the domain is changing. Density of disc-nodes in triangulation 
is $\phi=0.407$. } 
\label{166domain}
\end{figure}

\begin{figure}[!tbp]
\centering
\subfigure[$t=7$]{\includegraphics[width=0.32\textwidth]{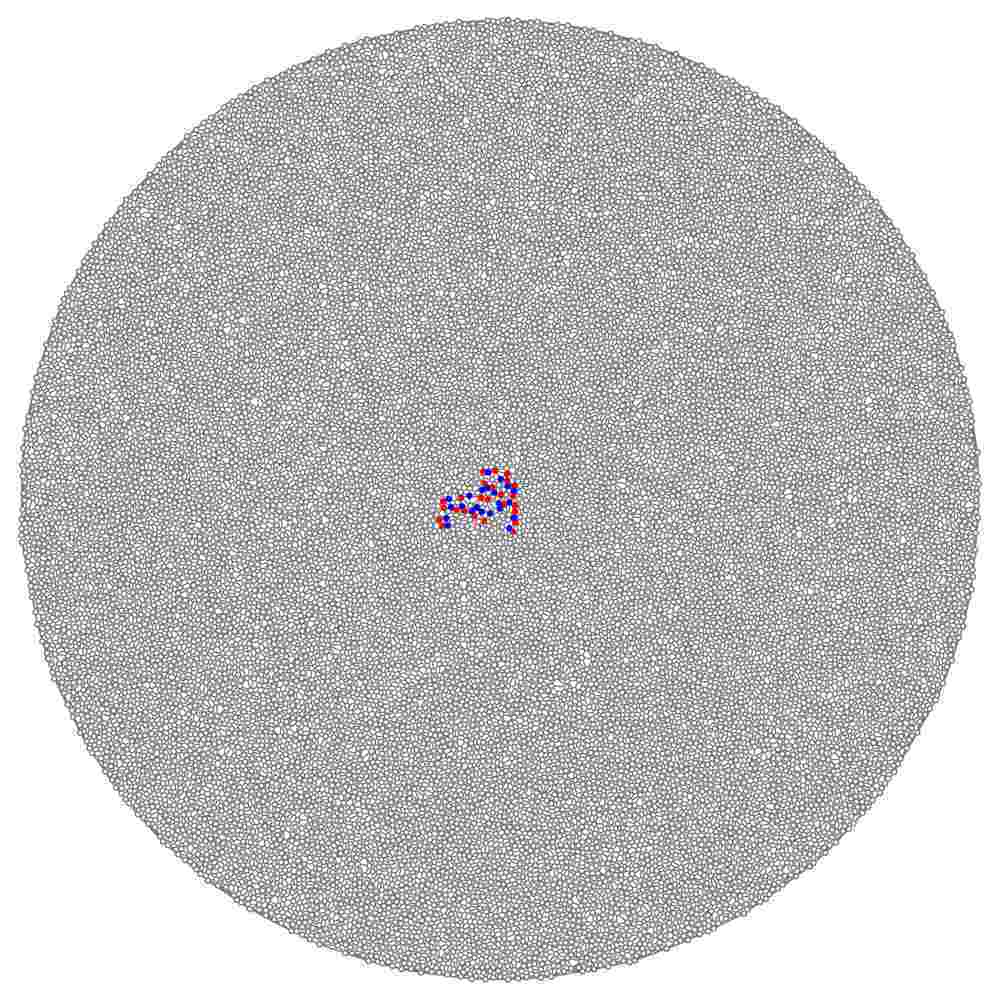}}
\subfigure[$t=23$]{\includegraphics[width=0.32\textwidth]{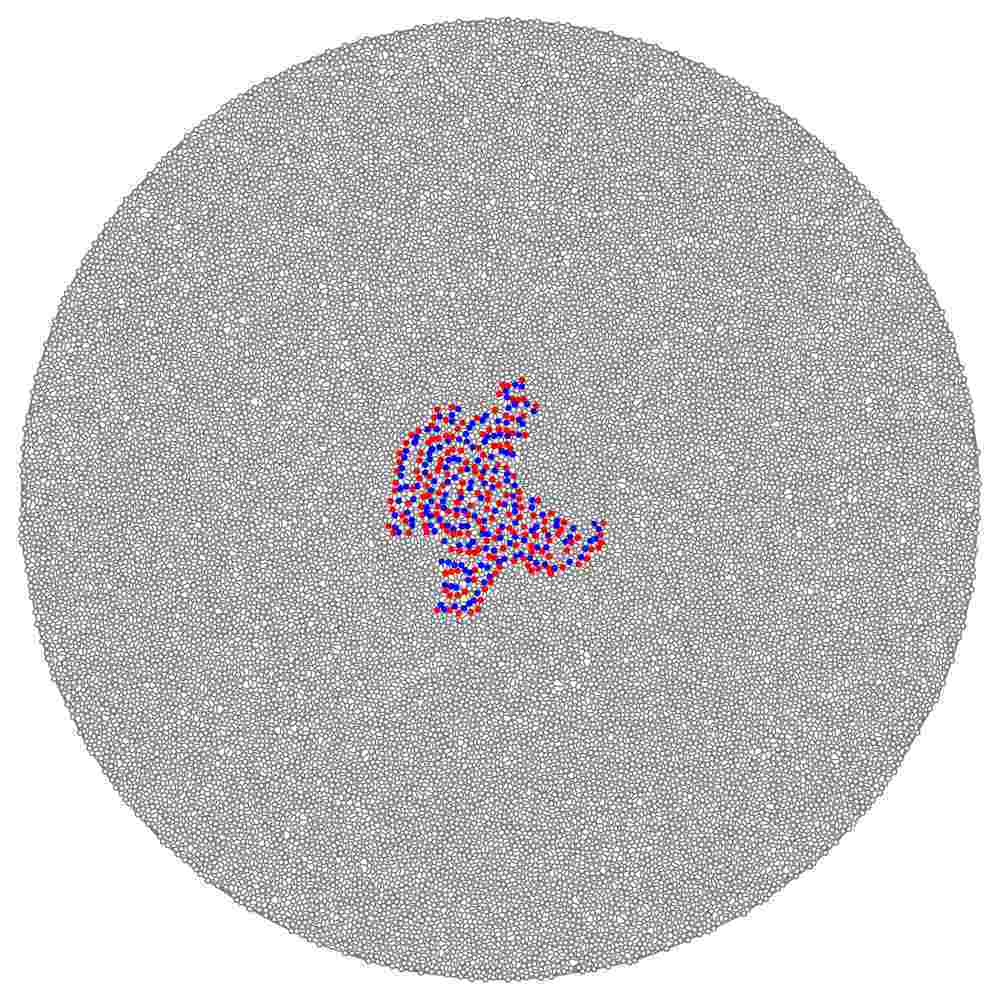}}
\subfigure[$t=33$]{\includegraphics[width=0.32\textwidth]{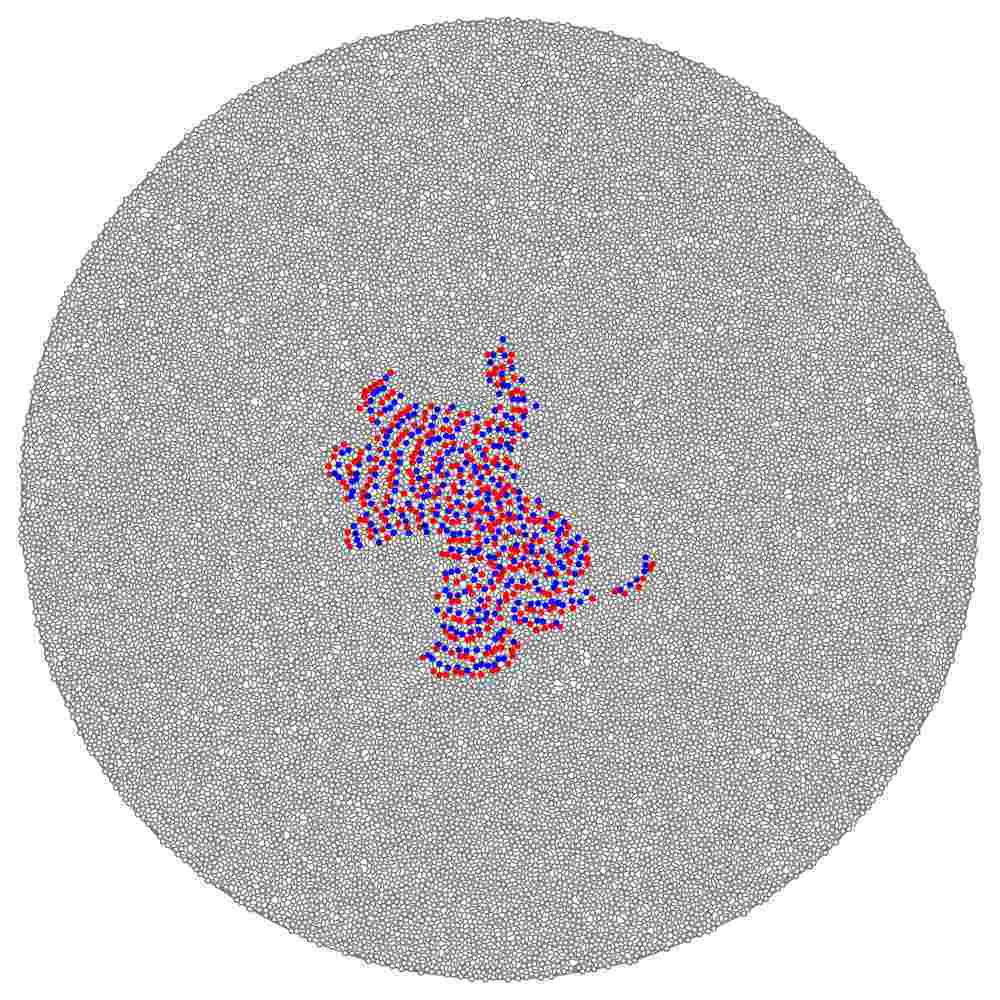}}
\subfigure[$t=53$]{\includegraphics[width=0.32\textwidth]{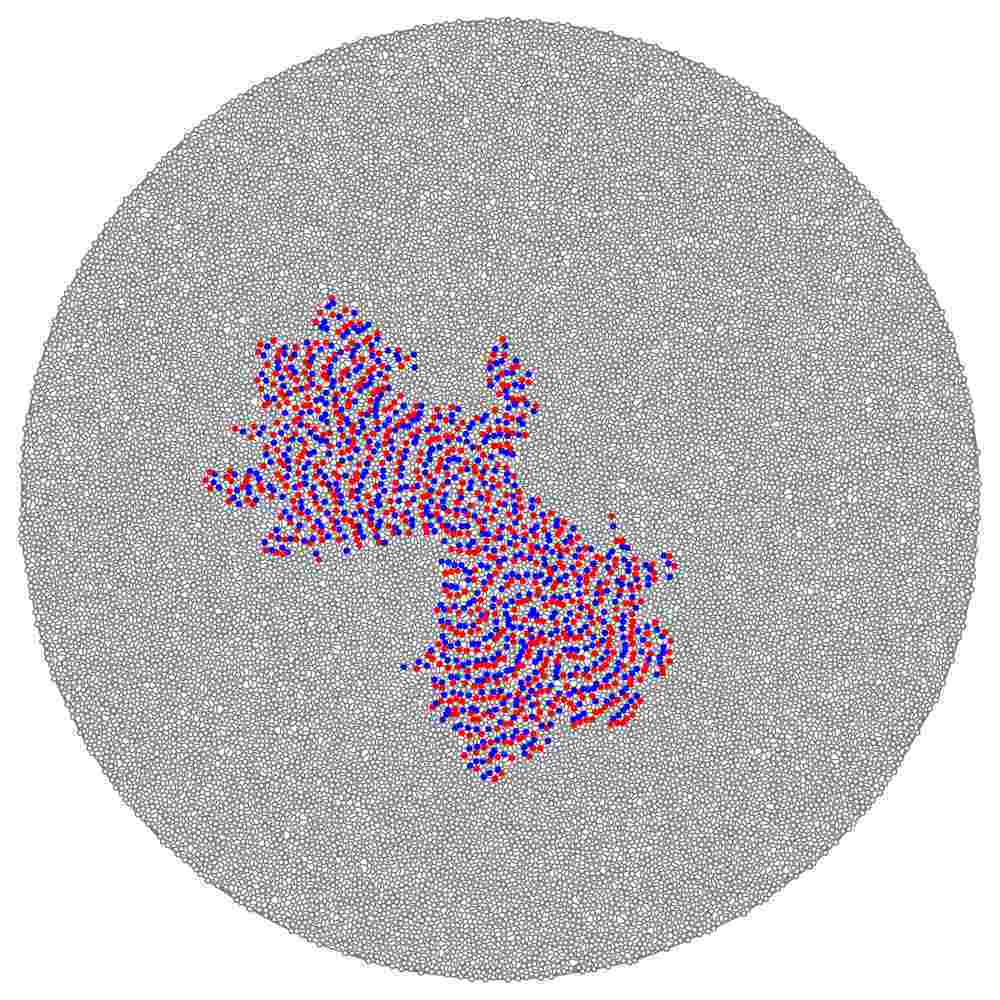}}
\subfigure[$t=119$]{\includegraphics[width=0.32\textwidth]{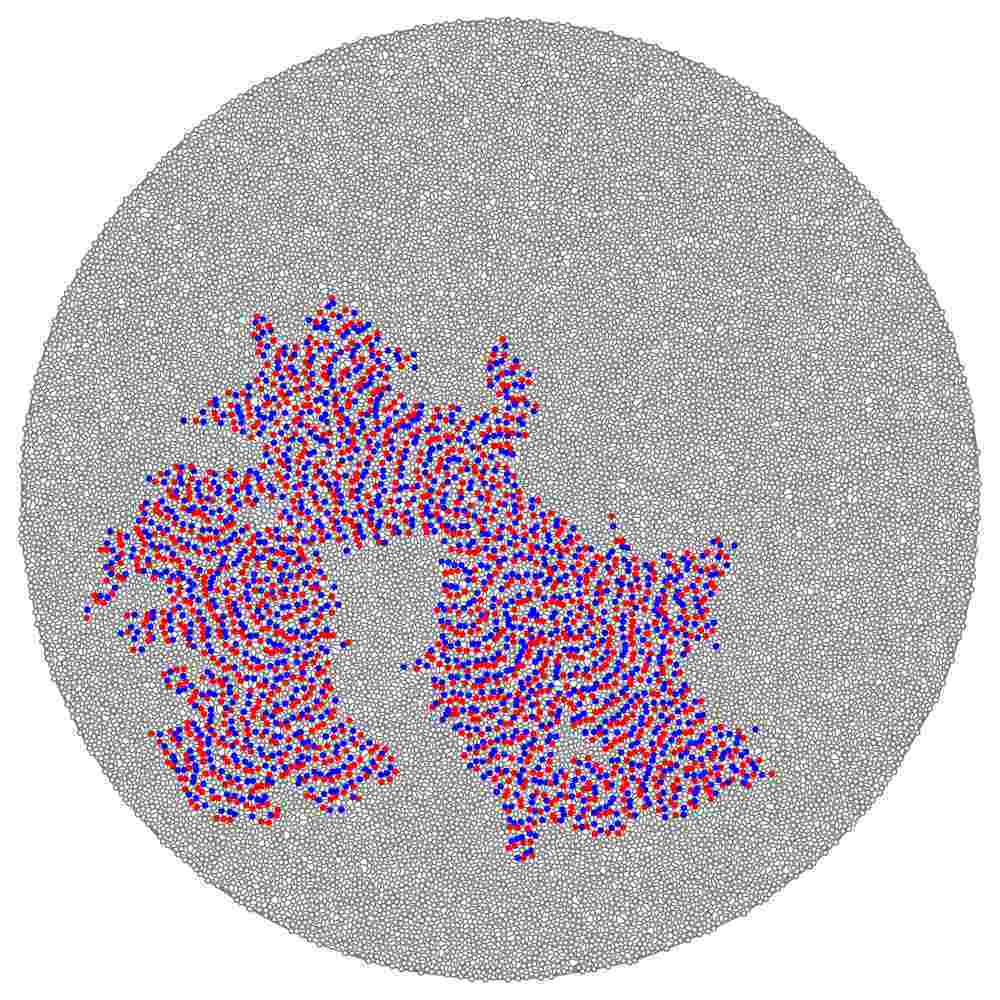}}
\subfigure[$t=278$]{\includegraphics[width=0.32\textwidth]{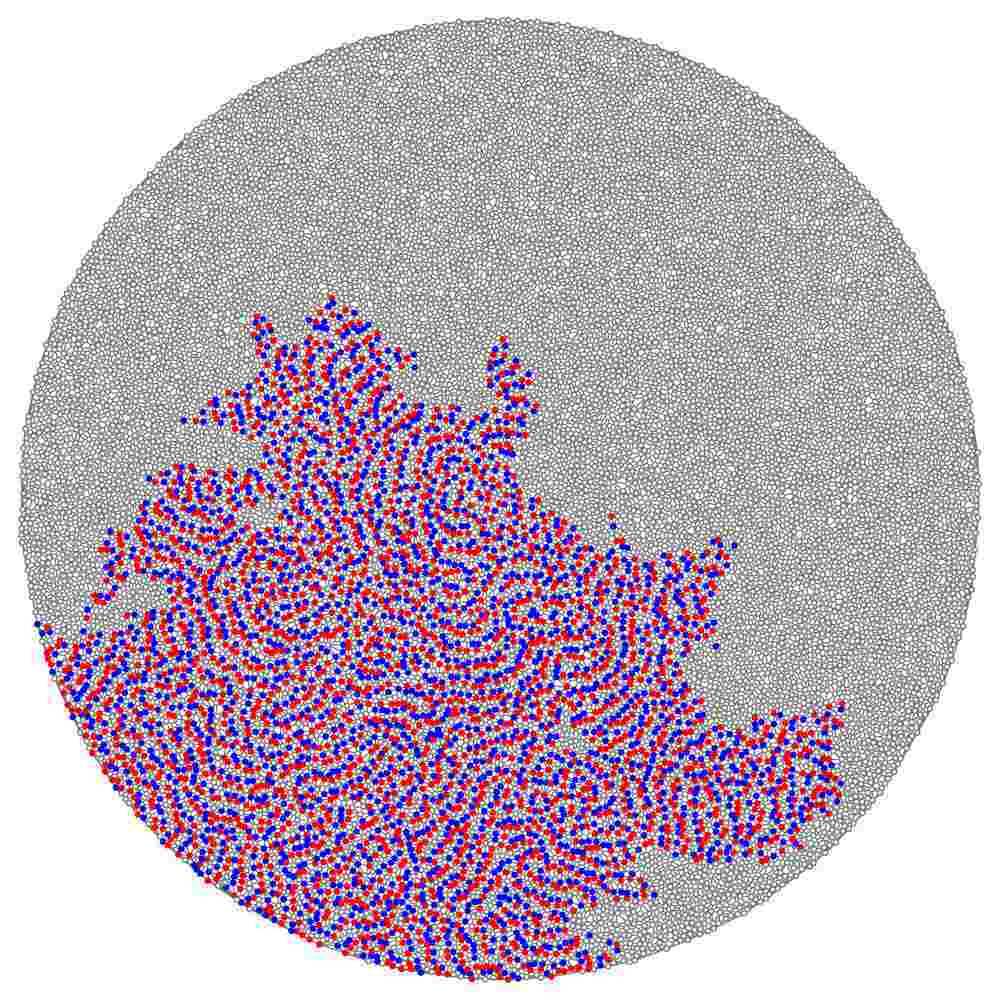}}
\caption{In triangulation, density of disc-nodes is $\phi=0.407$, excited by rule (\ref{eq2}) with $\epsilon=0.17$ a local perturbation leads to localised excitations and slowly propagating domains, which may occupy significant 
number of nodes.} 
\label{166domainbig}
\end{figure}

With $\epsilon$ exceeding 0.17 the automaton approaches regime of sub-excitability. Single site excitation 
no longer leads to formation of a circular wave. However we observe travelling and stationary localized 
excitations, distant analogous to wave-fragments in sub-excitable Belousov-Zhabotinsky medium~\cite{delacycostello_2005}.  
Initial local perturbations lead to propagating localized excitations. The excitations 
later can form a localized, and not changing it is outer shape, domain of activity,
see example in Fig.~\ref{166domain}, or a slowly growing domain, which may eventually occupy the whole 
triangulation (Fig.~\ref{166domainbig}).   Usually, two or three wave-fragments are formed near the site
of initial activation of resting triangulation. These wave-fragments travel outward the perturbation loci. 
Some time after their initiation the wave-fragment backfires few excited micro-localizations which may 
form generators of wave-fragments. Thus the domain of excitation is born.  Sometimes generators of 
wave-fragments emerge due to folding of the `wings' of a wave-segment backwards. Folded parts of wave-fragment 
interact with each other and thus they produce the generator.

\begin{figure}[!tbp]
\centering
\subfigure[]{
\includegraphics[width=0.224\textwidth]{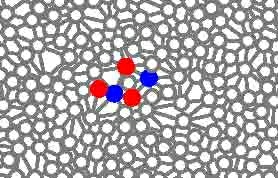}
\includegraphics[width=0.224\textwidth]{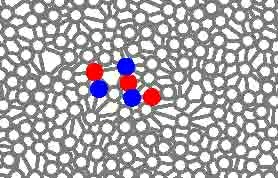}
\includegraphics[width=0.224\textwidth]{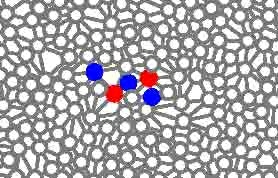}
}
\subfigure[]{
\includegraphics[width=0.175\textwidth]{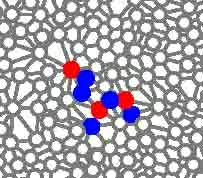}
\includegraphics[width=0.175\textwidth]{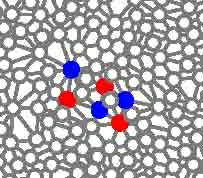}
\includegraphics[width=0.175\textwidth]{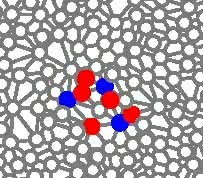}
}
\subfigure[]{
\includegraphics[width=0.14\textwidth]{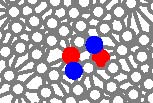}
\includegraphics[width=0.14\textwidth]{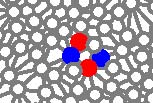}
\includegraphics[width=0.14\textwidth]{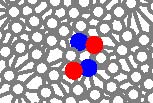}
}
\caption{Examples of oscillators:  (a) and (b) two oscillators for excitability $\epsilon = 0.2$,
(c)~minimal oscillator, $\epsilon=0.21$. Each sub-figure presents three consecutive states
of an oscillator. Density of disc-nodes in triangulation 
is $\phi=0.407$.} 
\label{oscillators}
\end{figure}

The growing domains guided by mobile localisations are typical for excitability values $0.17 < \epsilon < 0.2$. 
When $\epsilon$ exceed 0.2 the domains seize propagating. The excitation can still persist but in 
a minuscule quantity. A random initial configuration, where every node gets one of three states equiprobably, 
is either transformed to a totally resting state or to a resting configuration with one or two 
tiny oscillators. Examples of most common oscillators are shown in Fig.~\ref{oscillators}. 

What are degrees of nodes occupied by non-resting states of the oscillators? 
We stimulated triangulations with  random configurations of excited and refractory states (each state is assigned with 
probability $\frac{1}{3}$), waited till all activity but minimal oscillator ceases and recorded 
degrees of nodes occupied by the oscillator states. We found that nodes occupied by excited or refractory states in 
the minimal oscillator (Fig.~\ref{oscillators}c) have 4, 6, 7 or 8 neighbours. In larger 
oscillators nodes have degrees 4,5,6,7,8 and 4,5,6,7,8,9.     

\begin{proposition}
Diversity of node degrees is a necessary requirement for existence of an oscillator in 
a sub-excitable triangulation.
\end{proposition}

The oscillators survive for $0.2 \leq \epsilon < 0.25$. For $\epsilon \geq 0.25$ no excitation persist at all.

\begin{finding}
Delaunay excitable automata governed by rules of relative excitability with dimensionless threshold of excitation $\epsilon$ 
exhibit the following phenomena: \\
\begin{tabular}{l|l}
$\epsilon$                      & phenomenon \\ \hline
0                               & global oscillations \\
$0 < \epsilon < 0.09$           & classical excitation waves \\ 
$0.09 \leq \epsilon < 0.11$     & waves are reflected from edges \\
$0.11 \leq \epsilon < 0.17$     & waves backfire with excitation \\
$0.17 \leq \epsilon < 0.2$      & localized wave-fragments lead growing tips of branching domains \\
$0.2 \leq \epsilon < 0.25$      & only tiny oscillating domains are present \\
$0.25 \leq \epsilon$            & no excitation persists \\
\end{tabular}
\end{finding}

\begin{figure}[!tbp]
\centering
\includegraphics[width=0.99\textwidth]{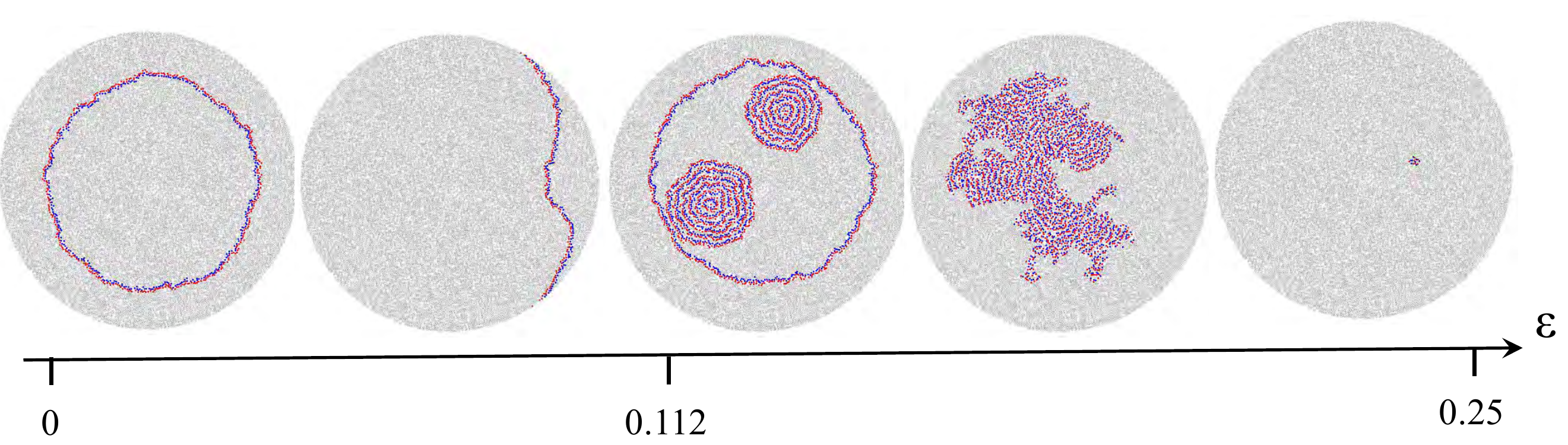}
\caption{Visualisation of basic types of excitation activity parameterised by relative excitability threshold $\epsilon$.}
\label{parameterisation}
\end{figure}

The finding is illustrated in Fig.~\ref{parameterisation}. Is there any structural meaning of $\epsilon$-boundaries between different classes of excitable triangulations?  Value $\epsilon=0.09$ corresponds to a situation when a node with 11 
neighbours has at least one excited neighbour.  Nodes with degree 11 are rare species in Delaunay triangulations (Fig.~\ref{degreedistribution}). However when such nodes are deprived, i.e. for $\epsilon > 0.09$, from a chance to be excited at all by a minimal perturbation waves of excitation start to be reflected by  edges of triangulation.  

Excitation wave-fronts backfire when we make nodes with degree 9 non-excitable by raising excitability 
to over $0.11$. This is because $\epsilon=0.11$ describes a situation of a node with 9 neighbours which has 
just one neighbour excited. 

Localizations emerge in triangulations when $\epsilon \geq 0.17$. In principle this value of $\epsilon$ corresponds
to two situations: a node with six neighbours has one excited neighbour and a node with 12 neighbours has two excited neighbours. 
The latter situation is not important because nodes with 12 neighbours are extremely rare. Thus, we can 
propose that excitation becomes localized when six-neighbour nodes are activated by at least two neighbours. 
Two excited neighbours is a minimal requirement for existence of localizations in 
orthogonal  automaton networks with eight-cell neighbourhoods~\cite{adamatzky_2001}. In a regular network, or
cellular automaton, a travelling localization preserves its shape and velocity vector as long as necessary.

Delaunay triangulations are irregular therefore localized wave-fragments do not survive for a long time. These travelling 
localizations frequently change directions of their motion. Soon after its birth a localization either perishes or expands, 
backfires and forms a slowlty growing cluster of activity (Fig.~\ref{166domainbig}). Any excitation ceases to sustain when $\epsilon$ exceeds 0.25, which corresponds to one of four,  two of eight, and three of twelve excited neighbours.

Does density of disc-nodes affect structure of behavioural space? 

For relative excitability 
$0 < \epsilon < 0.09$ decrease of density $\phi$ increases chances of wave-generators to be formed 
from a simple propagating wave-front. This  possibly happens due to increased inhomogeneity of 
node degrees and increasing role of each particular node in excitation propagation.
In case of $0.09 \leq \epsilon < 0.11$ decrease of $\phi$ increases opportunities for
wave-generators to form when a wave collides to edge of triangulation. 

For excitability $0.11 \leq \epsilon < 0.17$ time of backfiring and formation of generators 
behind wave-fronts, is proportional to density $\phi$, higher the $\phi$ the longer it takes for a 
singular wave-front to backfire. And finally, chances that an excitation domain growing from a single
perturbation, in triangulations with excitability $0.17 \leq \epsilon < 0.2$, occupies the whole 
triangulation are inversely proportional to density of disc-nodes in the triangulation.

\section{Discussion}
\label{discussion}

We defined excitable automata on Delaunay triangulation and demonstrated how to control a 
space-time dynamics of excitation on the triangulation using absolute and relative excitability 
thresholds. We uncovered several interesting phenomena ranging from reflection of excitation waves
by edge of triangulation to branching domains of activity guided by travelling localised excitations. 

Growing domains of localized excitation may be analogous to patterns of excitation in Belousov-Zhabotinsky 
reaction in micro-emulsion~\cite{kaminaga_2005}. Our findings on reflection of waves by edge of triangulation
support previously published results on active waves reflection in spatially inhomogeneous non-linear 
media~\cite{matsuoka_1998} and reflection of reaction-diffusion waves by no-flux boundary due to consumption 
of reactant between a wave and a boundary~\cite{mikhailov_2006}. Further studies of wave speed up along 
edge of triangulation could be enhanced by additional techniques of characterising topological and dynamical
properties of complex networks, e.g. a diversity entropy proposed in~\cite{viana_2010}. 

We believe our findings will contribute towards designs of novel computing substrates in non-crystalline 
structure. Also, automaton interpretation of activity dynamics on Delaunay triangulation 
can make a viable model of automaton-network approaches to design of 
nano-computing devices~\cite{isokawa_2007}.

\section{Acknowledgment}

The work is part of the European project 248992 funded under 7th FWP (Seventh Framework Programme) FET Proactive 3: Bio-Chemistry-Based Information Technology CHEM-IT (ICT-2009.8.3).

Many thanks to Julian Holley for commenting on the paper.

\end{document}

%% file: figs/dist7.tex
\setlength{\unitlength}{0.240900pt}
\ifx\plotpoint\undefined\newsavebox{\plotpoint}\fi
\sbox{\plotpoint}{\rule[-0.200pt]{0.400pt}{0.400pt}}%
\begin{picture}(1500,900)(0,0)
\font\gnuplot=cmr8 at 8pt
\gnuplot
\sbox{\plotpoint}{\rule[-0.200pt]{0.400pt}{0.400pt}}%
\put(211.0,131.0){\rule[-0.200pt]{293.657pt}{0.400pt}}
\put(211.0,131.0){\rule[-0.200pt]{4.818pt}{0.400pt}}
\put(191,131){\makebox(0,0)[r]{ 0}}
\put(1410.0,131.0){\rule[-0.200pt]{4.818pt}{0.400pt}}
\put(211.0,277.0){\rule[-0.200pt]{293.657pt}{0.400pt}}
\put(211.0,277.0){\rule[-0.200pt]{4.818pt}{0.400pt}}
\put(191,277){\makebox(0,0)[r]{ 0.1}}
\put(1410.0,277.0){\rule[-0.200pt]{4.818pt}{0.400pt}}
\put(211.0,423.0){\rule[-0.200pt]{293.657pt}{0.400pt}}
\put(211.0,423.0){\rule[-0.200pt]{4.818pt}{0.400pt}}
\put(191,423){\makebox(0,0)[r]{ 0.2}}
\put(1410.0,423.0){\rule[-0.200pt]{4.818pt}{0.400pt}}
\put(211.0,568.0){\rule[-0.200pt]{293.657pt}{0.400pt}}
\put(211.0,568.0){\rule[-0.200pt]{4.818pt}{0.400pt}}
\put(191,568){\makebox(0,0)[r]{ 0.3}}
\put(1410.0,568.0){\rule[-0.200pt]{4.818pt}{0.400pt}}
\put(211.0,714.0){\rule[-0.200pt]{202.115pt}{0.400pt}}
\put(1410.0,714.0){\rule[-0.200pt]{4.818pt}{0.400pt}}
\put(211.0,714.0){\rule[-0.200pt]{4.818pt}{0.400pt}}
\put(191,714){\makebox(0,0)[r]{ 0.4}}
\put(1410.0,714.0){\rule[-0.200pt]{4.818pt}{0.400pt}}
\put(211.0,860.0){\rule[-0.200pt]{293.657pt}{0.400pt}}
\put(211.0,860.0){\rule[-0.200pt]{4.818pt}{0.400pt}}
\put(191,860){\makebox(0,0)[r]{ 0.5}}
\put(1410.0,860.0){\rule[-0.200pt]{4.818pt}{0.400pt}}
\put(211.0,131.0){\rule[-0.200pt]{0.400pt}{175.616pt}}
\put(211.0,131.0){\rule[-0.200pt]{0.400pt}{4.818pt}}
\put(211,90){\makebox(0,0){ 0}}
\put(211.0,840.0){\rule[-0.200pt]{0.400pt}{4.818pt}}
\put(298.0,131.0){\rule[-0.200pt]{0.400pt}{175.616pt}}
\put(298.0,131.0){\rule[-0.200pt]{0.400pt}{4.818pt}}
\put(298,90){\makebox(0,0){ 1}}
\put(298.0,840.0){\rule[-0.200pt]{0.400pt}{4.818pt}}
\put(385.0,131.0){\rule[-0.200pt]{0.400pt}{175.616pt}}
\put(385.0,131.0){\rule[-0.200pt]{0.400pt}{4.818pt}}
\put(385,90){\makebox(0,0){ 2}}
\put(385.0,840.0){\rule[-0.200pt]{0.400pt}{4.818pt}}
\put(472.0,131.0){\rule[-0.200pt]{0.400pt}{175.616pt}}
\put(472.0,131.0){\rule[-0.200pt]{0.400pt}{4.818pt}}
\put(472,90){\makebox(0,0){ 3}}
\put(472.0,840.0){\rule[-0.200pt]{0.400pt}{4.818pt}}
\put(559.0,131.0){\rule[-0.200pt]{0.400pt}{175.616pt}}
\put(559.0,131.0){\rule[-0.200pt]{0.400pt}{4.818pt}}
\put(559,90){\makebox(0,0){ 4}}
\put(559.0,840.0){\rule[-0.200pt]{0.400pt}{4.818pt}}
\put(646.0,131.0){\rule[-0.200pt]{0.400pt}{175.616pt}}
\put(646.0,131.0){\rule[-0.200pt]{0.400pt}{4.818pt}}
\put(646,90){\makebox(0,0){ 5}}
\put(646.0,840.0){\rule[-0.200pt]{0.400pt}{4.818pt}}
\put(733.0,131.0){\rule[-0.200pt]{0.400pt}{175.616pt}}
\put(733.0,131.0){\rule[-0.200pt]{0.400pt}{4.818pt}}
\put(733,90){\makebox(0,0){ 6}}
\put(733.0,840.0){\rule[-0.200pt]{0.400pt}{4.818pt}}
\put(821.0,131.0){\rule[-0.200pt]{0.400pt}{175.616pt}}
\put(821.0,131.0){\rule[-0.200pt]{0.400pt}{4.818pt}}
\put(821,90){\makebox(0,0){ 7}}
\put(821.0,840.0){\rule[-0.200pt]{0.400pt}{4.818pt}}
\put(908.0,131.0){\rule[-0.200pt]{0.400pt}{175.616pt}}
\put(908.0,131.0){\rule[-0.200pt]{0.400pt}{4.818pt}}
\put(908,90){\makebox(0,0){ 8}}
\put(908.0,840.0){\rule[-0.200pt]{0.400pt}{4.818pt}}
\put(995.0,131.0){\rule[-0.200pt]{0.400pt}{175.616pt}}
\put(995.0,131.0){\rule[-0.200pt]{0.400pt}{4.818pt}}
\put(995,90){\makebox(0,0){ 9}}
\put(995.0,840.0){\rule[-0.200pt]{0.400pt}{4.818pt}}
\put(1082.0,131.0){\rule[-0.200pt]{0.400pt}{131.290pt}}
\put(1082.0,840.0){\rule[-0.200pt]{0.400pt}{4.818pt}}
\put(1082.0,131.0){\rule[-0.200pt]{0.400pt}{4.818pt}}
\put(1082,90){\makebox(0,0){ 10}}
\put(1082.0,840.0){\rule[-0.200pt]{0.400pt}{4.818pt}}
\put(1169.0,131.0){\rule[-0.200pt]{0.400pt}{131.290pt}}
\put(1169.0,840.0){\rule[-0.200pt]{0.400pt}{4.818pt}}
\put(1169.0,131.0){\rule[-0.200pt]{0.400pt}{4.818pt}}
\put(1169,90){\makebox(0,0){ 11}}
\put(1169.0,840.0){\rule[-0.200pt]{0.400pt}{4.818pt}}
\put(1256.0,131.0){\rule[-0.200pt]{0.400pt}{131.290pt}}
\put(1256.0,840.0){\rule[-0.200pt]{0.400pt}{4.818pt}}
\put(1256.0,131.0){\rule[-0.200pt]{0.400pt}{4.818pt}}
\put(1256,90){\makebox(0,0){ 12}}
\put(1256.0,840.0){\rule[-0.200pt]{0.400pt}{4.818pt}}
\put(1343.0,131.0){\rule[-0.200pt]{0.400pt}{131.290pt}}
\put(1343.0,840.0){\rule[-0.200pt]{0.400pt}{4.818pt}}
\put(1343.0,131.0){\rule[-0.200pt]{0.400pt}{4.818pt}}
\put(1343,90){\makebox(0,0){ 13}}
\put(1343.0,840.0){\rule[-0.200pt]{0.400pt}{4.818pt}}
\put(1430.0,131.0){\rule[-0.200pt]{0.400pt}{175.616pt}}
\put(1430.0,131.0){\rule[-0.200pt]{0.400pt}{4.818pt}}
\put(1430,90){\makebox(0,0){ 14}}
\put(1430.0,840.0){\rule[-0.200pt]{0.400pt}{4.818pt}}
\put(211.0,131.0){\rule[-0.200pt]{0.400pt}{175.616pt}}
\put(211.0,131.0){\rule[-0.200pt]{293.657pt}{0.400pt}}
\put(1430.0,131.0){\rule[-0.200pt]{0.400pt}{175.616pt}}
\put(211.0,860.0){\rule[-0.200pt]{293.657pt}{0.400pt}}
\put(70,495){\makebox(0,0){$p(d)$}}
\put(820,29){\makebox(0,0){$d$}}
\put(1270,820){\makebox(0,0)[r]{$\scriptstyle \phi=0.0027$}}
\put(1290.0,820.0){\rule[-0.200pt]{24.090pt}{0.400pt}}
\put(211,131){\usebox{\plotpoint}}
\put(298,130.67){\rule{20.958pt}{0.400pt}}
\multiput(298.00,130.17)(43.500,1.000){2}{\rule{10.479pt}{0.400pt}}
\multiput(385.00,132.58)(0.948,0.498){89}{\rule{0.857pt}{0.120pt}}
\multiput(385.00,131.17)(85.222,46.000){2}{\rule{0.428pt}{0.400pt}}
\multiput(472.58,178.00)(0.499,1.273){171}{\rule{0.120pt}{1.116pt}}
\multiput(471.17,178.00)(87.000,218.683){2}{\rule{0.400pt}{0.558pt}}
\multiput(559.58,399.00)(0.499,0.909){171}{\rule{0.120pt}{0.826pt}}
\multiput(558.17,399.00)(87.000,156.285){2}{\rule{0.400pt}{0.413pt}}
\multiput(646.00,555.92)(1.249,-0.498){67}{\rule{1.094pt}{0.120pt}}
\multiput(646.00,556.17)(84.729,-35.000){2}{\rule{0.547pt}{0.400pt}}
\multiput(733.58,518.25)(0.499,-1.007){173}{\rule{0.120pt}{0.905pt}}
\multiput(732.17,520.12)(88.000,-175.123){2}{\rule{0.400pt}{0.452pt}}
\multiput(821.58,342.05)(0.499,-0.765){171}{\rule{0.120pt}{0.711pt}}
\multiput(820.17,343.52)(87.000,-131.523){2}{\rule{0.400pt}{0.356pt}}
\multiput(908.00,210.92)(0.764,-0.499){111}{\rule{0.711pt}{0.120pt}}
\multiput(908.00,211.17)(85.525,-57.000){2}{\rule{0.355pt}{0.400pt}}
\multiput(995.00,153.92)(2.204,-0.496){37}{\rule{1.840pt}{0.119pt}}
\multiput(995.00,154.17)(83.181,-20.000){2}{\rule{0.920pt}{0.400pt}}
\multiput(1082.00,133.95)(19.216,-0.447){3}{\rule{11.700pt}{0.108pt}}
\multiput(1082.00,134.17)(62.716,-3.000){2}{\rule{5.850pt}{0.400pt}}
\put(211,131){\raisebox{-.8pt}{\makebox(0,0){$\scriptstyle \Diamond$}}}
\put(298,131){\raisebox{-.8pt}{\makebox(0,0){$\scriptstyle \Diamond$}}}
\put(385,132){\raisebox{-.8pt}{\makebox(0,0){$\scriptstyle \Diamond$}}}
\put(472,178){\raisebox{-.8pt}{\makebox(0,0){$\scriptstyle \Diamond$}}}
\put(559,399){\raisebox{-.8pt}{\makebox(0,0){$\scriptstyle \Diamond$}}}
\put(646,557){\raisebox{-.8pt}{\makebox(0,0){$\scriptstyle \Diamond$}}}
\put(733,522){\raisebox{-.8pt}{\makebox(0,0){$\scriptstyle \Diamond$}}}
\put(821,345){\raisebox{-.8pt}{\makebox(0,0){$\scriptstyle \Diamond$}}}
\put(908,212){\raisebox{-.8pt}{\makebox(0,0){$\scriptstyle \Diamond$}}}
\put(995,155){\raisebox{-.8pt}{\makebox(0,0){$\scriptstyle \Diamond$}}}
\put(1082,135){\raisebox{-.8pt}{\makebox(0,0){$\scriptstyle \Diamond$}}}
\put(1169,132){\raisebox{-.8pt}{\makebox(0,0){$\scriptstyle \Diamond$}}}
\put(1340,820){\raisebox{-.8pt}{\makebox(0,0){$\scriptstyle \Diamond$}}}
\put(211.0,131.0){\rule[-0.200pt]{20.958pt}{0.400pt}}
\put(1270,779){\makebox(0,0)[r]{$\scriptstyle \phi=0.027$}}
\multiput(1290,779)(20.756,0.000){5}{\usebox{\plotpoint}}
\put(1390,779){\usebox{\plotpoint}}
\put(211,131){\usebox{\plotpoint}}
\multiput(211,131)(20.756,0.000){5}{\usebox{\plotpoint}}
\multiput(298,131)(20.756,0.000){4}{\usebox{\plotpoint}}
\multiput(385,131)(20.413,3.754){4}{\usebox{\plotpoint}}
\multiput(472,147)(10.518,17.893){8}{\usebox{\plotpoint}}
\multiput(559,295)(7.073,19.513){13}{\usebox{\plotpoint}}
\multiput(646,535)(19.256,7.747){4}{\usebox{\plotpoint}}
\multiput(733,570)(9.721,-18.338){9}{\usebox{\plotpoint}}
\multiput(821,404)(10.209,-18.071){9}{\usebox{\plotpoint}}
\multiput(908,250)(14.846,-14.505){6}{\usebox{\plotpoint}}
\multiput(995,165)(19.886,-5.943){4}{\usebox{\plotpoint}}
\multiput(1082,139)(20.689,-1.665){4}{\usebox{\plotpoint}}
\multiput(1169,132)(20.754,-0.239){4}{\usebox{\plotpoint}}
\multiput(1256,131)(20.756,0.000){5}{\usebox{\plotpoint}}
\put(1343,131){\usebox{\plotpoint}}
\put(211,131){\makebox(0,0){$\scriptstyle +$}}
\put(298,131){\makebox(0,0){$\scriptstyle +$}}
\put(385,131){\makebox(0,0){$\scriptstyle +$}}
\put(472,147){\makebox(0,0){$\scriptstyle +$}}
\put(559,295){\makebox(0,0){$\scriptstyle +$}}
\put(646,535){\makebox(0,0){$\scriptstyle +$}}
\put(733,570){\makebox(0,0){$\scriptstyle +$}}
\put(821,404){\makebox(0,0){$\scriptstyle +$}}
\put(908,250){\makebox(0,0){$\scriptstyle +$}}
\put(995,165){\makebox(0,0){$\scriptstyle +$}}
\put(1082,139){\makebox(0,0){$\scriptstyle +$}}
\put(1169,132){\makebox(0,0){$\scriptstyle +$}}
\put(1256,131){\makebox(0,0){$\scriptstyle +$}}
\put(1343,131){\makebox(0,0){$\scriptstyle +$}}
\put(1340,779){\makebox(0,0){$\scriptstyle +$}}
\sbox{\plotpoint}{\rule[-0.400pt]{0.800pt}{0.800pt}}%
\sbox{\plotpoint}{\rule[-0.200pt]{0.400pt}{0.400pt}}%
\put(1270,738){\makebox(0,0)[r]{$\scriptstyle \phi=0.136$}}
\sbox{\plotpoint}{\rule[-0.400pt]{0.800pt}{0.800pt}}%
\put(1290.0,738.0){\rule[-0.400pt]{24.090pt}{0.800pt}}
\put(211,131){\usebox{\plotpoint}}
\multiput(385.00,132.38)(14.193,0.560){3}{\rule{14.120pt}{0.135pt}}
\multiput(385.00,129.34)(57.693,5.000){2}{\rule{7.060pt}{0.800pt}}
\multiput(473.41,136.00)(0.501,0.638){167}{\rule{0.121pt}{1.221pt}}
\multiput(470.34,136.00)(87.000,108.466){2}{\rule{0.800pt}{0.610pt}}
\multiput(560.41,247.00)(0.501,1.681){167}{\rule{0.121pt}{2.876pt}}
\multiput(557.34,247.00)(87.000,285.031){2}{\rule{0.800pt}{1.438pt}}
\multiput(647.41,538.00)(0.501,0.511){167}{\rule{0.121pt}{1.018pt}}
\multiput(644.34,538.00)(87.000,86.886){2}{\rule{0.800pt}{0.509pt}}
\multiput(734.41,618.66)(0.501,-1.135){169}{\rule{0.121pt}{2.009pt}}
\multiput(731.34,622.83)(88.000,-194.830){2}{\rule{0.800pt}{1.005pt}}
\multiput(822.41,419.92)(0.501,-1.096){167}{\rule{0.121pt}{1.947pt}}
\multiput(819.34,423.96)(87.000,-185.959){2}{\rule{0.800pt}{0.974pt}}
\multiput(908.00,236.09)(0.536,-0.501){155}{\rule{1.059pt}{0.121pt}}
\multiput(908.00,236.34)(84.801,-81.000){2}{\rule{0.530pt}{0.800pt}}
\multiput(995.00,155.09)(2.029,-0.505){37}{\rule{3.364pt}{0.122pt}}
\multiput(995.00,155.34)(80.019,-22.000){2}{\rule{1.682pt}{0.800pt}}
\put(1082,131.84){\rule{20.958pt}{0.800pt}}
\multiput(1082.00,133.34)(43.500,-3.000){2}{\rule{10.479pt}{0.800pt}}
\put(1169,129.84){\rule{20.958pt}{0.800pt}}
\multiput(1169.00,130.34)(43.500,-1.000){2}{\rule{10.479pt}{0.800pt}}
\put(211.0,131.0){\rule[-0.400pt]{41.917pt}{0.800pt}}
\put(211,131){\raisebox{-.8pt}{\makebox(0,0){$\scriptstyle \Box$}}}
\put(298,131){\raisebox{-.8pt}{\makebox(0,0){$\scriptstyle \Box$}}}
\put(385,131){\raisebox{-.8pt}{\makebox(0,0){$\scriptstyle \Box$}}}
\put(472,136){\raisebox{-.8pt}{\makebox(0,0){$\scriptstyle \Box$}}}
\put(559,247){\raisebox{-.8pt}{\makebox(0,0){$\scriptstyle \Box$}}}
\put(646,538){\raisebox{-.8pt}{\makebox(0,0){$\scriptstyle \Box$}}}
\put(733,627){\raisebox{-.8pt}{\makebox(0,0){$\scriptstyle \Box$}}}
\put(821,428){\raisebox{-.8pt}{\makebox(0,0){$\scriptstyle \Box$}}}
\put(908,238){\raisebox{-.8pt}{\makebox(0,0){$\scriptstyle \Box$}}}
\put(995,157){\raisebox{-.8pt}{\makebox(0,0){$\scriptstyle \Box$}}}
\put(1082,135){\raisebox{-.8pt}{\makebox(0,0){$\scriptstyle \Box$}}}
\put(1169,132){\raisebox{-.8pt}{\makebox(0,0){$\scriptstyle \Box$}}}
\put(1256,131){\raisebox{-.8pt}{\makebox(0,0){$\scriptstyle \Box$}}}
\put(1343,131){\raisebox{-.8pt}{\makebox(0,0){$\scriptstyle \Box$}}}
\put(1430,131){\raisebox{-.8pt}{\makebox(0,0){$\scriptstyle \Box$}}}
\put(1340,738){\raisebox{-.8pt}{\makebox(0,0){$\scriptstyle \Box$}}}
\put(1256.0,131.0){\rule[-0.400pt]{41.917pt}{0.800pt}}
\sbox{\plotpoint}{\rule[-0.500pt]{1.000pt}{1.000pt}}%
\sbox{\plotpoint}{\rule[-0.200pt]{0.400pt}{0.400pt}}%
\put(1270,697){\makebox(0,0)[r]{$\scriptstyle \phi=0.407$}}
\sbox{\plotpoint}{\rule[-0.500pt]{1.000pt}{1.000pt}}%
\multiput(1290,697)(20.756,0.000){5}{\usebox{\plotpoint}}
\put(1390,697){\usebox{\plotpoint}}
\put(211,131){\usebox{\plotpoint}}
\multiput(211,131)(20.756,0.000){5}{\usebox{\plotpoint}}
\multiput(298,131)(20.756,0.000){4}{\usebox{\plotpoint}}
\multiput(385,131)(20.756,0.000){4}{\usebox{\plotpoint}}
\multiput(472,131)(18.858,8.670){5}{\usebox{\plotpoint}}
\multiput(559,171)(5.061,20.129){17}{\usebox{\plotpoint}}
\multiput(646,517)(6.409,19.741){13}{\usebox{\plotpoint}}
\multiput(733,785)(5.244,-20.082){17}{\usebox{\plotpoint}}
\multiput(821,448)(6.563,-19.690){13}{\usebox{\plotpoint}}
\multiput(908,187)(17.816,-10.648){5}{\usebox{\plotpoint}}
\multiput(995,135)(20.734,-0.953){5}{\usebox{\plotpoint}}
\multiput(1082,131)(20.756,0.000){4}{\usebox{\plotpoint}}
\multiput(1169,131)(20.756,0.000){4}{\usebox{\plotpoint}}
\put(1256,131){\usebox{\plotpoint}}
\put(211,131){\makebox(0,0){$\scriptstyle \times$}}
\put(298,131){\makebox(0,0){$\scriptstyle \times$}}
\put(385,131){\makebox(0,0){$\scriptstyle \times$}}
\put(472,131){\makebox(0,0){$\scriptstyle \times$}}
\put(559,171){\makebox(0,0){$\scriptstyle \times$}}
\put(646,517){\makebox(0,0){$\scriptstyle \times$}}
\put(733,785){\makebox(0,0){$\scriptstyle \times$}}
\put(821,448){\makebox(0,0){$\scriptstyle \times$}}
\put(908,187){\makebox(0,0){$\scriptstyle \times$}}
\put(995,135){\makebox(0,0){$\scriptstyle \times$}}
\put(1082,131){\makebox(0,0){$\scriptstyle \times$}}
\put(1169,131){\makebox(0,0){$\scriptstyle \times$}}
\put(1256,131){\makebox(0,0){$\scriptstyle \times$}}
\put(1340,697){\makebox(0,0){$\scriptstyle \times$}}
\sbox{\plotpoint}{\rule[-0.200pt]{0.400pt}{0.400pt}}%
\put(211.0,131.0){\rule[-0.200pt]{0.400pt}{175.616pt}}
\put(211.0,131.0){\rule[-0.200pt]{293.657pt}{0.400pt}}
\put(1430.0,131.0){\rule[-0.200pt]{0.400pt}{175.616pt}}
\put(211.0,860.0){\rule[-0.200pt]{293.657pt}{0.400pt}}
\end{picture}

%% file: figs/deviation3.tex
\setlength{\unitlength}{0.240900pt}
\ifx\plotpoint\undefined\newsavebox{\plotpoint}\fi
\sbox{\plotpoint}{\rule[-0.200pt]{0.400pt}{0.400pt}}%
\begin{picture}(1500,900)(0,0)
\font\gnuplot=cmr10 at 8pt
\gnuplot
\sbox{\plotpoint}{\rule[-0.200pt]{0.400pt}{0.400pt}}%
\put(185.0,105.0){\rule[-0.200pt]{303.293pt}{0.400pt}}
\put(185.0,105.0){\rule[-0.200pt]{4.818pt}{0.400pt}}
\put(169,105){\makebox(0,0)[r]{ 0}}
\put(1424.0,105.0){\rule[-0.200pt]{4.818pt}{0.400pt}}
\put(185.0,258.0){\rule[-0.200pt]{303.293pt}{0.400pt}}
\put(185.0,258.0){\rule[-0.200pt]{4.818pt}{0.400pt}}
\put(169,258){\makebox(0,0)[r]{ 0.01}}
\put(1424.0,258.0){\rule[-0.200pt]{4.818pt}{0.400pt}}
\put(185.0,410.0){\rule[-0.200pt]{303.293pt}{0.400pt}}
\put(185.0,410.0){\rule[-0.200pt]{4.818pt}{0.400pt}}
\put(169,410){\makebox(0,0)[r]{ 0.02}}
\put(1424.0,410.0){\rule[-0.200pt]{4.818pt}{0.400pt}}
\put(185.0,563.0){\rule[-0.200pt]{303.293pt}{0.400pt}}
\put(185.0,563.0){\rule[-0.200pt]{4.818pt}{0.400pt}}
\put(169,563){\makebox(0,0)[r]{ 0.03}}
\put(1424.0,563.0){\rule[-0.200pt]{4.818pt}{0.400pt}}
\put(185.0,715.0){\rule[-0.200pt]{303.293pt}{0.400pt}}
\put(185.0,715.0){\rule[-0.200pt]{4.818pt}{0.400pt}}
\put(169,715){\makebox(0,0)[r]{ 0.04}}
\put(1424.0,715.0){\rule[-0.200pt]{4.818pt}{0.400pt}}
\put(185.0,868.0){\rule[-0.200pt]{303.293pt}{0.400pt}}
\put(185.0,868.0){\rule[-0.200pt]{4.818pt}{0.400pt}}
\put(169,868){\makebox(0,0)[r]{ 0.05}}
\put(1424.0,868.0){\rule[-0.200pt]{4.818pt}{0.400pt}}
\put(185.0,105.0){\rule[-0.200pt]{0.400pt}{183.807pt}}
\put(185.0,105.0){\rule[-0.200pt]{0.400pt}{4.818pt}}
\put(185,72){\makebox(0,0){ 0}}
\put(185.0,848.0){\rule[-0.200pt]{0.400pt}{4.818pt}}
\put(275.0,105.0){\rule[-0.200pt]{0.400pt}{183.807pt}}
\put(275.0,105.0){\rule[-0.200pt]{0.400pt}{4.818pt}}
\put(275,72){\makebox(0,0){ 1}}
\put(275.0,848.0){\rule[-0.200pt]{0.400pt}{4.818pt}}
\put(365.0,105.0){\rule[-0.200pt]{0.400pt}{183.807pt}}
\put(365.0,105.0){\rule[-0.200pt]{0.400pt}{4.818pt}}
\put(365,72){\makebox(0,0){ 2}}
\put(365.0,848.0){\rule[-0.200pt]{0.400pt}{4.818pt}}
\put(455.0,105.0){\rule[-0.200pt]{0.400pt}{183.807pt}}
\put(455.0,105.0){\rule[-0.200pt]{0.400pt}{4.818pt}}
\put(455,72){\makebox(0,0){ 3}}
\put(455.0,848.0){\rule[-0.200pt]{0.400pt}{4.818pt}}
\put(545.0,105.0){\rule[-0.200pt]{0.400pt}{183.807pt}}
\put(545.0,105.0){\rule[-0.200pt]{0.400pt}{4.818pt}}
\put(545,72){\makebox(0,0){ 4}}
\put(545.0,848.0){\rule[-0.200pt]{0.400pt}{4.818pt}}
\put(635.0,105.0){\rule[-0.200pt]{0.400pt}{183.807pt}}
\put(635.0,105.0){\rule[-0.200pt]{0.400pt}{4.818pt}}
\put(635,72){\makebox(0,0){ 5}}
\put(635.0,848.0){\rule[-0.200pt]{0.400pt}{4.818pt}}
\put(725.0,105.0){\rule[-0.200pt]{0.400pt}{183.807pt}}
\put(725.0,105.0){\rule[-0.200pt]{0.400pt}{4.818pt}}
\put(725,72){\makebox(0,0){ 6}}
\put(725.0,848.0){\rule[-0.200pt]{0.400pt}{4.818pt}}
\put(815.0,105.0){\rule[-0.200pt]{0.400pt}{183.807pt}}
\put(815.0,105.0){\rule[-0.200pt]{0.400pt}{4.818pt}}
\put(815,72){\makebox(0,0){ 7}}
\put(815.0,848.0){\rule[-0.200pt]{0.400pt}{4.818pt}}
\put(904.0,105.0){\rule[-0.200pt]{0.400pt}{183.807pt}}
\put(904.0,105.0){\rule[-0.200pt]{0.400pt}{4.818pt}}
\put(904,72){\makebox(0,0){ 8}}
\put(904.0,848.0){\rule[-0.200pt]{0.400pt}{4.818pt}}
\put(994.0,105.0){\rule[-0.200pt]{0.400pt}{183.807pt}}
\put(994.0,105.0){\rule[-0.200pt]{0.400pt}{4.818pt}}
\put(994,72){\makebox(0,0){ 9}}
\put(994.0,848.0){\rule[-0.200pt]{0.400pt}{4.818pt}}
\put(1084.0,105.0){\rule[-0.200pt]{0.400pt}{183.807pt}}
\put(1084.0,105.0){\rule[-0.200pt]{0.400pt}{4.818pt}}
\put(1084,72){\makebox(0,0){ 10}}
\put(1084.0,848.0){\rule[-0.200pt]{0.400pt}{4.818pt}}
\put(1174.0,105.0){\rule[-0.200pt]{0.400pt}{147.190pt}}
\put(1174.0,848.0){\rule[-0.200pt]{0.400pt}{4.818pt}}
\put(1174.0,105.0){\rule[-0.200pt]{0.400pt}{4.818pt}}
\put(1174,72){\makebox(0,0){ 11}}
\put(1174.0,848.0){\rule[-0.200pt]{0.400pt}{4.818pt}}
\put(1264.0,105.0){\rule[-0.200pt]{0.400pt}{147.190pt}}
\put(1264.0,848.0){\rule[-0.200pt]{0.400pt}{4.818pt}}
\put(1264.0,105.0){\rule[-0.200pt]{0.400pt}{4.818pt}}
\put(1264,72){\makebox(0,0){ 12}}
\put(1264.0,848.0){\rule[-0.200pt]{0.400pt}{4.818pt}}
\put(1354.0,105.0){\rule[-0.200pt]{0.400pt}{147.190pt}}
\put(1354.0,848.0){\rule[-0.200pt]{0.400pt}{4.818pt}}
\put(1354.0,105.0){\rule[-0.200pt]{0.400pt}{4.818pt}}
\put(1354,72){\makebox(0,0){ 13}}
\put(1354.0,848.0){\rule[-0.200pt]{0.400pt}{4.818pt}}
\put(1444.0,105.0){\rule[-0.200pt]{0.400pt}{183.807pt}}
\put(1444.0,105.0){\rule[-0.200pt]{0.400pt}{4.818pt}}
\put(1444,72){\makebox(0,0){ 14}}
\put(1444.0,848.0){\rule[-0.200pt]{0.400pt}{4.818pt}}
\put(185.0,105.0){\rule[-0.200pt]{0.400pt}{183.807pt}}
\put(185.0,105.0){\rule[-0.200pt]{303.293pt}{0.400pt}}
\put(1444.0,105.0){\rule[-0.200pt]{0.400pt}{183.807pt}}
\put(185.0,868.0){\rule[-0.200pt]{303.293pt}{0.400pt}}
\put(56,486){\makebox(0,0){$\sigma$}}
\put(814,23){\makebox(0,0){$d$}}
\put(1312,832){\makebox(0,0)[r]{$\scriptstyle \phi=0.0027$}}
\put(1328.0,832.0){\rule[-0.200pt]{20.236pt}{0.400pt}}
\put(185,105){\usebox{\plotpoint}}
\multiput(275.00,105.58)(0.981,0.498){89}{\rule{0.883pt}{0.120pt}}
\multiput(275.00,104.17)(88.168,46.000){2}{\rule{0.441pt}{0.400pt}}
\multiput(365.58,151.00)(0.499,1.269){177}{\rule{0.120pt}{1.113pt}}
\multiput(364.17,151.00)(90.000,225.689){2}{\rule{0.400pt}{0.557pt}}
\multiput(455.58,379.00)(0.499,1.186){177}{\rule{0.120pt}{1.047pt}}
\multiput(454.17,379.00)(90.000,210.828){2}{\rule{0.400pt}{0.523pt}}
\multiput(545.58,592.00)(0.499,1.124){177}{\rule{0.120pt}{0.998pt}}
\multiput(544.17,592.00)(90.000,199.929){2}{\rule{0.400pt}{0.499pt}}
\put(635,792.17){\rule{18.100pt}{0.400pt}}
\multiput(635.00,793.17)(52.433,-2.000){2}{\rule{9.050pt}{0.400pt}}
\multiput(725.58,788.41)(0.499,-0.957){177}{\rule{0.120pt}{0.864pt}}
\multiput(724.17,790.21)(90.000,-170.206){2}{\rule{0.400pt}{0.432pt}}
\multiput(815.58,616.13)(0.499,-1.041){175}{\rule{0.120pt}{0.931pt}}
\multiput(814.17,618.07)(89.000,-183.067){2}{\rule{0.400pt}{0.466pt}}
\multiput(904.58,431.98)(0.499,-0.784){177}{\rule{0.120pt}{0.727pt}}
\multiput(903.17,433.49)(90.000,-139.492){2}{\rule{0.400pt}{0.363pt}}
\multiput(994.58,291.39)(0.499,-0.661){177}{\rule{0.120pt}{0.629pt}}
\multiput(993.17,292.69)(90.000,-117.695){2}{\rule{0.400pt}{0.314pt}}
\multiput(1084.00,173.92)(1.895,-0.496){45}{\rule{1.600pt}{0.120pt}}
\multiput(1084.00,174.17)(86.679,-24.000){2}{\rule{0.800pt}{0.400pt}}
\put(185,105){\raisebox{-.8pt}{\makebox(0,0){$\scriptstyle \Diamond$}}}
\put(275,105){\raisebox{-.8pt}{\makebox(0,0){$\scriptstyle \Diamond$}}}
\put(365,151){\raisebox{-.8pt}{\makebox(0,0){$\scriptstyle \Diamond$}}}
\put(455,379){\raisebox{-.8pt}{\makebox(0,0){$\scriptstyle \Diamond$}}}
\put(545,592){\raisebox{-.8pt}{\makebox(0,0){$\scriptstyle \Diamond$}}}
\put(635,794){\raisebox{-.8pt}{\makebox(0,0){$\scriptstyle \Diamond$}}}
\put(725,792){\raisebox{-.8pt}{\makebox(0,0){$\scriptstyle \Diamond$}}}
\put(815,620){\raisebox{-.8pt}{\makebox(0,0){$\scriptstyle \Diamond$}}}
\put(904,435){\raisebox{-.8pt}{\makebox(0,0){$\scriptstyle \Diamond$}}}
\put(994,294){\raisebox{-.8pt}{\makebox(0,0){$\scriptstyle \Diamond$}}}
\put(1084,175){\raisebox{-.8pt}{\makebox(0,0){$\scriptstyle \Diamond$}}}
\put(1174,151){\raisebox{-.8pt}{\makebox(0,0){$\scriptstyle \Diamond$}}}
\put(1370,832){\raisebox{-.8pt}{\makebox(0,0){$\scriptstyle \Diamond$}}}
\put(185.0,105.0){\rule[-0.200pt]{21.681pt}{0.400pt}}
\put(1312,799){\makebox(0,0)[r]{$\scriptstyle \phi=0.027$}}
\multiput(1328,799)(20.756,0.000){5}{\usebox{\plotpoint}}
\put(1412,799){\usebox{\plotpoint}}
\put(185,105){\usebox{\plotpoint}}
\multiput(185,105)(20.756,0.000){5}{\usebox{\plotpoint}}
\multiput(275,105)(20.735,0.922){4}{\usebox{\plotpoint}}
\multiput(365,109)(18.808,8.777){5}{\usebox{\plotpoint}}
\multiput(455,151)(15.685,13.593){6}{\usebox{\plotpoint}}
\multiput(545,229)(16.032,13.182){5}{\usebox{\plotpoint}}
\multiput(635,303)(20.473,3.412){5}{\usebox{\plotpoint}}
\multiput(725,318)(18.398,-9.608){5}{\usebox{\plotpoint}}
\multiput(815,271)(17.030,-11.864){5}{\usebox{\plotpoint}}
\multiput(904,209)(18.888,-8.604){5}{\usebox{\plotpoint}}
\multiput(994,168)(19.819,-6.166){4}{\usebox{\plotpoint}}
\multiput(1084,140)(20.261,-4.503){5}{\usebox{\plotpoint}}
\multiput(1174,120)(20.674,-1.838){4}{\usebox{\plotpoint}}
\multiput(1264,112)(20.724,-1.151){4}{\usebox{\plotpoint}}
\put(1354,107){\usebox{\plotpoint}}
\put(185,105){\makebox(0,0){$\scriptstyle \scriptstyle +$}}
\put(275,105){\makebox(0,0){$\scriptstyle \scriptstyle +$}}
\put(365,109){\makebox(0,0){$\scriptstyle \scriptstyle +$}}
\put(455,151){\makebox(0,0){$\scriptstyle \scriptstyle +$}}
\put(545,229){\makebox(0,0){$\scriptstyle \scriptstyle +$}}
\put(635,303){\makebox(0,0){$\scriptstyle \scriptstyle +$}}
\put(725,318){\makebox(0,0){$\scriptstyle \scriptstyle +$}}
\put(815,271){\makebox(0,0){$\scriptstyle \scriptstyle +$}}
\put(904,209){\makebox(0,0){$\scriptstyle \scriptstyle +$}}
\put(994,168){\makebox(0,0){$\scriptstyle \scriptstyle +$}}
\put(1084,140){\makebox(0,0){$\scriptstyle \scriptstyle +$}}
\put(1174,120){\makebox(0,0){$\scriptstyle \scriptstyle +$}}
\put(1264,112){\makebox(0,0){$\scriptstyle \scriptstyle +$}}
\put(1354,107){\makebox(0,0){$\scriptstyle \scriptstyle +$}}
\put(1370,799){\makebox(0,0){$\scriptstyle \scriptstyle +$}}
\sbox{\plotpoint}{\rule[-0.400pt]{0.800pt}{0.800pt}}%
\sbox{\plotpoint}{\rule[-0.200pt]{0.400pt}{0.400pt}}%
\put(1312,766){\makebox(0,0)[r]{$\scriptstyle \phi=0.136$}}
\sbox{\plotpoint}{\rule[-0.400pt]{0.800pt}{0.800pt}}%
\put(1328.0,766.0){\rule[-0.400pt]{20.236pt}{0.800pt}}
\put(185,105){\usebox{\plotpoint}}
\multiput(365.00,106.40)(4.428,0.512){15}{\rule{6.745pt}{0.123pt}}
\multiput(365.00,103.34)(75.999,11.000){2}{\rule{3.373pt}{0.800pt}}
\multiput(455.00,117.41)(1.106,0.502){75}{\rule{1.956pt}{0.121pt}}
\multiput(455.00,114.34)(85.940,41.000){2}{\rule{0.978pt}{0.800pt}}
\multiput(545.00,158.41)(1.577,0.504){51}{\rule{2.683pt}{0.121pt}}
\multiput(545.00,155.34)(84.432,29.000){2}{\rule{1.341pt}{0.800pt}}
\multiput(635.00,187.41)(1.577,0.504){51}{\rule{2.683pt}{0.121pt}}
\multiput(635.00,184.34)(84.432,29.000){2}{\rule{1.341pt}{0.800pt}}
\multiput(725.00,213.09)(1.229,-0.503){67}{\rule{2.146pt}{0.121pt}}
\multiput(725.00,213.34)(85.546,-37.000){2}{\rule{1.073pt}{0.800pt}}
\multiput(815.00,176.09)(2.076,-0.505){37}{\rule{3.436pt}{0.122pt}}
\multiput(815.00,176.34)(81.868,-22.000){2}{\rule{1.718pt}{0.800pt}}
\multiput(904.00,154.09)(1.838,-0.504){43}{\rule{3.080pt}{0.121pt}}
\multiput(904.00,154.34)(83.607,-25.000){2}{\rule{1.540pt}{0.800pt}}
\multiput(994.00,129.09)(3.147,-0.508){23}{\rule{5.000pt}{0.122pt}}
\multiput(994.00,129.34)(79.622,-15.000){2}{\rule{2.500pt}{0.800pt}}
\multiput(1084.00,114.08)(7.745,-0.526){7}{\rule{10.486pt}{0.127pt}}
\multiput(1084.00,114.34)(68.236,-7.000){2}{\rule{5.243pt}{0.800pt}}
\put(1174,105.84){\rule{21.681pt}{0.800pt}}
\multiput(1174.00,107.34)(45.000,-3.000){2}{\rule{10.840pt}{0.800pt}}
\put(1264,103.84){\rule{21.681pt}{0.800pt}}
\multiput(1264.00,104.34)(45.000,-1.000){2}{\rule{10.840pt}{0.800pt}}
\put(185.0,105.0){\rule[-0.400pt]{43.362pt}{0.800pt}}
\put(185,105){\raisebox{-.8pt}{\makebox(0,0){$\scriptstyle \Box$}}}
\put(275,105){\raisebox{-.8pt}{\makebox(0,0){$\scriptstyle \Box$}}}
\put(365,105){\raisebox{-.8pt}{\makebox(0,0){$\scriptstyle \Box$}}}
\put(455,116){\raisebox{-.8pt}{\makebox(0,0){$\scriptstyle \Box$}}}
\put(545,157){\raisebox{-.8pt}{\makebox(0,0){$\scriptstyle \Box$}}}
\put(635,186){\raisebox{-.8pt}{\makebox(0,0){$\scriptstyle \Box$}}}
\put(725,215){\raisebox{-.8pt}{\makebox(0,0){$\scriptstyle \Box$}}}
\put(815,178){\raisebox{-.8pt}{\makebox(0,0){$\scriptstyle \Box$}}}
\put(904,156){\raisebox{-.8pt}{\makebox(0,0){$\scriptstyle \Box$}}}
\put(994,131){\raisebox{-.8pt}{\makebox(0,0){$\scriptstyle \Box$}}}
\put(1084,116){\raisebox{-.8pt}{\makebox(0,0){$\scriptstyle \Box$}}}
\put(1174,109){\raisebox{-.8pt}{\makebox(0,0){$\scriptstyle \Box$}}}
\put(1264,106){\raisebox{-.8pt}{\makebox(0,0){$\scriptstyle \Box$}}}
\put(1354,105){\raisebox{-.8pt}{\makebox(0,0){$\scriptstyle \Box$}}}
\put(1444,105){\raisebox{-.8pt}{\makebox(0,0){$\scriptstyle \Box$}}}
\put(1370,766){\raisebox{-.8pt}{\makebox(0,0){$\scriptstyle \Box$}}}
\put(1354.0,105.0){\rule[-0.400pt]{21.681pt}{0.800pt}}
\sbox{\plotpoint}{\rule[-0.500pt]{1.000pt}{1.000pt}}%
\sbox{\plotpoint}{\rule[-0.200pt]{0.400pt}{0.400pt}}%
\put(1312,733){\makebox(0,0)[r]{$\scriptstyle \phi=0.407$}}
\sbox{\plotpoint}{\rule[-0.500pt]{1.000pt}{1.000pt}}%
\multiput(1328,733)(20.756,0.000){5}{\usebox{\plotpoint}}
\put(1412,733){\usebox{\plotpoint}}
\put(185,105){\usebox{\plotpoint}}
\multiput(185,105)(20.756,0.000){5}{\usebox{\plotpoint}}
\multiput(275,105)(20.756,0.000){4}{\usebox{\plotpoint}}
\multiput(365,105)(20.750,0.461){5}{\usebox{\plotpoint}}
\multiput(455,107)(20.395,3.852){4}{\usebox{\plotpoint}}
\multiput(545,124)(20.352,4.070){4}{\usebox{\plotpoint}}
\multiput(635,142)(19.755,6.366){5}{\usebox{\plotpoint}}
\multiput(725,171)(20.055,-5.348){4}{\usebox{\plotpoint}}
\multiput(815,147)(20.387,-3.894){5}{\usebox{\plotpoint}}
\multiput(904,130)(20.352,-4.070){4}{\usebox{\plotpoint}}
\multiput(994,112)(20.724,-1.151){5}{\usebox{\plotpoint}}
\multiput(1084,107)(20.754,-0.231){4}{\usebox{\plotpoint}}
\multiput(1174,106)(20.754,-0.231){4}{\usebox{\plotpoint}}
\put(1264,105){\usebox{\plotpoint}}
\put(185,105){\makebox(0,0){$\scriptstyle \times$}}
\put(275,105){\makebox(0,0){$\scriptstyle \times$}}
\put(365,105){\makebox(0,0){$\scriptstyle \times$}}
\put(455,107){\makebox(0,0){$\scriptstyle \times$}}
\put(545,124){\makebox(0,0){$\scriptstyle \times$}}
\put(635,142){\makebox(0,0){$\scriptstyle \times$}}
\put(725,171){\makebox(0,0){$\scriptstyle \times$}}
\put(815,147){\makebox(0,0){$\scriptstyle \times$}}
\put(904,130){\makebox(0,0){$\scriptstyle \times$}}
\put(994,112){\makebox(0,0){$\scriptstyle \times$}}
\put(1084,107){\makebox(0,0){$\scriptstyle \times$}}
\put(1174,106){\makebox(0,0){$\scriptstyle \times$}}
\put(1264,105){\makebox(0,0){$\scriptstyle \times$}}
\put(1370,733){\makebox(0,0){$\scriptstyle \times$}}
\sbox{\plotpoint}{\rule[-0.200pt]{0.400pt}{0.400pt}}%
\put(185.0,105.0){\rule[-0.200pt]{0.400pt}{183.807pt}}
\put(185.0,105.0){\rule[-0.200pt]{303.293pt}{0.400pt}}
\put(1444.0,105.0){\rule[-0.200pt]{0.400pt}{183.807pt}}
\put(185.0,868.0){\rule[-0.200pt]{303.293pt}{0.400pt}}
\end{picture}

%% file: figs/activation1.tex
\setlength{\unitlength}{0.240900pt}
\ifx\plotpoint\undefined\newsavebox{\plotpoint}\fi
\sbox{\plotpoint}{\rule[-0.200pt]{0.400pt}{0.400pt}}%
\begin{picture}(1500,900)(0,0)
\font\gnuplot=cmr10 at 8pt
\gnuplot
\sbox{\plotpoint}{\rule[-0.200pt]{0.400pt}{0.400pt}}%
\put(169.0,105.0){\rule[-0.200pt]{4.818pt}{0.400pt}}
\put(153,105){\makebox(0,0)[r]{ 0}}
\put(1424.0,105.0){\rule[-0.200pt]{4.818pt}{0.400pt}}
\put(169.0,181.0){\rule[-0.200pt]{4.818pt}{0.400pt}}
\put(153,181){\makebox(0,0)[r]{ 0.1}}
\put(1424.0,181.0){\rule[-0.200pt]{4.818pt}{0.400pt}}
\put(169.0,258.0){\rule[-0.200pt]{4.818pt}{0.400pt}}
\put(153,258){\makebox(0,0)[r]{ 0.2}}
\put(1424.0,258.0){\rule[-0.200pt]{4.818pt}{0.400pt}}
\put(169.0,334.0){\rule[-0.200pt]{4.818pt}{0.400pt}}
\put(153,334){\makebox(0,0)[r]{ 0.3}}
\put(1424.0,334.0){\rule[-0.200pt]{4.818pt}{0.400pt}}
\put(169.0,410.0){\rule[-0.200pt]{4.818pt}{0.400pt}}
\put(153,410){\makebox(0,0)[r]{ 0.4}}
\put(1424.0,410.0){\rule[-0.200pt]{4.818pt}{0.400pt}}
\put(169.0,487.0){\rule[-0.200pt]{4.818pt}{0.400pt}}
\put(153,487){\makebox(0,0)[r]{ 0.5}}
\put(1424.0,487.0){\rule[-0.200pt]{4.818pt}{0.400pt}}
\put(169.0,563.0){\rule[-0.200pt]{4.818pt}{0.400pt}}
\put(153,563){\makebox(0,0)[r]{ 0.6}}
\put(1424.0,563.0){\rule[-0.200pt]{4.818pt}{0.400pt}}
\put(169.0,639.0){\rule[-0.200pt]{4.818pt}{0.400pt}}
\put(153,639){\makebox(0,0)[r]{ 0.7}}
\put(1424.0,639.0){\rule[-0.200pt]{4.818pt}{0.400pt}}
\put(169.0,715.0){\rule[-0.200pt]{4.818pt}{0.400pt}}
\put(153,715){\makebox(0,0)[r]{ 0.8}}
\put(1424.0,715.0){\rule[-0.200pt]{4.818pt}{0.400pt}}
\put(169.0,792.0){\rule[-0.200pt]{4.818pt}{0.400pt}}
\put(153,792){\makebox(0,0)[r]{ 0.9}}
\put(1424.0,792.0){\rule[-0.200pt]{4.818pt}{0.400pt}}
\put(169.0,868.0){\rule[-0.200pt]{4.818pt}{0.400pt}}
\put(153,868){\makebox(0,0)[r]{ 1}}
\put(1424.0,868.0){\rule[-0.200pt]{4.818pt}{0.400pt}}
\put(169.0,105.0){\rule[-0.200pt]{0.400pt}{4.818pt}}
\put(169,72){\makebox(0,0){ 0}}
\put(169.0,848.0){\rule[-0.200pt]{0.400pt}{4.818pt}}
\put(297.0,105.0){\rule[-0.200pt]{0.400pt}{4.818pt}}
\put(297,72){\makebox(0,0){ 20}}
\put(297.0,848.0){\rule[-0.200pt]{0.400pt}{4.818pt}}
\put(424.0,105.0){\rule[-0.200pt]{0.400pt}{4.818pt}}
\put(424,72){\makebox(0,0){ 40}}
\put(424.0,848.0){\rule[-0.200pt]{0.400pt}{4.818pt}}
\put(552.0,105.0){\rule[-0.200pt]{0.400pt}{4.818pt}}
\put(552,72){\makebox(0,0){ 60}}
\put(552.0,848.0){\rule[-0.200pt]{0.400pt}{4.818pt}}
\put(679.0,105.0){\rule[-0.200pt]{0.400pt}{4.818pt}}
\put(679,72){\makebox(0,0){ 80}}
\put(679.0,848.0){\rule[-0.200pt]{0.400pt}{4.818pt}}
\put(807.0,105.0){\rule[-0.200pt]{0.400pt}{4.818pt}}
\put(807,72){\makebox(0,0){ 100}}
\put(807.0,848.0){\rule[-0.200pt]{0.400pt}{4.818pt}}
\put(934.0,105.0){\rule[-0.200pt]{0.400pt}{4.818pt}}
\put(934,72){\makebox(0,0){ 120}}
\put(934.0,848.0){\rule[-0.200pt]{0.400pt}{4.818pt}}
\put(1062.0,105.0){\rule[-0.200pt]{0.400pt}{4.818pt}}
\put(1062,72){\makebox(0,0){ 140}}
\put(1062.0,848.0){\rule[-0.200pt]{0.400pt}{4.818pt}}
\put(1189.0,105.0){\rule[-0.200pt]{0.400pt}{4.818pt}}
\put(1189,72){\makebox(0,0){ 160}}
\put(1189.0,848.0){\rule[-0.200pt]{0.400pt}{4.818pt}}
\put(1317.0,105.0){\rule[-0.200pt]{0.400pt}{4.818pt}}
\put(1317,72){\makebox(0,0){ 180}}
\put(1317.0,848.0){\rule[-0.200pt]{0.400pt}{4.818pt}}
\put(1444.0,105.0){\rule[-0.200pt]{0.400pt}{4.818pt}}
\put(1444,72){\makebox(0,0){ 200}}
\put(1444.0,848.0){\rule[-0.200pt]{0.400pt}{4.818pt}}
\put(169.0,105.0){\rule[-0.200pt]{0.400pt}{183.807pt}}
\put(169.0,105.0){\rule[-0.200pt]{307.147pt}{0.400pt}}
\put(1444.0,105.0){\rule[-0.200pt]{0.400pt}{183.807pt}}
\put(169.0,868.0){\rule[-0.200pt]{307.147pt}{0.400pt}}
\put(56,486){\makebox(0,0){$\alpha$}}
\put(806,23){\makebox(0,0){$t$}}
\put(1312,832){\makebox(0,0)[r]{$\scriptstyle \phi=0.136, \eta=2$ }}
\put(1328.0,832.0){\rule[-0.200pt]{20.236pt}{0.400pt}}
\put(169,106){\usebox{\plotpoint}}
\put(169,105.67){\rule{1.445pt}{0.400pt}}
\multiput(169.00,105.17)(3.000,1.000){2}{\rule{0.723pt}{0.400pt}}
\put(175,106.67){\rule{1.686pt}{0.400pt}}
\multiput(175.00,106.17)(3.500,1.000){2}{\rule{0.843pt}{0.400pt}}
\put(182,107.67){\rule{1.445pt}{0.400pt}}
\multiput(182.00,107.17)(3.000,1.000){2}{\rule{0.723pt}{0.400pt}}
\put(188,108.67){\rule{1.686pt}{0.400pt}}
\multiput(188.00,108.17)(3.500,1.000){2}{\rule{0.843pt}{0.400pt}}
\put(195,110.17){\rule{1.300pt}{0.400pt}}
\multiput(195.00,109.17)(3.302,2.000){2}{\rule{0.650pt}{0.400pt}}
\put(201,112.17){\rule{1.300pt}{0.400pt}}
\multiput(201.00,111.17)(3.302,2.000){2}{\rule{0.650pt}{0.400pt}}
\put(207,114.17){\rule{1.500pt}{0.400pt}}
\multiput(207.00,113.17)(3.887,2.000){2}{\rule{0.750pt}{0.400pt}}
\multiput(214.00,116.61)(1.132,0.447){3}{\rule{0.900pt}{0.108pt}}
\multiput(214.00,115.17)(4.132,3.000){2}{\rule{0.450pt}{0.400pt}}
\put(220,119.17){\rule{1.300pt}{0.400pt}}
\multiput(220.00,118.17)(3.302,2.000){2}{\rule{0.650pt}{0.400pt}}
\multiput(226.00,121.61)(1.355,0.447){3}{\rule{1.033pt}{0.108pt}}
\multiput(226.00,120.17)(4.855,3.000){2}{\rule{0.517pt}{0.400pt}}
\multiput(233.00,124.61)(1.132,0.447){3}{\rule{0.900pt}{0.108pt}}
\multiput(233.00,123.17)(4.132,3.000){2}{\rule{0.450pt}{0.400pt}}
\multiput(239.00,127.61)(1.355,0.447){3}{\rule{1.033pt}{0.108pt}}
\multiput(239.00,126.17)(4.855,3.000){2}{\rule{0.517pt}{0.400pt}}
\multiput(246.00,130.60)(0.774,0.468){5}{\rule{0.700pt}{0.113pt}}
\multiput(246.00,129.17)(4.547,4.000){2}{\rule{0.350pt}{0.400pt}}
\multiput(252.00,134.61)(1.132,0.447){3}{\rule{0.900pt}{0.108pt}}
\multiput(252.00,133.17)(4.132,3.000){2}{\rule{0.450pt}{0.400pt}}
\multiput(258.00,137.60)(0.920,0.468){5}{\rule{0.800pt}{0.113pt}}
\multiput(258.00,136.17)(5.340,4.000){2}{\rule{0.400pt}{0.400pt}}
\multiput(265.00,141.61)(1.132,0.447){3}{\rule{0.900pt}{0.108pt}}
\multiput(265.00,140.17)(4.132,3.000){2}{\rule{0.450pt}{0.400pt}}
\multiput(271.00,144.59)(0.599,0.477){7}{\rule{0.580pt}{0.115pt}}
\multiput(271.00,143.17)(4.796,5.000){2}{\rule{0.290pt}{0.400pt}}
\multiput(277.00,149.60)(0.920,0.468){5}{\rule{0.800pt}{0.113pt}}
\multiput(277.00,148.17)(5.340,4.000){2}{\rule{0.400pt}{0.400pt}}
\multiput(284.00,153.59)(0.599,0.477){7}{\rule{0.580pt}{0.115pt}}
\multiput(284.00,152.17)(4.796,5.000){2}{\rule{0.290pt}{0.400pt}}
\multiput(290.00,158.59)(0.710,0.477){7}{\rule{0.660pt}{0.115pt}}
\multiput(290.00,157.17)(5.630,5.000){2}{\rule{0.330pt}{0.400pt}}
\multiput(297.00,163.59)(0.599,0.477){7}{\rule{0.580pt}{0.115pt}}
\multiput(297.00,162.17)(4.796,5.000){2}{\rule{0.290pt}{0.400pt}}
\multiput(303.00,168.59)(0.599,0.477){7}{\rule{0.580pt}{0.115pt}}
\multiput(303.00,167.17)(4.796,5.000){2}{\rule{0.290pt}{0.400pt}}
\multiput(309.00,173.59)(0.710,0.477){7}{\rule{0.660pt}{0.115pt}}
\multiput(309.00,172.17)(5.630,5.000){2}{\rule{0.330pt}{0.400pt}}
\multiput(316.00,178.59)(0.491,0.482){9}{\rule{0.500pt}{0.116pt}}
\multiput(316.00,177.17)(4.962,6.000){2}{\rule{0.250pt}{0.400pt}}
\multiput(322.00,184.59)(0.599,0.477){7}{\rule{0.580pt}{0.115pt}}
\multiput(322.00,183.17)(4.796,5.000){2}{\rule{0.290pt}{0.400pt}}
\multiput(328.00,189.59)(0.581,0.482){9}{\rule{0.567pt}{0.116pt}}
\multiput(328.00,188.17)(5.824,6.000){2}{\rule{0.283pt}{0.400pt}}
\multiput(335.00,195.59)(0.491,0.482){9}{\rule{0.500pt}{0.116pt}}
\multiput(335.00,194.17)(4.962,6.000){2}{\rule{0.250pt}{0.400pt}}
\multiput(341.00,201.59)(0.581,0.482){9}{\rule{0.567pt}{0.116pt}}
\multiput(341.00,200.17)(5.824,6.000){2}{\rule{0.283pt}{0.400pt}}
\multiput(348.00,207.59)(0.491,0.482){9}{\rule{0.500pt}{0.116pt}}
\multiput(348.00,206.17)(4.962,6.000){2}{\rule{0.250pt}{0.400pt}}
\multiput(354.59,213.00)(0.482,0.581){9}{\rule{0.116pt}{0.567pt}}
\multiput(353.17,213.00)(6.000,5.824){2}{\rule{0.400pt}{0.283pt}}
\multiput(360.00,220.59)(0.581,0.482){9}{\rule{0.567pt}{0.116pt}}
\multiput(360.00,219.17)(5.824,6.000){2}{\rule{0.283pt}{0.400pt}}
\multiput(367.59,226.00)(0.482,0.581){9}{\rule{0.116pt}{0.567pt}}
\multiput(366.17,226.00)(6.000,5.824){2}{\rule{0.400pt}{0.283pt}}
\multiput(373.59,233.00)(0.482,0.581){9}{\rule{0.116pt}{0.567pt}}
\multiput(372.17,233.00)(6.000,5.824){2}{\rule{0.400pt}{0.283pt}}
\multiput(379.00,240.59)(0.492,0.485){11}{\rule{0.500pt}{0.117pt}}
\multiput(379.00,239.17)(5.962,7.000){2}{\rule{0.250pt}{0.400pt}}
\multiput(386.59,247.00)(0.482,0.581){9}{\rule{0.116pt}{0.567pt}}
\multiput(385.17,247.00)(6.000,5.824){2}{\rule{0.400pt}{0.283pt}}
\multiput(392.00,254.59)(0.492,0.485){11}{\rule{0.500pt}{0.117pt}}
\multiput(392.00,253.17)(5.962,7.000){2}{\rule{0.250pt}{0.400pt}}
\multiput(399.59,261.00)(0.482,0.671){9}{\rule{0.116pt}{0.633pt}}
\multiput(398.17,261.00)(6.000,6.685){2}{\rule{0.400pt}{0.317pt}}
\multiput(405.59,269.00)(0.482,0.581){9}{\rule{0.116pt}{0.567pt}}
\multiput(404.17,269.00)(6.000,5.824){2}{\rule{0.400pt}{0.283pt}}
\multiput(411.59,276.00)(0.485,0.569){11}{\rule{0.117pt}{0.557pt}}
\multiput(410.17,276.00)(7.000,6.844){2}{\rule{0.400pt}{0.279pt}}
\multiput(418.59,284.00)(0.482,0.671){9}{\rule{0.116pt}{0.633pt}}
\multiput(417.17,284.00)(6.000,6.685){2}{\rule{0.400pt}{0.317pt}}
\multiput(424.59,292.00)(0.482,0.671){9}{\rule{0.116pt}{0.633pt}}
\multiput(423.17,292.00)(6.000,6.685){2}{\rule{0.400pt}{0.317pt}}
\multiput(430.59,300.00)(0.485,0.645){11}{\rule{0.117pt}{0.614pt}}
\multiput(429.17,300.00)(7.000,7.725){2}{\rule{0.400pt}{0.307pt}}
\multiput(437.59,309.00)(0.482,0.671){9}{\rule{0.116pt}{0.633pt}}
\multiput(436.17,309.00)(6.000,6.685){2}{\rule{0.400pt}{0.317pt}}
\multiput(443.59,317.00)(0.485,0.721){11}{\rule{0.117pt}{0.671pt}}
\multiput(442.17,317.00)(7.000,8.606){2}{\rule{0.400pt}{0.336pt}}
\multiput(450.59,327.00)(0.482,0.671){9}{\rule{0.116pt}{0.633pt}}
\multiput(449.17,327.00)(6.000,6.685){2}{\rule{0.400pt}{0.317pt}}
\multiput(456.59,335.00)(0.482,0.762){9}{\rule{0.116pt}{0.700pt}}
\multiput(455.17,335.00)(6.000,7.547){2}{\rule{0.400pt}{0.350pt}}
\multiput(462.59,344.00)(0.485,0.645){11}{\rule{0.117pt}{0.614pt}}
\multiput(461.17,344.00)(7.000,7.725){2}{\rule{0.400pt}{0.307pt}}
\multiput(469.59,353.00)(0.482,0.762){9}{\rule{0.116pt}{0.700pt}}
\multiput(468.17,353.00)(6.000,7.547){2}{\rule{0.400pt}{0.350pt}}
\multiput(475.59,362.00)(0.482,0.762){9}{\rule{0.116pt}{0.700pt}}
\multiput(474.17,362.00)(6.000,7.547){2}{\rule{0.400pt}{0.350pt}}
\multiput(481.59,371.00)(0.485,0.569){11}{\rule{0.117pt}{0.557pt}}
\multiput(480.17,371.00)(7.000,6.844){2}{\rule{0.400pt}{0.279pt}}
\multiput(488.59,379.00)(0.482,0.852){9}{\rule{0.116pt}{0.767pt}}
\multiput(487.17,379.00)(6.000,8.409){2}{\rule{0.400pt}{0.383pt}}
\multiput(494.59,389.00)(0.485,0.721){11}{\rule{0.117pt}{0.671pt}}
\multiput(493.17,389.00)(7.000,8.606){2}{\rule{0.400pt}{0.336pt}}
\multiput(501.59,399.00)(0.482,0.852){9}{\rule{0.116pt}{0.767pt}}
\multiput(500.17,399.00)(6.000,8.409){2}{\rule{0.400pt}{0.383pt}}
\multiput(507.59,409.00)(0.482,0.762){9}{\rule{0.116pt}{0.700pt}}
\multiput(506.17,409.00)(6.000,7.547){2}{\rule{0.400pt}{0.350pt}}
\multiput(513.59,418.00)(0.485,0.645){11}{\rule{0.117pt}{0.614pt}}
\multiput(512.17,418.00)(7.000,7.725){2}{\rule{0.400pt}{0.307pt}}
\multiput(520.59,427.00)(0.482,0.852){9}{\rule{0.116pt}{0.767pt}}
\multiput(519.17,427.00)(6.000,8.409){2}{\rule{0.400pt}{0.383pt}}
\multiput(526.59,437.00)(0.482,0.762){9}{\rule{0.116pt}{0.700pt}}
\multiput(525.17,437.00)(6.000,7.547){2}{\rule{0.400pt}{0.350pt}}
\multiput(532.59,446.00)(0.485,0.569){11}{\rule{0.117pt}{0.557pt}}
\multiput(531.17,446.00)(7.000,6.844){2}{\rule{0.400pt}{0.279pt}}
\multiput(539.59,454.00)(0.482,0.762){9}{\rule{0.116pt}{0.700pt}}
\multiput(538.17,454.00)(6.000,7.547){2}{\rule{0.400pt}{0.350pt}}
\multiput(545.59,463.00)(0.485,0.721){11}{\rule{0.117pt}{0.671pt}}
\multiput(544.17,463.00)(7.000,8.606){2}{\rule{0.400pt}{0.336pt}}
\multiput(552.59,473.00)(0.482,0.852){9}{\rule{0.116pt}{0.767pt}}
\multiput(551.17,473.00)(6.000,8.409){2}{\rule{0.400pt}{0.383pt}}
\multiput(558.59,483.00)(0.482,0.852){9}{\rule{0.116pt}{0.767pt}}
\multiput(557.17,483.00)(6.000,8.409){2}{\rule{0.400pt}{0.383pt}}
\multiput(564.59,493.00)(0.485,0.645){11}{\rule{0.117pt}{0.614pt}}
\multiput(563.17,493.00)(7.000,7.725){2}{\rule{0.400pt}{0.307pt}}
\multiput(571.59,502.00)(0.482,0.852){9}{\rule{0.116pt}{0.767pt}}
\multiput(570.17,502.00)(6.000,8.409){2}{\rule{0.400pt}{0.383pt}}
\multiput(577.59,512.00)(0.482,0.852){9}{\rule{0.116pt}{0.767pt}}
\multiput(576.17,512.00)(6.000,8.409){2}{\rule{0.400pt}{0.383pt}}
\multiput(583.59,522.00)(0.485,0.721){11}{\rule{0.117pt}{0.671pt}}
\multiput(582.17,522.00)(7.000,8.606){2}{\rule{0.400pt}{0.336pt}}
\multiput(590.59,532.00)(0.482,0.852){9}{\rule{0.116pt}{0.767pt}}
\multiput(589.17,532.00)(6.000,8.409){2}{\rule{0.400pt}{0.383pt}}
\multiput(596.59,542.00)(0.485,0.645){11}{\rule{0.117pt}{0.614pt}}
\multiput(595.17,542.00)(7.000,7.725){2}{\rule{0.400pt}{0.307pt}}
\multiput(603.59,551.00)(0.482,0.762){9}{\rule{0.116pt}{0.700pt}}
\multiput(602.17,551.00)(6.000,7.547){2}{\rule{0.400pt}{0.350pt}}
\multiput(609.59,560.00)(0.482,0.852){9}{\rule{0.116pt}{0.767pt}}
\multiput(608.17,560.00)(6.000,8.409){2}{\rule{0.400pt}{0.383pt}}
\multiput(615.59,570.00)(0.485,0.645){11}{\rule{0.117pt}{0.614pt}}
\multiput(614.17,570.00)(7.000,7.725){2}{\rule{0.400pt}{0.307pt}}
\multiput(622.59,579.00)(0.482,0.852){9}{\rule{0.116pt}{0.767pt}}
\multiput(621.17,579.00)(6.000,8.409){2}{\rule{0.400pt}{0.383pt}}
\multiput(628.59,589.00)(0.482,0.762){9}{\rule{0.116pt}{0.700pt}}
\multiput(627.17,589.00)(6.000,7.547){2}{\rule{0.400pt}{0.350pt}}
\multiput(634.59,598.00)(0.485,0.721){11}{\rule{0.117pt}{0.671pt}}
\multiput(633.17,598.00)(7.000,8.606){2}{\rule{0.400pt}{0.336pt}}
\multiput(641.59,608.00)(0.482,0.762){9}{\rule{0.116pt}{0.700pt}}
\multiput(640.17,608.00)(6.000,7.547){2}{\rule{0.400pt}{0.350pt}}
\multiput(647.59,617.00)(0.485,0.721){11}{\rule{0.117pt}{0.671pt}}
\multiput(646.17,617.00)(7.000,8.606){2}{\rule{0.400pt}{0.336pt}}
\multiput(654.59,627.00)(0.482,0.943){9}{\rule{0.116pt}{0.833pt}}
\multiput(653.17,627.00)(6.000,9.270){2}{\rule{0.400pt}{0.417pt}}
\multiput(660.59,638.00)(0.482,0.943){9}{\rule{0.116pt}{0.833pt}}
\multiput(659.17,638.00)(6.000,9.270){2}{\rule{0.400pt}{0.417pt}}
\multiput(666.59,649.00)(0.485,0.721){11}{\rule{0.117pt}{0.671pt}}
\multiput(665.17,649.00)(7.000,8.606){2}{\rule{0.400pt}{0.336pt}}
\multiput(673.59,659.00)(0.482,0.762){9}{\rule{0.116pt}{0.700pt}}
\multiput(672.17,659.00)(6.000,7.547){2}{\rule{0.400pt}{0.350pt}}
\multiput(679.59,668.00)(0.482,0.762){9}{\rule{0.116pt}{0.700pt}}
\multiput(678.17,668.00)(6.000,7.547){2}{\rule{0.400pt}{0.350pt}}
\multiput(685.59,677.00)(0.485,0.645){11}{\rule{0.117pt}{0.614pt}}
\multiput(684.17,677.00)(7.000,7.725){2}{\rule{0.400pt}{0.307pt}}
\multiput(692.59,686.00)(0.482,0.852){9}{\rule{0.116pt}{0.767pt}}
\multiput(691.17,686.00)(6.000,8.409){2}{\rule{0.400pt}{0.383pt}}
\multiput(698.59,696.00)(0.485,0.721){11}{\rule{0.117pt}{0.671pt}}
\multiput(697.17,696.00)(7.000,8.606){2}{\rule{0.400pt}{0.336pt}}
\multiput(705.59,706.00)(0.482,0.762){9}{\rule{0.116pt}{0.700pt}}
\multiput(704.17,706.00)(6.000,7.547){2}{\rule{0.400pt}{0.350pt}}
\multiput(711.59,715.00)(0.482,0.671){9}{\rule{0.116pt}{0.633pt}}
\multiput(710.17,715.00)(6.000,6.685){2}{\rule{0.400pt}{0.317pt}}
\multiput(717.59,723.00)(0.485,0.569){11}{\rule{0.117pt}{0.557pt}}
\multiput(716.17,723.00)(7.000,6.844){2}{\rule{0.400pt}{0.279pt}}
\multiput(724.59,731.00)(0.482,0.762){9}{\rule{0.116pt}{0.700pt}}
\multiput(723.17,731.00)(6.000,7.547){2}{\rule{0.400pt}{0.350pt}}
\multiput(730.59,740.00)(0.482,0.762){9}{\rule{0.116pt}{0.700pt}}
\multiput(729.17,740.00)(6.000,7.547){2}{\rule{0.400pt}{0.350pt}}
\multiput(736.59,749.00)(0.485,0.645){11}{\rule{0.117pt}{0.614pt}}
\multiput(735.17,749.00)(7.000,7.725){2}{\rule{0.400pt}{0.307pt}}
\multiput(743.59,758.00)(0.482,0.671){9}{\rule{0.116pt}{0.633pt}}
\multiput(742.17,758.00)(6.000,6.685){2}{\rule{0.400pt}{0.317pt}}
\multiput(749.59,766.00)(0.485,0.569){11}{\rule{0.117pt}{0.557pt}}
\multiput(748.17,766.00)(7.000,6.844){2}{\rule{0.400pt}{0.279pt}}
\multiput(756.59,774.00)(0.482,0.671){9}{\rule{0.116pt}{0.633pt}}
\multiput(755.17,774.00)(6.000,6.685){2}{\rule{0.400pt}{0.317pt}}
\multiput(762.59,782.00)(0.482,0.671){9}{\rule{0.116pt}{0.633pt}}
\multiput(761.17,782.00)(6.000,6.685){2}{\rule{0.400pt}{0.317pt}}
\multiput(768.59,790.00)(0.485,0.569){11}{\rule{0.117pt}{0.557pt}}
\multiput(767.17,790.00)(7.000,6.844){2}{\rule{0.400pt}{0.279pt}}
\multiput(775.59,798.00)(0.482,0.762){9}{\rule{0.116pt}{0.700pt}}
\multiput(774.17,798.00)(6.000,7.547){2}{\rule{0.400pt}{0.350pt}}
\multiput(781.59,807.00)(0.482,0.671){9}{\rule{0.116pt}{0.633pt}}
\multiput(780.17,807.00)(6.000,6.685){2}{\rule{0.400pt}{0.317pt}}
\multiput(787.00,815.59)(0.492,0.485){11}{\rule{0.500pt}{0.117pt}}
\multiput(787.00,814.17)(5.962,7.000){2}{\rule{0.250pt}{0.400pt}}
\multiput(794.59,822.00)(0.482,0.671){9}{\rule{0.116pt}{0.633pt}}
\multiput(793.17,822.00)(6.000,6.685){2}{\rule{0.400pt}{0.317pt}}
\multiput(800.00,830.59)(0.492,0.485){11}{\rule{0.500pt}{0.117pt}}
\multiput(800.00,829.17)(5.962,7.000){2}{\rule{0.250pt}{0.400pt}}
\multiput(807.59,837.00)(0.482,0.581){9}{\rule{0.116pt}{0.567pt}}
\multiput(806.17,837.00)(6.000,5.824){2}{\rule{0.400pt}{0.283pt}}
\multiput(813.00,844.59)(0.491,0.482){9}{\rule{0.500pt}{0.116pt}}
\multiput(813.00,843.17)(4.962,6.000){2}{\rule{0.250pt}{0.400pt}}
\multiput(819.00,850.59)(0.710,0.477){7}{\rule{0.660pt}{0.115pt}}
\multiput(819.00,849.17)(5.630,5.000){2}{\rule{0.330pt}{0.400pt}}
\multiput(826.00,855.60)(0.774,0.468){5}{\rule{0.700pt}{0.113pt}}
\multiput(826.00,854.17)(4.547,4.000){2}{\rule{0.350pt}{0.400pt}}
\multiput(832.00,859.61)(1.132,0.447){3}{\rule{0.900pt}{0.108pt}}
\multiput(832.00,858.17)(4.132,3.000){2}{\rule{0.450pt}{0.400pt}}
\put(838,862.17){\rule{1.500pt}{0.400pt}}
\multiput(838.00,861.17)(3.887,2.000){2}{\rule{0.750pt}{0.400pt}}
\put(845,864.17){\rule{1.300pt}{0.400pt}}
\multiput(845.00,863.17)(3.302,2.000){2}{\rule{0.650pt}{0.400pt}}
\put(851,865.67){\rule{1.686pt}{0.400pt}}
\multiput(851.00,865.17)(3.500,1.000){2}{\rule{0.843pt}{0.400pt}}
\put(858,866.67){\rule{1.445pt}{0.400pt}}
\multiput(858.00,866.17)(3.000,1.000){2}{\rule{0.723pt}{0.400pt}}
\put(1312,799){\makebox(0,0)[r]{$\scriptstyle \phi=0.407, \eta=2$}}
\multiput(1328,799)(20.756,0.000){5}{\usebox{\plotpoint}}
\put(1412,799){\usebox{\plotpoint}}
\put(169,105){\usebox{\plotpoint}}
\put(169.00,105.00){\usebox{\plotpoint}}
\put(189.59,107.23){\usebox{\plotpoint}}
\put(210.12,110.00){\usebox{\plotpoint}}
\put(230.66,112.67){\usebox{\plotpoint}}
\put(250.93,116.82){\usebox{\plotpoint}}
\put(270.95,121.98){\usebox{\plotpoint}}
\put(290.98,127.28){\usebox{\plotpoint}}
\put(310.41,134.40){\usebox{\plotpoint}}
\put(330.10,140.90){\usebox{\plotpoint}}
\put(349.63,147.82){\usebox{\plotpoint}}
\put(368.83,155.61){\usebox{\plotpoint}}
\put(387.82,163.91){\usebox{\plotpoint}}
\put(406.88,171.94){\usebox{\plotpoint}}
\put(425.63,180.81){\usebox{\plotpoint}}
\put(444.01,190.43){\usebox{\plotpoint}}
\put(462.28,200.16){\usebox{\plotpoint}}
\put(480.24,210.50){\usebox{\plotpoint}}
\put(498.41,220.52){\usebox{\plotpoint}}
\put(515.91,231.66){\usebox{\plotpoint}}
\put(533.41,242.80){\usebox{\plotpoint}}
\put(551.17,253.52){\usebox{\plotpoint}}
\put(568.14,265.37){\usebox{\plotpoint}}
\put(585.12,277.21){\usebox{\plotpoint}}
\put(602.36,288.64){\usebox{\plotpoint}}
\put(619.34,300.48){\usebox{\plotpoint}}
\put(635.69,313.20){\usebox{\plotpoint}}
\put(652.71,325.08){\usebox{\plotpoint}}
\put(668.89,338.06){\usebox{\plotpoint}}
\put(685.06,351.05){\usebox{\plotpoint}}
\put(702.09,362.92){\usebox{\plotpoint}}
\put(717.76,376.44){\usebox{\plotpoint}}
\put(733.91,389.26){\usebox{\plotpoint}}
\put(750.32,401.94){\usebox{\plotpoint}}
\put(766.22,415.22){\usebox{\plotpoint}}
\put(781.88,428.73){\usebox{\plotpoint}}
\put(797.88,441.88){\usebox{\plotpoint}}
\put(813.95,454.95){\usebox{\plotpoint}}
\put(829.55,468.55){\usebox{\plotpoint}}
\put(845.14,482.14){\usebox{\plotpoint}}
\put(860.97,495.47){\usebox{\plotpoint}}
\put(876.32,509.42){\usebox{\plotpoint}}
\put(891.90,523.07){\usebox{\plotpoint}}
\put(907.89,536.20){\usebox{\plotpoint}}
\put(922.97,550.40){\usebox{\plotpoint}}
\put(938.67,563.89){\usebox{\plotpoint}}
\put(954.08,577.77){\usebox{\plotpoint}}
\put(969.83,591.20){\usebox{\plotpoint}}
\put(985.18,605.15){\usebox{\plotpoint}}
\put(1001.04,618.53){\usebox{\plotpoint}}
\put(1016.90,631.92){\usebox{\plotpoint}}
\put(1032.72,645.27){\usebox{\plotpoint}}
\put(1048.59,658.65){\usebox{\plotpoint}}
\put(1064.69,671.69){\usebox{\plotpoint}}
\put(1080.17,685.41){\usebox{\plotpoint}}
\put(1095.25,699.61){\usebox{\plotpoint}}
\put(1111.79,712.13){\usebox{\plotpoint}}
\put(1127.58,725.47){\usebox{\plotpoint}}
\put(1144.03,738.02){\usebox{\plotpoint}}
\put(1160.57,750.55){\usebox{\plotpoint}}
\put(1177.23,762.88){\usebox{\plotpoint}}
\put(1193.87,775.25){\usebox{\plotpoint}}
\put(1210.92,787.09){\usebox{\plotpoint}}
\put(1227.63,799.36){\usebox{\plotpoint}}
\put(1245.16,810.44){\usebox{\plotpoint}}
\put(1263.32,820.47){\usebox{\plotpoint}}
\put(1280.98,831.28){\usebox{\plotpoint}}
\put(1299.14,841.22){\usebox{\plotpoint}}
\put(1317.75,850.37){\usebox{\plotpoint}}
\put(1336.90,858.30){\usebox{\plotpoint}}
\put(1356.68,864.56){\usebox{\plotpoint}}
\put(1377.05,868.00){\usebox{\plotpoint}}
\put(1380,868){\usebox{\plotpoint}}
\sbox{\plotpoint}{\rule[-0.400pt]{0.800pt}{0.800pt}}%
\sbox{\plotpoint}{\rule[-0.200pt]{0.400pt}{0.400pt}}%
\put(1312,766){\makebox(0,0)[r]{$\scriptstyle \phi=0.136, \eta=1$}}
\sbox{\plotpoint}{\rule[-0.400pt]{0.800pt}{0.800pt}}%
\put(1328.0,766.0){\rule[-0.400pt]{20.236pt}{0.800pt}}
\put(169,106){\usebox{\plotpoint}}
\put(169,105.84){\rule{1.445pt}{0.800pt}}
\multiput(169.00,104.34)(3.000,3.000){2}{\rule{0.723pt}{0.800pt}}
\put(175,108.84){\rule{1.686pt}{0.800pt}}
\multiput(175.00,107.34)(3.500,3.000){2}{\rule{0.843pt}{0.800pt}}
\multiput(182.00,113.39)(0.462,0.536){5}{\rule{1.000pt}{0.129pt}}
\multiput(182.00,110.34)(3.924,6.000){2}{\rule{0.500pt}{0.800pt}}
\multiput(188.00,119.40)(0.475,0.526){7}{\rule{1.000pt}{0.127pt}}
\multiput(188.00,116.34)(4.924,7.000){2}{\rule{0.500pt}{0.800pt}}
\multiput(196.39,125.00)(0.536,0.685){5}{\rule{0.129pt}{1.267pt}}
\multiput(193.34,125.00)(6.000,5.371){2}{\rule{0.800pt}{0.633pt}}
\multiput(202.39,133.00)(0.536,0.797){5}{\rule{0.129pt}{1.400pt}}
\multiput(199.34,133.00)(6.000,6.094){2}{\rule{0.800pt}{0.700pt}}
\multiput(208.40,142.00)(0.526,0.738){7}{\rule{0.127pt}{1.343pt}}
\multiput(205.34,142.00)(7.000,7.213){2}{\rule{0.800pt}{0.671pt}}
\multiput(215.39,152.00)(0.536,1.132){5}{\rule{0.129pt}{1.800pt}}
\multiput(212.34,152.00)(6.000,8.264){2}{\rule{0.800pt}{0.900pt}}
\multiput(221.39,164.00)(0.536,1.132){5}{\rule{0.129pt}{1.800pt}}
\multiput(218.34,164.00)(6.000,8.264){2}{\rule{0.800pt}{0.900pt}}
\multiput(227.40,176.00)(0.526,0.913){7}{\rule{0.127pt}{1.571pt}}
\multiput(224.34,176.00)(7.000,8.738){2}{\rule{0.800pt}{0.786pt}}
\multiput(234.39,188.00)(0.536,1.467){5}{\rule{0.129pt}{2.200pt}}
\multiput(231.34,188.00)(6.000,10.434){2}{\rule{0.800pt}{1.100pt}}
\multiput(240.40,203.00)(0.526,1.088){7}{\rule{0.127pt}{1.800pt}}
\multiput(237.34,203.00)(7.000,10.264){2}{\rule{0.800pt}{0.900pt}}
\multiput(247.39,217.00)(0.536,1.467){5}{\rule{0.129pt}{2.200pt}}
\multiput(244.34,217.00)(6.000,10.434){2}{\rule{0.800pt}{1.100pt}}
\multiput(253.39,232.00)(0.536,1.579){5}{\rule{0.129pt}{2.333pt}}
\multiput(250.34,232.00)(6.000,11.157){2}{\rule{0.800pt}{1.167pt}}
\multiput(259.40,248.00)(0.526,1.438){7}{\rule{0.127pt}{2.257pt}}
\multiput(256.34,248.00)(7.000,13.315){2}{\rule{0.800pt}{1.129pt}}
\multiput(266.39,266.00)(0.536,1.913){5}{\rule{0.129pt}{2.733pt}}
\multiput(263.34,266.00)(6.000,13.327){2}{\rule{0.800pt}{1.367pt}}
\multiput(272.39,285.00)(0.536,1.802){5}{\rule{0.129pt}{2.600pt}}
\multiput(269.34,285.00)(6.000,12.604){2}{\rule{0.800pt}{1.300pt}}
\multiput(278.40,303.00)(0.526,1.526){7}{\rule{0.127pt}{2.371pt}}
\multiput(275.34,303.00)(7.000,14.078){2}{\rule{0.800pt}{1.186pt}}
\multiput(285.39,322.00)(0.536,2.025){5}{\rule{0.129pt}{2.867pt}}
\multiput(282.34,322.00)(6.000,14.050){2}{\rule{0.800pt}{1.433pt}}
\multiput(291.40,342.00)(0.526,1.789){7}{\rule{0.127pt}{2.714pt}}
\multiput(288.34,342.00)(7.000,16.366){2}{\rule{0.800pt}{1.357pt}}
\multiput(298.39,364.00)(0.536,2.248){5}{\rule{0.129pt}{3.133pt}}
\multiput(295.34,364.00)(6.000,15.497){2}{\rule{0.800pt}{1.567pt}}
\multiput(304.39,386.00)(0.536,2.248){5}{\rule{0.129pt}{3.133pt}}
\multiput(301.34,386.00)(6.000,15.497){2}{\rule{0.800pt}{1.567pt}}
\multiput(310.40,408.00)(0.526,1.964){7}{\rule{0.127pt}{2.943pt}}
\multiput(307.34,408.00)(7.000,17.892){2}{\rule{0.800pt}{1.471pt}}
\multiput(317.39,432.00)(0.536,2.472){5}{\rule{0.129pt}{3.400pt}}
\multiput(314.34,432.00)(6.000,16.943){2}{\rule{0.800pt}{1.700pt}}
\multiput(323.39,456.00)(0.536,2.472){5}{\rule{0.129pt}{3.400pt}}
\multiput(320.34,456.00)(6.000,16.943){2}{\rule{0.800pt}{1.700pt}}
\multiput(329.40,480.00)(0.526,1.876){7}{\rule{0.127pt}{2.829pt}}
\multiput(326.34,480.00)(7.000,17.129){2}{\rule{0.800pt}{1.414pt}}
\multiput(336.39,503.00)(0.536,2.583){5}{\rule{0.129pt}{3.533pt}}
\multiput(333.34,503.00)(6.000,17.666){2}{\rule{0.800pt}{1.767pt}}
\multiput(342.40,528.00)(0.526,2.052){7}{\rule{0.127pt}{3.057pt}}
\multiput(339.34,528.00)(7.000,18.655){2}{\rule{0.800pt}{1.529pt}}
\multiput(349.39,553.00)(0.536,2.583){5}{\rule{0.129pt}{3.533pt}}
\multiput(346.34,553.00)(6.000,17.666){2}{\rule{0.800pt}{1.767pt}}
\multiput(355.39,578.00)(0.536,2.806){5}{\rule{0.129pt}{3.800pt}}
\multiput(352.34,578.00)(6.000,19.113){2}{\rule{0.800pt}{1.900pt}}
\multiput(361.40,605.00)(0.526,1.964){7}{\rule{0.127pt}{2.943pt}}
\multiput(358.34,605.00)(7.000,17.892){2}{\rule{0.800pt}{1.471pt}}
\multiput(368.39,629.00)(0.536,2.583){5}{\rule{0.129pt}{3.533pt}}
\multiput(365.34,629.00)(6.000,17.666){2}{\rule{0.800pt}{1.767pt}}
\multiput(374.39,654.00)(0.536,2.472){5}{\rule{0.129pt}{3.400pt}}
\multiput(371.34,654.00)(6.000,16.943){2}{\rule{0.800pt}{1.700pt}}
\multiput(380.40,678.00)(0.526,1.876){7}{\rule{0.127pt}{2.829pt}}
\multiput(377.34,678.00)(7.000,17.129){2}{\rule{0.800pt}{1.414pt}}
\multiput(387.39,701.00)(0.536,2.360){5}{\rule{0.129pt}{3.267pt}}
\multiput(384.34,701.00)(6.000,16.220){2}{\rule{0.800pt}{1.633pt}}
\multiput(393.40,724.00)(0.526,1.876){7}{\rule{0.127pt}{2.829pt}}
\multiput(390.34,724.00)(7.000,17.129){2}{\rule{0.800pt}{1.414pt}}
\multiput(400.39,747.00)(0.536,1.913){5}{\rule{0.129pt}{2.733pt}}
\multiput(397.34,747.00)(6.000,13.327){2}{\rule{0.800pt}{1.367pt}}
\multiput(406.39,766.00)(0.536,1.802){5}{\rule{0.129pt}{2.600pt}}
\multiput(403.34,766.00)(6.000,12.604){2}{\rule{0.800pt}{1.300pt}}
\multiput(412.40,784.00)(0.526,1.438){7}{\rule{0.127pt}{2.257pt}}
\multiput(409.34,784.00)(7.000,13.315){2}{\rule{0.800pt}{1.129pt}}
\multiput(419.39,802.00)(0.536,1.467){5}{\rule{0.129pt}{2.200pt}}
\multiput(416.34,802.00)(6.000,10.434){2}{\rule{0.800pt}{1.100pt}}
\multiput(425.39,817.00)(0.536,1.467){5}{\rule{0.129pt}{2.200pt}}
\multiput(422.34,817.00)(6.000,10.434){2}{\rule{0.800pt}{1.100pt}}
\multiput(431.40,832.00)(0.526,0.913){7}{\rule{0.127pt}{1.571pt}}
\multiput(428.34,832.00)(7.000,8.738){2}{\rule{0.800pt}{0.786pt}}
\multiput(438.39,844.00)(0.536,0.909){5}{\rule{0.129pt}{1.533pt}}
\multiput(435.34,844.00)(6.000,6.817){2}{\rule{0.800pt}{0.767pt}}
\multiput(443.00,855.39)(0.574,0.536){5}{\rule{1.133pt}{0.129pt}}
\multiput(443.00,852.34)(4.648,6.000){2}{\rule{0.567pt}{0.800pt}}
\multiput(450.00,861.39)(0.462,0.536){5}{\rule{1.000pt}{0.129pt}}
\multiput(450.00,858.34)(3.924,6.000){2}{\rule{0.500pt}{0.800pt}}
\put(456,865.34){\rule{1.445pt}{0.800pt}}
\multiput(456.00,864.34)(3.000,2.000){2}{\rule{0.723pt}{0.800pt}}
\put(462.0,868.0){\rule[-0.400pt]{1.686pt}{0.800pt}}
\sbox{\plotpoint}{\rule[-0.500pt]{1.000pt}{1.000pt}}%
\sbox{\plotpoint}{\rule[-0.200pt]{0.400pt}{0.400pt}}%
\put(1312,733){\makebox(0,0)[r]{$\scriptstyle \phi=0.407, \eta=1$}}
\sbox{\plotpoint}{\rule[-0.500pt]{1.000pt}{1.000pt}}%
\multiput(1328,733)(20.756,0.000){5}{\usebox{\plotpoint}}
\put(1412,733){\usebox{\plotpoint}}
\put(169,106){\usebox{\plotpoint}}
\put(169.00,106.00){\usebox{\plotpoint}}
\put(189.46,109.42){\usebox{\plotpoint}}
\put(208.70,116.97){\usebox{\plotpoint}}
\put(226.64,127.36){\usebox{\plotpoint}}
\put(243.72,139.05){\usebox{\plotpoint}}
\put(259.11,152.95){\usebox{\plotpoint}}
\put(273.93,167.42){\usebox{\plotpoint}}
\put(287.69,182.92){\usebox{\plotpoint}}
\put(300.76,199.02){\usebox{\plotpoint}}
\put(313.31,215.55){\usebox{\plotpoint}}
\put(325.09,232.63){\usebox{\plotpoint}}
\put(336.70,249.83){\usebox{\plotpoint}}
\put(348.09,267.16){\usebox{\plotpoint}}
\put(358.03,285.38){\usebox{\plotpoint}}
\put(368.72,303.16){\usebox{\plotpoint}}
\put(378.29,321.57){\usebox{\plotpoint}}
\put(388.57,339.56){\usebox{\plotpoint}}
\put(398.33,357.85){\usebox{\plotpoint}}
\put(407.14,376.63){\usebox{\plotpoint}}
\put(416.82,394.97){\usebox{\plotpoint}}
\put(425.97,413.59){\usebox{\plotpoint}}
\put(434.98,432.26){\usebox{\plotpoint}}
\put(443.57,451.14){\usebox{\plotpoint}}
\put(452.51,469.86){\usebox{\plotpoint}}
\put(460.99,488.80){\usebox{\plotpoint}}
\put(470.06,507.47){\usebox{\plotpoint}}
\put(478.23,526.54){\usebox{\plotpoint}}
\put(486.81,545.44){\usebox{\plotpoint}}
\put(494.71,564.63){\usebox{\plotpoint}}
\put(502.78,583.75){\usebox{\plotpoint}}
\put(510.25,603.11){\usebox{\plotpoint}}
\put(518.64,622.09){\usebox{\plotpoint}}
\put(526.55,641.27){\usebox{\plotpoint}}
\put(534.92,660.26){\usebox{\plotpoint}}
\put(543.66,679.09){\usebox{\plotpoint}}
\put(553.17,697.53){\usebox{\plotpoint}}
\put(562.12,716.25){\usebox{\plotpoint}}
\put(572.28,734.34){\usebox{\plotpoint}}
\put(582.22,752.56){\usebox{\plotpoint}}
\put(593.56,769.93){\usebox{\plotpoint}}
\put(605.55,786.83){\usebox{\plotpoint}}
\put(617.45,803.80){\usebox{\plotpoint}}
\put(631.02,819.52){\usebox{\plotpoint}}
\put(645.43,834.43){\usebox{\plotpoint}}
\put(661.12,847.93){\usebox{\plotpoint}}
\put(678.27,859.51){\usebox{\plotpoint}}
\put(697.59,866.86){\usebox{\plotpoint}}
\put(711,868){\usebox{\plotpoint}}
\sbox{\plotpoint}{\rule[-0.200pt]{0.400pt}{0.400pt}}%
\put(169.0,105.0){\rule[-0.200pt]{0.400pt}{183.807pt}}
\put(169.0,105.0){\rule[-0.200pt]{307.147pt}{0.400pt}}
\put(1444.0,105.0){\rule[-0.200pt]{0.400pt}{183.807pt}}
\put(169.0,868.0){\rule[-0.200pt]{307.147pt}{0.400pt}}
\end{picture}